\documentclass[notitlepage]{article}

\usepackage{graphicx}
\usepackage{appendix}
\usepackage{newclude}
\usepackage{pdflscape}
\usepackage{enumitem}
\usepackage{pdfpages}
\usepackage{subfiles}
\usepackage{titlesec}

\usepackage[nohyperlinks]{acronym}

\graphicspath{ {./images/} }
\usepackage{amsmath}
\usepackage{amssymb}
\usepackage{mathabx}
\usepackage{relsize}

\usepackage[margin=1in]{geometry}
\usepackage{float}

\usepackage{sectsty}
\usepackage{comment}
\usepackage{parskip} % stops indentation of new paragraph
\usepackage{tabularx}
\usepackage{tabularray}
\usepackage[dvipsnames,table,xcdraw]{xcolor}
\usepackage{colortbl}
\usepackage{adjustbox}
\usepackage{longtable}
\usepackage{chngcntr}
\usepackage{array}

\usepackage{booktabs}
\counterwithin{figure}{section}
\usepackage{hyperref}
\hypersetup{colorlinks=true, linkcolor=black}
\usepackage{threeparttable}

\usepackage{tikz}
\usetikzlibrary{positioning, calc, shapes.geometric, shapes.multipart,
    shapes, arrows.meta, decorations.markings, decorations.pathreplacing, external, trees, matrix, backgrounds}
\tikzstyle{Arrow} = [
  thick,
  decoration={markings,
              mark=at position 1 with {\arrow[thick]{latex}}},
  shorten >= 3pt, 
  preaction = {decorate}
]
\usepackage[caption=false]{subfig}
\usepackage{subcaption}
\usepackage{authblk}

\usepackage{chngcntr}
\counterwithout{figure}{section}

\newcommand{\indep}{\perp \!\!\! \perp}

\newcolumntype{L}[1]{>{\raggedright\arraybackslash}p{#1}}
 
\newcolumntype{C}[1]{>{\centering\arraybackslash}p{#1}}
 
\newcolumntype{R}[1]{>{\raggedleft\arraybackslash}p{#1}}

\renewcommand{\arraystretch}{1.5}

\usepackage[
backend=biber,
style=numeric-comp,
sorting=none,
maxbibnames=3,
citetracker=true,
natbib=true,
giveninits,
url=false,
isbn=false,
doi=false,
maxnames=3,
minnames=3
]{biblatex}

\AtEveryBibitem{\clearfield{month}}
\AtEveryBibitem{\clearfield{day}}

\DeclareNameAlias{sortname}{given-family}
\DeclareNameAlias{author}{sortname}
\DeclareNameAlias{editor}{sortname}
\DeclareNameAlias{translator}{sortname}

\DeclareNameWrapperAlias{author}{sortname}
\DeclareNameWrapperAlias{editor}{sortname}
\DeclareNameWrapperAlias{translator}{sortname}

\renewrobustcmd*{\bibinitdelim}{\,}

\DeclareDelimFormat[biblist,bib]{nametitledelim}{\addcomma\space}

\DeclareFieldFormat*{title}{#1}

\renewbibmacro{in:}{%
  \ifentrytype{article}
    {}
    {\printtext{\bibstring{in}\intitlepunct}}}

\newcommand*{\volnumdelim}{\adddot}
\renewbibmacro*{volume+number+eid}{%
  \printfield{volume}%
  \setunit*{\volnumdelim}%
  \printfield{number}%
  \setunit{\addcomma\space}%
  \printfield{eid}}

\renewcommand*{\volnumdelim}{}
\DeclareFieldFormat[article, periodical]{number}{\mkbibparens{#1}}

\newcommand*{\jourdatedelim}{\addcomma\space}
\newcommand*{\jourserdelim}{\addcomma\space}
\newcommand*{\datevolnumdelim}{\newunitpunct}

\renewbibmacro*{journal+issuetitle}{%
  \usebibmacro{journal}%
  \setunit*{\jourdatedelim}%
  \iffieldundef{series}
    {}
    {\setunit*{\jourserdelim}%
     \printfield{series}%
     \setunit{\jourdatedelim}}%
  \usebibmacro{date}%
  \setunit{\datevolnumdelim}%
  \usebibmacro{volume+number+eid}%
  \setunit{\addcolon\space}%
  \usebibmacro{issue}%
  \newunit}

\DefineBibliographyStrings{english}{
  pages = {p\adddot},
}

\newacro{RR}{respiratory rate}
\newacro{MSM}{marginal structural model}
\newacro{DAG}{directed acyclic graph}
\newacro{OMOP}{Observational Medical Outcomes Partnership}
\newacro{CDM}{common data model}
\newacro{EHR}{electronic health records}
\newacro{ARDS}{acute respiratory distress syndrome}
\newacro{UCLH}{University College London Hospital}
\newacro{NHS}{National Health Service}
\newacro{IMV}{invasive mechanical ventilation}
\newacro{NIV}{non-invasive ventilation}
\newacro{IPW}{inverse probability weighting}
\newacro{IPTW}{inverse probability of treatment weighting}
\newacro{IPTMW}{inverse probability of treatment and monitoring weights}
\newacro{NDE}{no direct effect}
\newacro{ICU}{intensive care unit}
\newacro{LOCF}{last observation carried forward}
\newacro{PS}{pressure support}
\newacro{PEEP}{positive and expiratory pressure}
\newacro{PSV}{pressure support ventilation}
\newacro{CPAP}{continuous positive airway pressure}
\newacro{BiPAP}{Bi-level positive airway pressure}
\newacro{HFNC}{high-flow nasal canula}
\newacro{ARDS}{acute respiratory distress syndrome}
\newacro{PCR}{polymerase chain reaction}
\newacro{RCT}{randomised controlled trial}
\newacro{UCL}{University College London}
\newacro{UCLH}{University College London Hospitals}
\newacro{GCS}{Glasgow coma scale}
\newacro{ACPU}{alert, verbal, pain, unresponsive}
\newacro{TMLE}{targeted maximum likelihood estimation}
\newacro{AIPTW}{augmented inverse probability of treatment weights}
\newacro{MICE}{multiple imputation via chained equations}
\newacro{ATE}{average treatment effect}
\newacro{NPSEM}{non-parametric structural equation model}
\newacro{CC-HIC}{Critical Care Health Informatics Collaborative}
\newacro{BMI}{body mass index}
\newacro{IMD}{index of multiple deprivation}
\newacro{MTP}{modified treatment policy}
\newacro{ICE}{iterated conditional expectation}
\newacro{NICE}{non-iterated conditional expectation}

\title{Evaluating the impact of longitudinal treatment strategies in the presence of informative monitoring and time-dependent confounding}

\author[1]{Leah Pirondini}
\author[2]{Karla Diaz-Ordaz}
\author[3]{Edward Palmer}
\author[1]{Ruth H. Keogh}

\affil[1]{Department of Medical Statistics, London School of Hygiene and Tropical Medicine, London, UK}
\affil[2]{Department of Statistical Science, University College London, London, UK}
\affil[3]{Critical Care, Whittington Hospital, London, UK}

\begin{document}

\maketitle

\begin{abstract}
% **************
% abstract
% **************

Routinely-collected data from electronic health records (EHR) provide opportunities to study effects of longitudinal treatment strategies in real-world clinical settings. A challenge presented by EHR data is that frequency of covariate monitoring differs by patient, covariate type and over time, and may be informative about a patient's health status. Many causal inference methods assume measurements of covariates are observed at a common set of regular time points. In this paper we describe and evaluate methods for estimating causal effects of longitudinal treatments on time-to-event outcomes in the presence of informative monitoring of time-dependent confounders. We show how methods based on inverse probability weighting, G-computation and longitudinal targeted maximum likelihood estimation (TMLE) can be adapted to allow for informative monitoring by incorporating monitoring indicator variables as additional time-dependent confounders. We evaluate these methods using a simulation study, comparing against more simple approaches that ignore monitoring variables. We demonstrate that ignoring monitoring can result in biased estimates of treatment effects. The methods are illustrated through an investigation into the effect of early versus delayed initiation of invasive mechanical ventilation on mortality of intensive care patients using routinely-collected data from an intensive care unit. We consider static treatment strategies such as `always treat’ and `never treat’ but also generalise to treatment strategies that allow for flexibility in the exact initiation time and duration of treatment.

\end{abstract}

\section{Introduction}
\label{sec:intro}

% ------------------------------------------
% intro to problem of informative monitoring
% ------------------------------------------

Routinely-collected observational data originating from \ac{EHR} are increasingly used for research purposes to investigate causal effects of treatments or exposures on health outcomes. When causal effects of longitudinal treatment strategies are examined using observational data, time-varying confounding (where confounders of the treatment-outcome relationship are themselves affected by treatment at earlier time points \cite{clare_causal_2019, daniel_methods_2013}) is a particular issue. Statistical methods for estimating the causal effects of longitudinal treatment strategies in the presence of time-dependent confounding are well-established \cite{clare_causal_2019}. The most commonly used are \acp{MSM} estimated using \ac{IPW} \cite{robins_marginal_2000,hernan_marginal_2000} and G-computation \cite{robins_new_1986, daniel_gformula_2011}. More recently, \ac{TMLE} \cite{laan_targeted_2006, laan_targeted_2012, laan_targeted_2011} was proposed as a doubly robust and efficient alternative to \ac{IPW} estimation for evaluating the effects of both point treatments and longitudinal treatment strategies.

A key assumption implicit in most methods for the estimation of longitudinal treatment effects is that patients are monitored at regular time intervals, such that measurements of time-varying confounders are available at a common set of observation times for all patients, and that changes in treatment status occur only at the corresponding set of observation times. In practice, when using routinely-collected observational data, this assumption is likely to be unrealistic. Researchers have no control over the data collection process, as data are collected as part of routine clinical care. Our emphasis in this paper is on routinely collected data on individuals in a hospital setting, however the same issues apply to other \ac{EHR} data, such as those arising from a primary care setting. General features are that the frequency and timing of measurements recorded in the database will vary by patient (patients who are more sick are likely to be monitored more frequently); by variable type (for example, depending on whether monitoring requires a simple observation versus an invasive procedure or lab test); and over time (for example, patients may be monitored differently at admission and just before discharge). Monitoring is therefore associated with the patient's health status (``\textit{informative monitoring}''). Monitoring in a hospital setting is typically determined by the treating clinicians. Clinicians may decide to order more frequent measurements of a particular variable when worsening values of that variable (or other variables) have recently been observed. They may also decide to monitor a patient more frequently around the time that a treatment is being initiated or withdrawn.  When examining the impact of longitudinal treatment strategies using routinely collected data, monitoring can act as a type of time-varying confounder. Patients with frequent monitoring may differ from patients with infrequent monitoring, with respect to characteristics relating to disease severity, and these characteristics may also impact patient outcomes.

There is little practical guidance on how to handle the challenges of informative patient monitoring in the context of estimating causal effects of longitudinal treatment strategies. Existing work has tended to focus on estimating the effects of joint dynamic treatment and monitoring strategies. Dynamic treatment strategies involve rules that adapt the treatment to the status of the patient. For example \textit{treat when variable $L$ exceeds threshold $r$}. If the variable involved in the dynamic rule is monitored informatively, an intervention on the monitoring pattern may also be of interest, as more frequent monitoring allows for earlier detection of a value of $L$ exceeding the threshold of interest. Robins et al. \cite{robins_estimation_2008} developed theory to provide estimators for the effects of joint dynamic treatment and monitoring strategies. Neugebauer et al. \cite{neugebauer_identification_2017} built on these results by proposing \acp{NPSEM} to describe the observational data structure in the presence of informative monitoring, as well as proving some identifiability results to derive an estimator for the effect of a joint dynamic treatment and static monitoring regime. However, they did not describe estimation methods. Kreif et al. \cite{kreif_exploiting_2021} extended this work by describing an \ac{IPW} estimator, applying this method to estimate the effects of joint dynamic treatment and static monitoring interventions in a diabetes study based on \ac{EHR} data. These studies focused on dynamic treatment regimes. Hernan et al. \cite{hernan_observation_2009} considered the effects of joint static treatment and static monitoring interventions where it is assumed that treatment can only change when covariates are monitored, also describing a method based on \ac{IPW}. None of these papers evaluated the methods using simulations, nor did they explore or describe alternative methods to \ac{IPW}.

The focus of this paper is on estimating the causal effects of longitudinal treatment strategies on time-to-event outcomes using routinely-collected data, in the presence of time-dependent confounding, where these time-dependent confounders have been monitored irregularly and potentially informatively. We present three methods based on \ac{IPW}, G-computation, and \ac{TMLE}, evaluating and comparing methods using a simulation study. We focus on static treatment strategies, including both simple static treatment strategies such as \textit{always treat} or \textit{never treat} as well as more flexible static treatment strategies that allow for periods of time where treatment is allowed to follow its natural course. For these latter treatment strategies, we extend to the setting with informative monitoring  methods by Wanis et al. \cite{wanis_grace_2024}, .

% ------------------------------------------
% overview of paper sections
% ------------------------------------------

The paper is organised as follows. Section \ref{sec:motivating_example} describes our motivating example for this research. In section \ref{sec:notation_assumptions}, we describe the assumed data structure using a causal \ac{DAG} that incorporates nodes representing monitoring, and the assumptions required to estimate the causal effect of interest using such data. In section \ref{sec:methods} we introduce and describe methods for dealing with informative monitoring when estimating causal estimands of interest. We evaluate and compare these methods using a simulation study in section \ref{sec:simulation_study}. In section \ref{sec:application} we present results for our motivating example comparing different mechanical ventilation strategies (early versus delayed initiation of ventilation) on the outcome of mortality in intensive care patients. We conclude with a discussion in section \ref{sec:discussion}.

\section{Motivating example}
\label{sec:motivating_example}

This work is motivated by questions about the optimal timing of initiation of mechanical ventilation for intensive care patients. Many patients admitted to the \ac{ICU} require some form of ventilatory support, ranging from supplemental oxygen (received via a simple mask) and non-invasive ventilation (including \ac{CPAP} or \ac{BiPAP}) to \ac{IMV}. To receive \ac{IMV} a patient must be sedated, intubated (have a tube placed into their trachea) and physically attached to a mechanical ventilator (a device to generate air flow and hence expand the lungs). For many patients, the decision about the optimal time to initiate \ac{IMV} in the \ac{ICU} remains controversial \cite{torjesen_covid-19_2021,patel_alternatives_2020,lepper_mechanical_2020, papoutsi_effect_2021}. Intubation itself is a highly specialised intervention, and exposes the patient to the risks of loss of airway control and haemodynamic instability. Long-term \ac{IMV} poses myriad risks relating to direct lung injury, as well as the secondary impact of sedation, immobility and the potential to develop secondary organ failures. Delaying \ac{IMV} may lead to patients being more unwell at the point of intubation, which may serve to aggravate the above named risks. To date no \acp{RCT} comparing different timings of initiation of \ac{IMV} have been conducted. There is a growing body of research using observational data to compare the effects of early versus delayed initiation of \ac{IMV}, but many such studies \cite{kangelaris_timing_2016, bauer_association_2017, xixi_association_2022, gonzalez_impact_2022} do not appropriately account for time-varying confounding or are prone to immortal time bias. A recent study by Wanis et al. \cite{wanis_emulating_2023} presents results from a target trial emulation comparing early versus delayed intubation in intensive care patients, finding little difference in 30-day mortality after applying realistic eligibility criteria. 

In this study we use the \ac{CC-HIC} data: routinely-collected hospital data from \ac{UCLH} intensive care units \cite{CC-HIC, harris_critical_2018}, to investigate the effects of early versus delayed \ac{IMV} strategies on 30-day mortality of intensive care patients, accounting for the irregular and informative nature of monitoring of time-varying confounders. Decisions about when and for how long to ventilate a patient are likely to be highly specific to the patient's needs. Treatment strategies such as \textit{always ventilate}, \textit{never ventilate} or \textit{initiate ventilation on day three of ICU admission} are unlikely to be implemented in real-world clinical practice. More plausible ventilation strategies might allow for deviations when clinical improvement or deterioration occur by incorporating time windows when invasive ventilation may be both initiated and stopped. For example, \textit{initiate ventilation by day three of ICU admission, then allow ventilation to be stopped when deemed appropriate by treating clinicians}. In this paper we consider these types of flexible static treatment strategies. We use techniques described by Wanis et al. \cite{wanis_grace_2024}, extending to a setting with irregular monitoring.

\section{Observed data structure, assumptions and causal estimand}
\label{sec:notation_assumptions}

\subsection{Observed data structure}
\label{sec:observed_data}

We consider a study in which $n$ individuals are followed up from a specified time zero, which, in our motivating example, is admission to an intensive care unit, until the earlier of time of death and censoring time. We assume that there is an underlying grid of times at which, for each individual, measurements of covariates can potentially be observed and changes in treatment status can occur. This could be, say, minutes, hours, days, etc. The grid times are $k=0,1,...,K$, representing the complete set of possible observation times.

We let $A_k$ denote binary treatment status at observation time $k$, which is assumed to be known at all possible observation times. For each observation time $k$ we let $\mathbf{L}^*_{k}$ denote a vector of $P$ covariates, where each $L^*_{p,k+1}$ denotes the true level of covariate $p$ at observation time $k$. However, each covariate is not monitored at every possible observation time for every individual, and different covariates can have different monitoring patterns for the same individual. We therefore let $\mathbf{L}_{k}$ and $\mathbf{N}_{k}$ denote, respectively, the observed values of the covariates and binary indicators of whether the covariates are monitored at the next possible observation time $k+1$, with the $p$th element $N_p$ being linked to the corresponding covariate $L_p$. For a single covariate $p$, the value of $N_{p,k}$ indicates whether the corresponding covariate $L^*_{p,k+1}$ is monitored at the next possible observation time. We have used subscript $p$ to denote the specific covariate (for set of covariates $p=1,...,P$) to make explicit the fact that, for a given individual, each covariate can have its own monitoring pattern. For simplicity we drop the subscript $p$ going forward. The true underlying (but potentially unobserved) level of the covariate $L^*_{k}$ is linked to the observed level of the covariate $L_{k}$. If a decision is made to monitor $L^{*}_{k}$ (i.e. if $N_{k-1}=1$) we observe $L_{k} = L^{*}_{k}$, and otherwise (i.e. if $N_{k-1}=0$) $L^{*}_{k}$ is missing from the observed data and $L_{k}$ is defined as the last monitored value of the covariate. That is, $L_{k} = L_{j}$ where $j<k$ is the latest observation time when the covariate was monitored, i.e. the last time $N_{j-1}$ was equal to 1. 

We can think of each $N_k$ as representing a decision, made by clinical staff in our motivating example, to monitor the corresponding covariate at the next observation time, which might be affected by the patient's clinical status up to that point. $N_k$ may also indicate a missing value (i.e. that the covariate measurement was observed but not recorded in the \ac{EHR} data) and we discuss the difference between these interpretations, and their implications, in section \ref{sec:assumps}.

We assume that, by study design, $L_0$ is always monitored, so $L_0 = L_0^*$ for all individuals. $L_0$ may include a set of time-fixed covariates, as well as baseline values of all time-varying covariates. A bar over a time-dependent variable indicates the history, that is, $\Bar{A}_k = \{ A_0,A_1,...,A_k \}$, $\Bar{L}_k = \{ L_0,L_1,...,L_k \}$, $\Bar{L}^*_k = \{ L^*_0,L^*_1,...,L^*_k \}$ and $\Bar{N}_k = \{ N_0,N_1,...,N_k \}$.

We consider a time-to-event outcome, such as death, and we let $Y_k$ be a binary indicator of whether this outcome has occurred by the end of period $k$. The assumed ordering of variables is \newline $(L^*_0,L_0,A_0,N_0,Y_1,...,L^*_{K-1},L_{K-1},A_{K-1},Y_K)$ and we have $Y_0=0$ since all individuals are alive at time 0. For simplicity we assume that all censoring is administrative at time $K+1$, however, methods extend more generally to the situation where censoring can occur at any time. We comment on this further in the discussion.

\begin{figure}[H]
\begin{center}
\begin{tikzpicture}
% nodes
\node (1) at (0,0) {$L_0=L^*_0$};
\node (2) at (3,1.5) {$L^*_1$};
\node (3) at (6,1.5) {$L^*_2$};
\node (4) at (9,1.5) {$L^*_3$};
\node (5) at (12,1.5) {$L^*_4$};

\node (6) at (3.5,0) {$L_1$};
\node (7) at (6.5,0) {$L_2$};
\node (8) at (9.5,0) {$L_3$};
\node (9) at (12.5,0) {$L_4$};

\node (10) at (1.5,-3) {$A_0$};
\node (11) at (4.5,-3) {$A_1$};
\node (12) at (7.5,-3) {$A_2$};
\node (13) at (10.5,-3) {$A_3$};
\node (14) at (13.5,-3) {$A_4$};

\node (15) at (2.7,-1.5) {$N_0$};
\node (16) at (5.7,-1.5) {$N_1$};
\node (17) at (8.7,-1.5) {$N_2$};
\node (18) at (11.7,-1.5) {$N_3$};

\node (19) at (3,-4.5) {$Y_1$};
\node (20) at (6,-4.5) {$Y_2$};
\node (21) at (9,-4.5) {$Y_3$};
\node (22) at (12,-4.5) {$Y_4$};
\node (23) at (15,-4.5) {$Y_5$};

% straight arrows
% L* to L*
\draw[thick,Arrow] (1) -- (2);
\draw[thick,Arrow] (2) -- (3);
\draw[thick,Arrow] (3) -- (4);
\draw[thick,Arrow] (4) -- (5);

% A to A
\draw[thick,Arrow] (10) -- (11);
\draw[thick,Arrow] (11) -- (12);
\draw[thick,Arrow] (12) -- (13);
\draw[thick,Arrow] (13) -- (14);

% N to N
\draw[thick,Arrow] (15) -- (16);
\draw[thick,Arrow] (16) -- (17);
\draw[thick,Arrow] (17) -- (18);

% Y to Y
\draw[thick,Arrow] (19) -- (20);
\draw[thick,Arrow] (20) -- (21);
\draw[thick,Arrow] (21) -- (22);
\draw[thick,Arrow] (22) -- (23);

% L to A
\draw[thick,Arrow] (1) -- (10);
\draw[thick,Arrow] (6) -- (11);
\draw[thick,Arrow] (7) -- (12);
\draw[thick,Arrow] (8) -- (13);
\draw[thick,Arrow] (9) -- (14);

% N to A
\draw[thick,Arrow] (15) -- (11);
\draw[thick,Arrow] (16) -- (12);
\draw[thick,Arrow] (17) -- (13);
\draw[thick,Arrow] (18) -- (14);

% A to L*
\draw[thick,Arrow] (10) -- (2);
\draw[thick,Arrow] (11) -- (3);
\draw[thick,Arrow] (12) -- (4);
\draw[thick,Arrow] (13) -- (5);

% L to N
\draw[thick,Arrow] (1) -- (15);
\draw[thick,Arrow] (6) -- (16);
\draw[thick,Arrow] (7) -- (17);
\draw[thick,Arrow] (8) -- (18);

% A to N
\draw[thick,Arrow] (10) -- (15);
\draw[thick,Arrow] (11) -- (16);
\draw[thick,Arrow] (12) -- (17);
\draw[thick,Arrow] (13) -- (18);

% A to Y
\draw[thick,Arrow] (10) -- (19);
\draw[thick,Arrow] (11) -- (20);
\draw[thick,Arrow] (12) -- (21);
\draw[thick,Arrow] (13) -- (22);
\draw[thick,Arrow] (14) -- (23);

% Y to A
\draw[thick,Arrow] (19) -- (11);
\draw[thick,Arrow] (20) -- (12);
\draw[thick,Arrow] (21) -- (13);
\draw[thick,Arrow] (22) -- (14);

% dashed arrows (L to L, L* to L, N to L)
\draw[thick,Arrow,dashed] (1) -- (6);
\draw[thick,Arrow,dashed] (6) -- (7);
\draw[thick,Arrow,dashed] (7) -- (8);
\draw[thick,Arrow,dashed] (8) -- (9);
\draw[thick,Arrow,dashed] (2) -- (6);
\draw[thick,Arrow,dashed] (3) -- (7);
\draw[thick,Arrow,dashed] (4) -- (8);
\draw[thick,Arrow,dashed] (5) -- (9);
\draw[thick,Arrow,dashed] (15) -- (6);
\draw[thick,Arrow,dashed] (16) -- (7);
\draw[thick,Arrow,dashed] (17) -- (8);
\draw[thick,Arrow,dashed] (18) -- (9);

% curved arrows
\draw[thick,Arrow] (1) to[bend left=20] (19);
\draw[thick,Arrow] (2) to[bend left=15] (20);
\draw[thick,Arrow] (3) to[bend left=15] (21);
\draw[thick,Arrow] (4) to[bend left=15] (22);
\draw[thick,Arrow] (5) to[bend left=15] (23);

% grey arrows
\draw[thick,lightgray,arrows={-latex[angle=40:5pt,lightgray,fill=lightgray]}] (15) -- (2);
\draw[thick,lightgray,arrows={-latex[angle=40:5pt,lightgray,fill=lightgray]}] (16) -- (3);
\draw[thick,lightgray,arrows={-latex[angle=50:5pt,lightgray,fill=lightgray]}] (17) -- (4);
\draw[thick,lightgray,arrows={-latex[angle=50:5pt,lightgray,fill=lightgray]}] (18) -- (5);

\draw[thick, lightgray, arrows={-latex[angle=50:5pt,lightgray,fill=lightgray]}, bend right=15] (15) to (20);
\draw[thick, lightgray, arrows={-latex[angle=50:5pt,lightgray,fill=lightgray]}, bend right=15] (16) to (21);
\draw[thick, lightgray, arrows={-latex[angle=50:5pt,lightgray,fill=lightgray]}, bend right=15] (17) to (22);
\draw[thick, lightgray, arrows={-latex[angle=50:5pt,lightgray,fill=lightgray]}, bend right=15] (18) to (23);

\end{tikzpicture}
\caption{Causal DAG illustrating relationships between time-dependent treatment $A_k$, true underlying (potentially unobserved) time-dependent covariate $L^*_k$, observed time-dependent covariate $L_k$, monitoring indicator $N_k$ and outcome $Y_k$. Dashed arrows represent deterministic relationships. Grey arrows represent possible effects of monitoring on (unmeasured) covariate and outcome, which would not be present under the ``\textit{no direct effect}'' assumption.}
\label{fig:dag_im}
\end{center}
\end{figure}
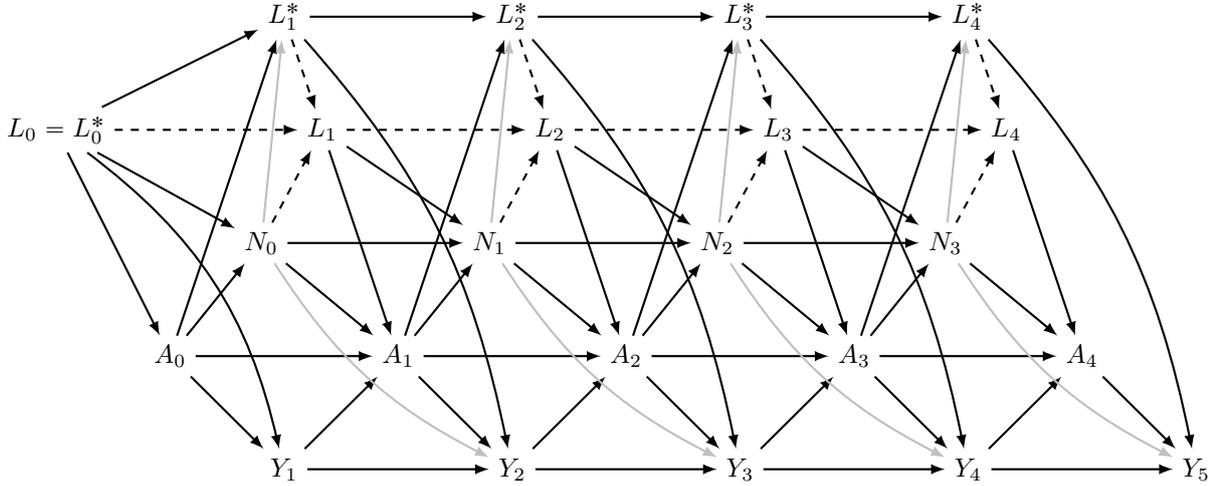

The causal \ac{DAG} in figure \ref{fig:dag_im} illustrates the assumed relationships between variables in a simplified scenario for a single covariate $L_k$ (and corresponding $N_k$) with five potential observation times.

We assume that the outcome $Y_k$ is potentially affected by treatment $A_k$ and the true, underlying level of the covariate (whether measured or not) $L^*_k$. The true covariate $L^*_k$ is in turn affected by past levels of treatment and past levels of $L^*$. There are no direct arrows from $L^*_k$ to $A_k$, because treatment decisions are assumed to be entirely mediated through the measured value $L_k$, i.e. the covariate $L^*_k$ can only impact treatment decisions if it is observed. The same is true for monitoring decisions $N_k$ and so there are no direct arrows from $L^*_k$ to $N_k$. However, both treatment and monitoring decisions can be affected by past levels of treatment, monitoring, and the \textit{observed} covariate. The dashed lines in the \ac{DAG} represent deterministic relationships, i.e. that $L_k = L^*_k$ when $N_{k-1} = 1$ and $L_k = L_{k-1}$ when $N_{k-1} = 0$. The grey arrows from $N_k$ to $L^*_{k+1}$ and from $N_k$ to $Y_k$ represent the possibility that monitoring itself affects the underlying level of the covariate or the outcome, for example if monitoring requires an invasive test with potentially adverse effects. In section \ref{sec:assumps} we discuss the situations in which it is possible to remove these grey arrows, and the implications of this assumption. The \ac{DAG} could be extended in various ways, for example, so that there are long term effects of $L$ on $A$ and vice versa, long term effects of $L^*$ and $A$ on $Y$, or long term effects of $L$ or $A$ on $N$. The assumptions encapsulated in the \ac{DAG} are made explicit in section \ref{sec:assumps}.

% **************************
% code for figures
% **************************

% *************************************************************
% *************************************************************

% ***********************************
% code for figure - portion of DAG
% ***********************************

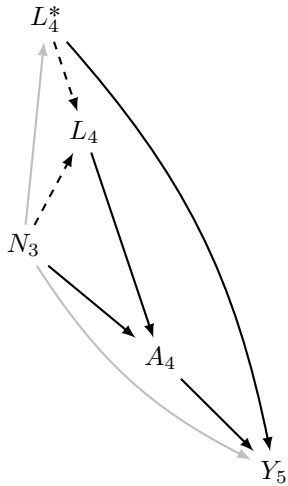
\begin{figure}[H]
\begin{center}
\begin{tikzpicture}

% nodes
\node (1) at (12,1.5) {$L^*_4$};
\node (2) at (12.5,0) {$L_4$};
\node (3) at (13.5,-3) {$A_4$};
\node (4) at (11.7,-1.5) {$N_3$};
\node (5) at (15,-4.5) {$Y_5$};

% L to A
\draw[thick,Arrow] (2) -- (3);

% N to A
\draw[thick,Arrow] (4) -- (3);

% A to Y
\draw[thick,Arrow] (3) -- (5);

% dashed arrow (N to L)
\draw[thick,Arrow,dashed] (4) -- (2);

% dashed arrow (L* to L)
\draw[thick,Arrow,dashed] (1) -- (2);

% curved arrow (L* to Y)
\draw[thick,Arrow] (1) to[bend left=15] (5);

% grey straight arrow (N to L*)
\draw[thick,lightgray,arrows={-latex[angle=50:5pt,lightgray,fill=lightgray]}] (4) -- (1);

% grey curved arrow (N to Y)
\draw[thick, lightgray, arrows={-latex[angle=50:5pt,lightgray,fill=lightgray]}, bend right=15] (4) to (5);

\end{tikzpicture}
\caption{Portion of the causal DAG in figure \ref{fig:dag_im} for a single time-point}
\label{fig:dag_im2}
\end{center}
\end{figure}

Figure \ref{fig:dag_im2} shows a portion of the causal \ac{DAG} for a single time-point. The following pathways exist in the \ac{DAG} connecting $A_4$ and $Y_4$:

\[
\begin{split}
  & (1) \; \; A_4 \rightarrow Y_5 \\
  & (2) \; \; A_4 \leftarrow L_4 \leftarrow L^*_4 \rightarrow Y_5 \\
  & (3) \; \; A_4 \leftarrow N_3 \rightarrow L_4 \leftarrow L^*_4 \rightarrow Y_5
\end{split}
\]

Adjustment for $L_4$ is required to block the backdoor pathway (2). However, in pathway (3), $L_4$ is a collider, therefore adjustment for $L_4$ opens this path, requiring additional adjustment for $N_3$. In addition, in the case where monitoring has a direct effect on the unmeasured covariate or the outcome, we also have pathways $A_4 \leftarrow N_3 \rightarrow Y_5$ and $A_4 \leftarrow N_3 \rightarrow L^*_4 \rightarrow Y_5$, further requiring adjustment for $N_3$. Adjustment for $L_4$ and $N_3$ is therefore sufficient to block all open backdoor pathways between $A_4$ and $Y_5$, and we do not require adjustment for $L^*_4$.

\subsection{Treatment strategies}
\label{sec:treatment_strategies}

In this paper we consider longitudinal static treatment strategies. Examples include simple static treatment strategies such as \textit{always treat}, \textit{never treat} or \textit{initiate treatment at time $t_0$ and remain treated thereafter}. However, motivated by our real-world data example, we also consider more flexible static treatment strategies that incorporate \textit{``grace-periods''}, which are periods where treatment is allowed to follow its natural course \cite{wanis_grace_2024}. Such treatment strategies are aimed at better reflecting realistic clinical decision making in the context of mechanical ventilation.

Throughout this paper we consider two flexible treatment strategies that incorporate grace periods: \textit{ventilate early} and \textit{wait to ventilate}. We describe these treatment strategies in more detail in sections \ref{sec:simulation_study} and \ref{sec:application}. Briefly, the \textit{ventilate early} strategy involves initiating ventilation at $k=0$ and remaining ventilated for at least $p_1$ consecutive periods (or until death or censoring), after which and for the remainder of follow-up, ventilation is allowed to follow its natural course. The \textit{wait to ventilate} strategy involves remaining unventilated for the first $q_1$ consecutive periods (or until death or censoring), after which, alive and uncensored individuals must initiate ventilation in any period $q_1 \leq k < q_2$, such that, after $p_2$ time periods, an alive and uncensored individual has received $p_1$ consecutive periods of ventilation, where $q_2 = p_2 - p_1$. In addition we also consider the simple static treatment strategies: \textit{always ventilate} and \textit{never ventilate}.

\subsection{Causal estimand}
\label{sec:causal_estimand}
We denote our longitudinal static treatment strategies using the letter $s$. The notation $\Bar{A}_k=\Bar{s}_k$ refers to the treatment history up to period $k$ that is compatible with the treatment strategy of interest.

We define the potential outcome for subject $i$, $Y_{k,i}^{\Bar{s}_k}$, as the failure status at the end of period $k$ that we would see for subject $i$ if, possibly contrary to fact, that subject had followed treatment strategy $\Bar{s}_k$ up to time $k$. We can then define our causal estimands of interest as the counterfactual probability of failure by time $k$, $\Pr[Y_{k}^{\Bar{s}_k}=1]$, and the causal contrast under two different treatment strategies $\Bar{s}_{k,1}$ and $\Bar{s}_{k,2}$:
\begin{equation}
    \Pr[Y_{k}^{\Bar{s}_{k,1}}=1] \; \text{vs.} \; \Pr[Y_{k}^{\Bar{s}_{k,2}}=1]
\end{equation}

In a discrete-time setting, we can write this causal estimand in terms of the counterfactual discrete-time hazards as

\begin{equation} 
    \Pr[Y^{\Bar{s}_k}_k=1] = 1 - \prod_{j=1}^k (1 - \Pr(Y^{\Bar{s}_j}_j=1|Y^{\Bar{s}_{j-1}}_{j-1}=0))
    \label{eq:counterfac_discrete_time_haz}
\end{equation}

where the counterfactual discrete-time hazard $\Pr(Y^{\Bar{s}_k}_k=1|Y^{\Bar{s}_{k-1}}_{k-1}=0)$ is the risk of failure during the period $(k-1,k]$ conditional on survival up to the end of period $k-1$, in a world in which individuals had, possibly contrary to fact, followed the hypothetical treatment strategy of interest up to period $k$.

\subsection{Identifiability assumptions}
\label{sec:assumps}

Identifiability of the causal estimand from the observed data relies on the assumptions of no interference, positivity, consistency, and conditional exchangeability, which in our case need to accommodate the informative nature of monitoring of covariates. The no interference assumption is that the counterfactual outcome for a given individual $Y_k^{\Bar{s}_k}$ does not depend on the treatment received by any other individuals. The positivity assumption is that, given a patient with treatment history $\Bar{A}_{k-1}$, monitoring history $\Bar{N}_{k-1}$, and observed covariate history $\Bar{L}_{k}$, there is a non-zero probability of observing any treatment level under the treatment strategy of interest at period $k$, $s_k$. That is, $\Pr(A_k=s_k|\Bar{N}_{k-1},\Bar{A}_{k-1},\Bar{L}_k) > 0  \; \; \text{for all} \; \; k=0,...,K$. Consistency means that the observed outcome is equal to the potential outcome under assignment to the treatment history actually received, i.e. $Y_k^{\Bar{s}_k} = Y_k \; \text{if} \; \Bar{A}_k=\Bar{s}_k$. The conditional exchangeability assumption is that the potential outcome is independent of the actual treatment assignment given past treatment, monitoring and \textit{observed} covariate history, i.e. $Y_k^{\Bar{s}_k} \indep A_k | \Bar{N}_{k},\Bar{A}_{k-1},\Bar{L}_{k-1} \; \; \text{for all} \; \; k=0,...,K$. 

We assume that conditional exchangeability holds without conditioning on $L^*_k$, and that the covariate $L^*_k$ only affects treatment decisions when observed. In practice in \ac{EHR} data we will not know whether a missing value of $L^*_k$ represents the fact that the patient was not monitored at time $k$ or that the patient was monitored but the observation is missing from the \ac{EHR} data. In the former case it is reasonable to assume that $L^*_k$, when unmonitored, cannot affect treatment decisions, and in the latter case we may be willing to assume that unrecorded measurements of $L^*_k$ would not be reviewed by treating clinicians and therefore could not inform treatment decisions.  

In addition to the above assumptions, Robins et al. \cite{robins_estimation_2008} proposed a \ac{NDE} assumption that monitoring itself has no direct effect on the unobserved level of the covariate $L^*_k$ or on the outcome $Y_k$ when treatment is set to $\Bar{s}_k$. The \ac{NDE} assumption is represented in the \ac{DAG} by lack of arrows between $N_k$ and $L^*_{k+1}$ and between $N_k$ and $Y_{k+1}$, i.e. by removal of the grey arrows, or any similar arrows connecting $N_k$ and $L^*_{j}$ or $N_k$ and $Y_j$ for $j>k$. This assumption is the focus of work by Neugebauer et al. \cite{neugebauer_identification_2017} and Kreif et al. \cite{kreif_exploiting_2021}, upon which this work builds. Their research focuses on evaluating the effects of joint dynamic treatment and static monitoring strategies, where exploiting the \ac{NDE} assumption allows for reduced data support for the monitoring strategy of interest. In our setting of estimating the effects of static treatment strategies, the \ac{NDE} assumption is not required, however, accounting for irregular monitoring of covariates is still necessary to avoid biased treatment effects. Violations of the \ac{NDE} assumption occur when monitoring directly affects the outcome or the true, underlying covariate. This can occur when monitoring requires an invasive procedure with potential adverse effects. However, in many instances monitoring requires only routine blood tests or measurement of vital signs.
\section{Estimation methods handling informative monitoring}
\label{sec:methods}

In this section we describe methods to estimate the causal estimands described in section \ref{sec:causal_estimand} using observational data of the form described in section \ref{sec:observed_data}, which is subject to informative monitoring of covariates, under the assumptions discussed in section \ref{sec:assumps}. We build upon work by Krief et al. \cite{kreif_exploiting_2021} who present an \ac{IPW}-based estimator. We describe methods based on \ac{IPW}, G-computation and \ac{TMLE}. To allow methods to estimate the effects of static \textit{grace-period} treatment strategies, we also build on work by Wanis et al. \cite{wanis_grace_2024}. Their work focuses on the setting of non-adherence to a particular medication, and does not consider the problem of informative monitoring. We show that their methods can also be used to estimate flexible ventilation strategies that allow for periods of time when treatment is allowed to follow its natural course, and also extend to the setting with informative monitoring of covariates.

\subsection{Inverse probability weighting}

The idea behind inverse probability weighting is to re-weight individuals, in this case using time-dependent weights, to create a pseudo-population in which the treatment allocation at a given time does not depend on past covariates or past monitoring decisions. 

\subsubsection{Calculating the weights}

We define the time-dependent weight at observation time $k$ for a given individual as:

\begin{equation}
    {W}_k = \prod_{j=0}^{k} \frac{f^{g}(A_j|\bar{N}_{j-1},\bar{A}_{j-1},\bar{L}_{j},Y_{j-1}=0)}{\Pr(A_j=a_j|\bar{N}_{j-1},\bar{A}_{j-1},\bar{L}_{j},Y_{j-1}=0)}
    \label{eq:weights}
\end{equation}

where the term in the denominator is the probability of the individual receiving their observed treatment $a_j$ at time $j$, given their (observed) covariate, treatment, and monitoring history. Note that we do not need to condition on the true but unobserved values $L^*_k$, as treatment decisions are mediated by the observed covariate $L_k$, however, we do need to condition on $N_k$ as well as $L_k$, even under the \ac{NDE} assumption, to block open pathways between $A_k$ and $Y_k$, as described at the end of section \ref{sec:observed_data}. 

The term in the numerator, $f^{g}(A_j|\bar{N}_{j-1},\bar{A}_{j-1},\bar{L}_{j},Y_{j-1}=0)$, serves to adapt the weights to accommodate different types of longitudinal treatment strategy. Under a simple strategy such as \textit{always treat}, an individual's treatment is forced to take value $A_j=1$ at all times, and hence $f^{g}(A_j|\bar{N}_{j-1},\bar{A}_{j-1},\bar{L}_{j},Y_{j-1}=0)=1$ for all $j$, and the time-dependent weight at time $k$ takes the form 

\begin{equation}
    {W}_k = \prod_{j=0}^{k} \frac{1}{\Pr(A_j=a_j|\bar{N}_{j-1},\bar{A}_{j-1},\bar{L}_{j},Y_{j-1}=0)}
\end{equation} 
Under more complex strategies the treatment status at a given time may be allowed to take its natural value. Wanis et al. \cite{wanis_grace_2024} refer to the periods where $A_j$ is allowed to take its natural value as \textit{grace-periods} and to these types of flexible treatment strategies as \textit{natural grace-period strategies}. The treatment strategies of interest in our motivating example of \textit{ventilate early} and \textit{wait to ventilate} are examples of natural grace-period strategies. In this case $f^{g}(A_j|\bar{N}_{j-1},\bar{A}_{j-1},\bar{L}_{j},Y_{j-1}=0)$ takes one of two forms depending on whether period $j$ is a \textit{grace-period} or not. For periods of $j$ when the treatment strategy of interest involves allowing $A_j$ to take its natural value (i.e. during a \textit{grace-period}) $f^{g}(A_j|\bar{N}_{j-1},\bar{A}_{j-1},\bar{L}_{j},Y_{j-1}=0) = \Pr(A_j=a_j|\bar{N}_{j-1},\bar{A}_{j-1},\bar{L}_{j},Y_{j-1}=0)$. As such, the contribution to the weight $w_k$ for time $j$ is equal to 1. But for periods where the treatment strategy involves setting $A_j=a_j$ ($a_j \in \{0,1\}$) we have $f^{g}(A_j|\bar{N}_{j-1},\bar{A}_{j-1},\bar{L}_{j},Y_{j-1}=0) = 1$ and the contribution to the weight $w_k$ for time $j$ is $1/\Pr(A_j=a_j|\bar{N}_{j-1},\bar{A}_{j-1},\bar{L}_{j},Y_{j-1}=0)$.

The weights described above are the unstabilised inverse probability weights. We can also estimate stabilised inverse probability weights by amending the term in the numerator, for non \textit{grace-periods}, to $\Pr(A_j=a_j|\bar{A}_{j-1},Y_{j-1}=0)$. 

\subsubsection{Estimating the estimand of interest}

We can estimate the counterfactual discrete-time hazard at each time-point $k$ non-parametrically, by calculating a weighted average of the observed outcome $Y_{i,k}$ within individuals who have survived up to the end of period $k-1$ and have followed the treatment strategy of interest up to period $k$, using our estimated inverse probability weights $\hat{w}_{i,k}$. This can be written

\begin{equation}
    \hat{\Pr}(Y^{\bar{s}_k}_k = 1|Y^{\bar{s}_{k-1}}_{k-1}=0) = 
    \frac{\sum \Big( \mathbb{I}(\bar{A}_k=\bar{s}_k) \; \; \hat{w}_{i,k} \times Y_{i,k} \times (1 - Y_{i,k-1}) \Big) }
         {\sum \Big( \mathbb{I}(\bar{A}_k=\bar{s}_k) \; \; \hat{w}_{i,k} \times (1 - Y_{i,k-1}) \Big) }    
\end{equation}

where the weights are set to zero once an individual deviates from the treatment strategy of interest, so that we average over only those individuals following the treatment strategy of interest up to period $k$. The estimated counterfactual discrete-time hazards can be plugged into equation (\ref{eq:counterfac_discrete_time_haz}) to obtain estimates of the counterfactual risk of failure by period $k$.

Estimating the counterfactual risk non-parametrically, as described above, may be impractical, particularly in a setting with many time-points, due to insufficient support in the data for each treatment strategy of interest. We may alternatively wish to specify a \ac{MSM}: a model for the counterfactual outcome. The \ac{MSM} must specify how the expectation of the counterfactual outcome at time $k$ depends on the history of treatment up to time $k$, $\bar{a}_k$, through a function $g(.)$. By using a \ac{MSM} we make more use of the available data, albeit at the cost of an additional modelling assumption that the \ac{MSM} is correctly specified. The MSM takes the form

\begin{equation}
    \Pr(Y_k^{\bar{a}_k}=1|Y^{\bar{a}_{k-1}}_{k-1}=0)=g \Bigl( \bar{a}_k; \boldsymbol{\beta} \Bigr)
\end{equation}

In our setting where $Y_k$ is a binary variable the function $g(.)$ is the inverse logit link function. For example we may specify that 
$ g \Bigl( \bar{a}_k; \boldsymbol{\beta} \Bigr)  = logit^{-1} \Bigl( \beta_0 + \sum_{j=0}^{k} \beta_j a_j \Bigr) $
i.e. the expected potential outcome at time $k$ depends on the full history of treatment up to time $k$ through main effect terms for treatment at each observation time from $j=0,...,k$. Other possibilities would be to include only later values of $a_k$ or to include interactions between treatment at different observation times.

The \ac{MSM} cannot be fitted directly to the observed data, instead we use the inverse-probability weights to construct a pseudo-population and fit the model $\mathrm{E}(Y_k|Y_{k-1}=0,\bar{A}_k) = logit^{-1} \Bigl( \bar{A}_k; \boldsymbol{\beta} \Bigr)$ to the pseudo-population data using inverse-probability weighted least squares. We can fit this model using pooled logistic regression treating each person-time as an observation. The observed data must be in long format and time $k$ is included in the model, for example, as a categorical variable or as a continuous variable modelled flexibly using a spline.

Under both the non-parametric and MSM IPW approaches, confidence intervals can be obtained using bootstrapping. For each bootstrap sample, the full \ac{IPW} procedure (i.e. estimating (a) the weights and (b) the estimand of interest using the re-weighted data either non-parametrically or via a \ac{MSM}) is performed. Standard errors and confidence intervals are obtained from the distribution of bootstrap samples.

\subsection{G-computation}
\label{sec:method_g_comp}

Based on the identifying assumptions described in section \ref{sec:notation_assumptions}, the g-formula estimator of $\mathrm{E}[Y_k^{\bar{s}_k}]$ for a specified treatment strategy $\bar{s}_k$ can be written as:

\begin{equation}
    \begin{split}
        \mathrm{E}[Y_k^{\bar{s}_k}] = &\sum_{j=1}^{k} \sum_{\bar{a}_j} \sum_{\bar{l}_j} \sum_{\bar{n}_{j-1}} \Pr (Y_j=1|\bar{A}_j = \bar{s}_j,\bar{L}_j = \bar{l}_j, \bar{N}_{j-1} = \bar{n}_{j-1}, Y_{j-1}=0) \\
        & \times \prod_{m=1}^j \Bigl\{ \Pr (Y_{m-1}=0 |  \bar{A}_{m-1} = \bar{s}_{m-1},\bar{L}_{m-1} = \bar{l}_{m-1}, \bar{N}_{m-2} = \bar{n}_{m-2}, Y_{m-2}=0)\\
        & \times f(l_m|\bar{a}_{m-1},\bar{l}_{m-1},\bar{n}_{m-1},y_{m-1}=0) \\
        &  \times f(n_{m-1}|\bar{a}_{m-1},\bar{l}_{m-1},\bar{n}_{m-2},y_{m-2}=0) \\
        &\times f^{g}(a_m|\bar{a}_{m-1},\bar{l}_{m},\bar{n}_{m-1},y_{m-1}=0) \Bigl\}
    \end{split}
    \label{eq:gform_nice}
\end{equation}

$\sum_{\bar{a}_j}$, $\sum_{\bar{l}_j}$ and $\sum_{\bar{n}_{j-1}}$ denote sums over all possible levels of treamtent, confounder, and monitoring history, respectively. For any continuous covariates we replace sums $\sum_{\bar{l}_j}$ with integrals $\int_{\bar{l}_j}$.

Within this version of the g-formula $f^{g}(a_k|\bar{n}_{k-1},\bar{a}_{k-1},\bar{l}_{k},y_{k-1}=0) 
= f(a_k|\bar{n}_{k-1},\bar{a}_{k-1},\bar{l}_{k},y_{k-1}=0)$ for periods $k$ when $A_k$ is set to a specified value (0 or 1) according to the treatment intervention; and $f^{g}(a_k|\bar{n}_{k-1},\bar{a}_{k-1},\bar{l}_{k},y_{k-1}=0) 
= 1$ for periods $k$ when $A_k$ is allowed to take its natural value under the treatment intervention.

To implement G-computation we must specify and fit models:

\begin{equation}
    \begin{split}
        &f(Y_k|\bar{A}_k,\bar{L}_k,\bar{N}_k,Y_{k-1}=0;\psi_Y) \\
        &f(L_k|\bar{A}_{k-1},\bar{L}_{k-1},\bar{N}_{k-1},Y_{k-1}=0;\psi_L) \\
        &f(A_k|\bar{A}_{k-1},\bar{L}_{k},\bar{N}_{k-1},Y_{k-1}=0;\psi_A) \\
        &f(N_k|\bar{A}_{k},\bar{L}_{k},\bar{N}_{k-1},Y_{k-1}=0;\psi_N)
    \end{split}
    \label{eq:conditional_models_gcomp}
\end{equation}

Using these models, the g-formula in (\ref{eq:gform_nice}) can be evaluated using Monte-Carlo integration, through simulation of the longitudinal confounders, monitoring nodes and outcome, under the treatment strategy of interest. That is, given maximum likelihood estimates of the parameters in the conditional models in (\ref{eq:conditional_models_gcomp}), we sequentially simulate, for each individual $i=1,...,n$, as follows
\begin{enumerate}
    \item Generate a simulated value $\tilde{L}_{i,0}$ using the fitted model for $L_0$: $\tilde{L}_{i,0} \sim f(L_{0};\hat{\psi}_L)$ (or, since $L_0$ does not depend on previous variables, sample $\tilde{L}_{i,0}$ from its empirical distribution, that is, set $\tilde{L}_{i,0} = {L}_{i,0}$).
    \item For non-grace-periods, set $A_{i,0}$ to the value compatible with the treatment intervention. For grace-periods, generate a simulated value $\tilde{A}_{i,0}$ ($=$ 0 or 1) using the fitted model for $A_0$ and setting $L_{i,0}$ to its simulated value $\tilde{L}_{i,0}$: $\tilde{A}_{i,0} \sim f(A_{0}|\tilde{L}_{i,0};\hat{\psi}_A)$
    \item Generate a simulated value $\tilde{N}_{i,0}$ ($=$ 0 or 1) using the fitted model for $N_0$, setting $L_{i,0}$ to its simulated value and $A_{i,0}$ to its value under the treatment intervention: $\tilde{N}_{i,0} \sim f(N_{0}|A_{i,0}=s_{i,0},\tilde{L}_{i,0};\hat{\psi}_N)$ (where, from the previous step, $s_{i,0}$ was set deterministically for non-grace periods and $s_{i,0} = \tilde{A}_{i,0}$ for grace-periods)
    \item Generate a simulated value $\tilde{Y}_{i,0}$ ($=$ 0 or 1) using the fitted model for $Y_0$, setting $L_{i,0}$ and $N_{i,0}$ to their simulated values and $A_{i,0}$ to its value under the treatment intervention: $\tilde{Y}_{i,0} \sim f(Y_{0}|A_{i,0}=s_{i,0},\tilde{L}_{i,0},\tilde{N}_{i,0} ;\hat{\psi}_Y)$
    \item Sequentially simulate $\tilde{L}_{i,k}$, $\tilde{A}_{i,k}$ (when $k$ is a grace-period), $\tilde{N}_{i,k}$ and $\tilde{Y}_{i,k}$ for $k=1,...,K$ under the treatment strategy of interest. Once $\tilde{Y}_{i,k}=1$ all future values of $\tilde{Y}_{i,j}$ ($j>k$) are also 1 and no additional simulated values of $\tilde{L}_{i,k}$, $\tilde{A}_{i,k}$, and $\tilde{N}_{i,k}$ are necessary.
\end{enumerate}

We then calculate the estimated counterfactual risk of failure by the end of period $k$ as
$
   \hat{E}(Y^{\bar{s}_k}_{k}) = \frac{1}{n}\sum_{i=1}^{n} \tilde Y_{i,k}
$. 

As with the \ac{IPW} method, confidence intervals can be obtained using bootstrapping, by performing the full G-computation procedure within each bootstrap sample.

\subsection{Targeted maximum likelihood estimation}

Targeted learning provides a doubly robust and efficient estimation methodology that can easily incorporate machine learning models to data-adaptively estimate nuisance parameters on which estimation of the estimand of interest relies. To describe the TMLE approach, we first note that the g-formula in (\ref{eq:gform_nice}) can be re-expressed as a series of iterated conditional expectations of the observed outcome \cite{petersen_targeted_2014}:

\begin{equation}
\begin{aligned}
\mathrm{E}\!\left(Y_k^{\bar{s}_k} \right)
&= \mathrm{E}\Bigg[
    \mathrm{E}\Bigg(
        \cdots
        \mathrm{E}\Bigg[
            \mathrm{E}\Big(
                Y_k
                \,\Big|\,
                \bar{A}_k = \bar{s}_k,
                \bar{L}_k,
                \bar{N}_{k-1},
                Y_{k-1}=0
            \Big)
\\
&\qquad\qquad\Big|\,
            \bar{A}_{k-1} = \bar{s}_{k-1},
            \bar{L}_{k-1},
            \bar{N}_{k-2},
            Y_{k-2}=0
        \Bigg]
        \cdots
\\
&\qquad\Big|\,
        A_0 = s_0,
        L_0
    \Bigg)
\Bigg]
\end{aligned}
\label{eq:gform_ice}
\end{equation}

Whereas the g-formula in (\ref{eq:gform_nice}) requires integrating out the time-dependent confounders and monitoring variables ($L_k$ and $N_k$), the g-formula in (\ref{eq:gform_ice}) requires sequentially marginalising the distribution of the outcome over current $L_k$ and $N_k$, given the intervention of interest up to $k$, and given the history of $L$ and $N$ up to time $k-1$. Therefore integration of the time-dependent confounders and monitoring variables is not required. The version of the g-formula in (\ref{eq:gform_ice}) has been referred to as the \ac{ICE} g-formula estimator, in contrast to the \ac{NICE} g-formula estimator in (\ref{eq:gform_nice}) \cite{wen_parametric_2021}. 

The g-formula expression in (\ref{eq:gform_ice}) can be evaluated using a variation of the G-computation approach described in the previous section, by specifying and fitting conditional models for the outcome, and then simulating from these according to the intervention of interest (starting with $Y_k$ and working backwards to $Y_0$). However, this approach relies on consistent estimation of the conditional expectations in each step, typically based on parametric regression models, which are susceptible to mis-specification, particularly in this setting with informative monitoring where the relationships between $Y_k$, $\bar{L}_k$, $\bar{A}_k$ and $\bar{N}_k$ are complex. Instead the longitudinal \ac{TMLE} estimator, developed by van der Laan and Gruber \cite{laan_targeted_2012}, involves incorporating an additional targeting step at each time-point, to de-bias the initial fits $\hat{\mathrm{E}}(Y_k|\bar{A}_k=\bar{s}_k,\bar{L}_k,\bar{N}_{k-1},Y_{k-1}=0)$. 

The targeting step involves making use of the same weights as are used in the IPW method. We start by estimating these time-dependent weights using the formula from (\ref{eq:weights}).

To estimate $E(Y^{\bar{s}_k}_k)$ we then carry out the following steps:

\begin{enumerate}
    \item Fit a model for the outcome $Y_k$ using the observed data conditional on past treatment, monitoring and observed covariate history up to time $k$ ($\bar{A}_k,\bar{L}_k,\bar{N}_{k-1}$) within subjects that remain alive in period $k-1$. 
    \item For each individual alive in period $k-1$, obtain initial predicted probabilities $\hat{Q}^{\bar{s}_k}_{k} = \hat{\Pr} (Y_k = 1|\bar{A}_k = \bar{s}_k,\bar{L}_k, \bar{N}_{k-1},Y_{k-1}=0)$ by plugging in to the fitted model in step (1) the observed values of $\bar{L}_k$ and $\bar{N}_{k-1}$ and the values of treatment compatible with the treatment strategy of interest, $\bar{A}_k=\bar{s}_k$. For each $j = 0,...,k$ this involves setting $A_{j}=a_{j}$ during non \textit{grace-periods} and letting $A_{j}$ take its natural value (according to the observed data) during \textit{grace-periods}. We note that whether an individual is in a \textit{grace-period} or not can depend on their own history of treatment and therefore the timings of \textit{grace-periods} are not necessarily common across individuals. 
    \item To improve the initial predictions with respect to the target parameter, fit an intercept-only logistic regression of the outcome $Y_k$, using the initial estimates $\hat{Q}^{\bar{s}_k}_{k}$ as an offset term and with weights $\hat{w}_{k}$. This model is fitted within subjects who remain alive in period $k-1$ and are following the treatment strategy of interest up to $k$. Obtain updated predictions from this model $\hat{Q}^{\bar{s}_k,*}_{k}$.
\end{enumerate}

Steps (1) to (3) are repeated for $j=k-1,...,0$, with the only difference being that the initial model in step (1) and the updated intercept-only logistic regression model in step (3) are fitted with the updated prediction of the conditional outcome from the previous iteration ($\hat{Q}^{\bar{s}_{j+1},*}_{j+1}$) as the outcome instead of the observed $Y_k$. In more detail, the steps are, for $j=k-1,...,0$:

\begin{enumerate}
\setcounter{enumi}{3}
    \item Fit a model for the outcome $\hat{Q}^{\bar{s}_{j+1},*}_{j+1}$ using the observed data conditional on past treatment, monitoring and observed covariate history ($\bar{A}_j,\bar{L}_j,\bar{N}_{j-1}$) within subjects that remain alive in period $j-1$. 
    \item For each individual obtain initial predicted probabilities $\hat{Q}^{\bar{s}_{j}}_{j}$ by plugging in to the fitted model in step (4) the observed values of $\bar{L}_j$ and $\bar{N}_{j-1}$ and the values of treatment compatible with the treatment strategy of interest, $\bar{A}_j=\bar{s}_j$. 
    \item Fit an intercept-only logistic regression of the outcome $\hat{Q}^{\bar{s}_{j+1},*}_{j+1}$, using the initial estimates $\hat{Q}^{\bar{s}_{j}}_{j}$ as an offset term and with weights $\hat{w}_{j}$ within subjects who remain alive and uncensored until period $j-1$ and are following the treatment strategy of interest up to $j$. Obtain updated predictions from this model $\hat{Q}^{\bar{s}_j,*}_{j}$ 

\end{enumerate}

Steps (4) to (6) are repeated from $j=k-1$ to $j=0$ until we obtain updated predictions $\hat{Q}^{s_0,*}_{0}$. These final predictions are averaged over all subjects to obtain the TMLE estimate of the target parameter of interest $\hat{E}(Y^{\bar{s}_k}_{k})=\frac{1}{n} \sum_{i=1}^n \hat{Q}^{s_0,*}_{i,0}$.

The above steps result in an estimate of the risk up to time $k$. Where we wish to estimate the full survival curve, we need to repeat the entire procedure for each $k=0,...,K$ using a subset of the data cut off at the end of each observation period $k$ to target each outcome $Y_k$. 

The \ac{TMLE} estimator provides the solution to the efficient influence curve equation,  making it a doubly robust locally efficient substitution estimator.  Because of this, we can obtain standard error estimates and confidence intervals based on the sample variance of the efficient influence function, which results in conservative 95\% confidence intervals \cite{petersen_targeted_2014}. The double-robustness property means that the estimator is consistent as long as either the outcome model or the treatment model is estimated consistently. The \ac{TMLE} estimator is also locally efficient, meaning that the estimator has minimal large sample variance among estimators that make the same model assumptions, when both the outcome and treatment models are correctly specified. Furthermore, \ac{TMLE} allows us to minimise the impact of model mis-specification, since both outcome and treatment models may be fitted using data-adaptive machine learning techniques while the final estimator still yields valid asymptotic properties for statistical inference \cite{laan_targeted_2011, bang_doubly_2005}.
\section{Simulation study}
\label{sec:simulation_study}

\subsection{Design}

To evaluate the performance of the methods described in section \ref{sec:methods} we conducted a simulation study, using the ADEMP framework set out by Morris et al. \cite{morris_using_2019}. R code for reproducing the simulation results is available on \href{https://github.com/leahpirondini/informative-monitoring}{GitHub}.

\subsubsection{Data-generating mechanism}
\label{sec:data_generating_mechanism}

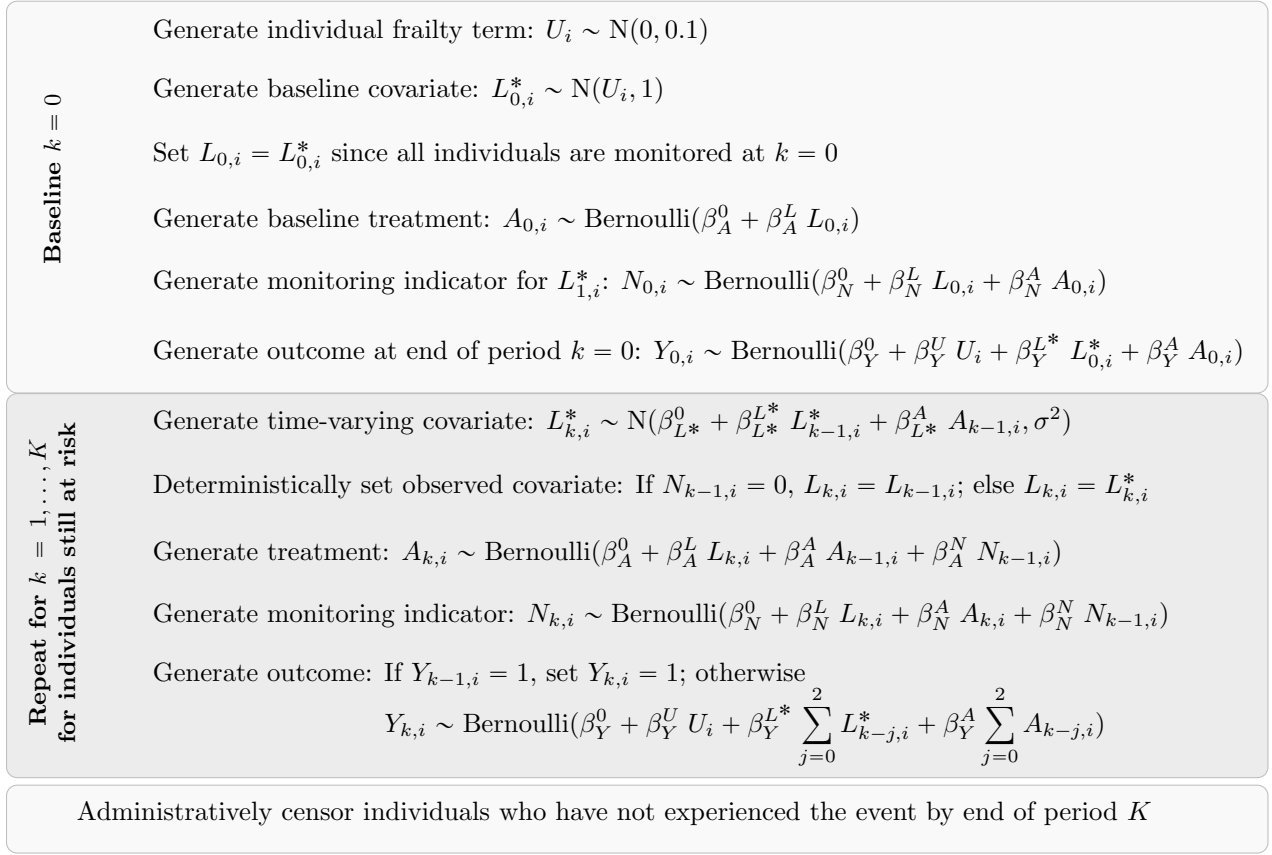
\begin{figure}[H]

\begin{tikzpicture}[
    lightgreybox/.style={
        fill=gray!5,
        draw=gray!50,
        rounded corners
    },
    darkgreybox/.style={
        fill=gray!15,
        draw=gray!50,
        rounded corners
    }
]

% =========================
% Matrix (single source of truth)
% =========================
\matrix (m) [
    matrix of nodes,
    nodes={anchor=west},
    row sep=2mm,
    anchor=north west
] at (0,0) {

Generate individual frailty term:
$U_i \sim \mathrm{N}(0,0.1)$ \\

Generate baseline covariate:
$L^*_{0,i} \sim \mathrm{N}(U_i,1)$ \\

Set $L_{0,i} = L^*_{0,i}$ since all individuals are monitored at $k=0$ \\

Generate baseline treatment:
$A_{0,i} \sim \mathrm{Bernoulli}(\beta_A^0 + \beta_A^L \; L_{0,i})$ \\

Generate monitoring indicator for $L^*_{1,i}$:
$N_{0,i} \sim \mathrm{Bernoulli}(\beta_N^0 + \beta_N^L \; L_{0,i} + \beta_N^A \; A_{0,i})$ \\

Generate outcome at end of period $k=0$:
$Y_{0,i} \sim \mathrm{Bernoulli}(\beta_Y^0 + \beta_Y^U \; U_i + \beta_Y^{L^*} \; L^*_{0,i} + \beta_Y^A \; A_{0,i})$ \\

Generate time-varying covariate:
$L^*_{k,i} \sim \mathrm{N}(\beta_{L^*}^0 + \beta_{L^*}^{L^*} \; L^*_{k-1,i} + \beta_{L^*}^A \; A_{k-1,i},\sigma^2)$ \\

Deterministically set observed covariate:
If $N_{k-1,i}=0$, $L_{k,i}=L_{k-1,i}$; else $L_{k,i}=L^*_{k,i}$ \\

Generate treatment:
$A_{k,i} \sim \mathrm{Bernoulli}(\beta_A^0 +\beta_A^L \; L_{k,i} + \beta_A^A \; A_{k-1,i} + \beta_A^N \; N_{k-1,i})$ \\

Generate monitoring indicator:
$N_{k,i} \sim \mathrm{Bernoulli}(\beta_N^0 + \beta_N^L \; L_{k,i} + \beta_N^A \; A_{k,i} + \beta_N^N \; N_{k-1,i})$ \\

Generate outcome:
\parbox[t]{11cm}{%
If $Y_{k-1,i}=1$, set $Y_{k,i}=1$; otherwise\\[-0.5ex]
$\displaystyle
Y_{k,i} \sim \mathrm{Bernoulli}(
\beta_Y^0 +\beta_Y^U \; U_i + \beta_Y^{L^*} \sum_{j=0}^2 L^*_{k-j,i} + \beta_Y^A \sum_{j=0}^2 A_{k-j,i} ) $ } \\
};

% =========================
% Labels (row-aligned)
% =========================
\node[rotate=90, font=\small\bfseries]
at ($(m-3-1.west)+(-1.2cm,-0.3cm)$)
{Baseline $k=0$};

\node[rotate=90, 
      font=\small\bfseries,
      text width=5cm,
      align=center]
at ($(m-9-1.west)+(-1.2cm,-0.4cm)$)
{Repeat for $k=1,\ldots,K$ \\ 
for individuals still at risk};

% =========================
% Bottom note (width tied to matrix)
% =========================
\node[
    anchor=north west,
    text width=16cm,
    align=left
] at ($(m-11-1.south west)+(-1cm,-0.2cm)$)
{Administratively censor individuals who have not experienced the event by end of period $K$};

% =========================
% Background boxes (auto-sized)
% =========================
\begin{scope}[on background layer]

\draw[lightgreybox]
  ($(m-1-1.north west)+(-1.8cm,0.1cm)$)
  rectangle
  ($(m-11-1.south east)+(0.5cm,5.1cm)$);

\draw[darkgreybox]
  ($(m-7-1.north west)+(-1.8cm,0cm)$)
  rectangle
  ($(m-11-1.south east)+(0.5cm,0cm)$);

\draw[lightgreybox]
  ($(m-11-1.south west)+(-1.8cm,-0.1cm)$)
  rectangle
  ($(m-11-1.south east)+(0.5cm,-1cm)$);

\end{scope}

\end{tikzpicture}

\caption{Data generating mechanism: steps to generate longitudinal data according to the \ac{DAG} in figure \ref{fig:dag_im2} (with the addition of an individual frailty term $U_i$) for individuals $i=1,...,n$.}
\label{fig:dat_gen_mech}
\end{figure}

We simulated data on $n$ individuals at 5 potential observation times ($k=0,1,...,4$) for a single, continuous covariate $L^*_k$, monitored informatively, with corresponding binary monitoring decision variable $N_k$ and observed covariate $L_k$, a binary treatment $A_k$, and a binary outcome $Y_k$ indicating whether the outcome has occurred by the end of period $k$. The data-generating process produces longitudinal data according to the \ac{DAG} in figure \ref{fig:dag_im2}, with the addition of a time-fixed continuous variable $U$ representing an individual frailty term. The steps to generate the longitudinal simulated data are set out in figure \ref{fig:dat_gen_mech}. We considered a main simulation scenario and then four alternative simulation scenarios in which parameter values were changed to vary the strengths of relationships between variables. Alternative scenarios are described in table \ref{tab:sim_scenario_desc} and parameter values for the five simulation scenarios are set out in supplementary table S1. For each scenario we generated 1000 simulated data sets each containing $n=3000$ individuals.

% --------------------------------------------------------------------
% --------------------------------------------------------------------
% FIGURE FOR DATA GENERATING PROCESS
% --------------------------------------------------------------------
% --------------------------------------------------------------------

% --------------------------------------------------------------------
% --------------------------------------------------------------------
% END OF FIGURE
% --------------------------------------------------------------------
% --------------------------------------------------------------------

% --------------------------------------------------------------------
% --------------------------------------------------------------------
% TABLE: SIMULATION SCENARIO DESCRIPTIONS
% --------------------------------------------------------------------
% --------------------------------------------------------------------

\begin{table}[H]
\centering
\begin{adjustbox}{width=0.95\textwidth}
\begin{tabular}{lll}
\rowcolor[HTML]{D9D9D9} 

\textbf{Scenario number} & 
\textbf{Scenario description} & 
\textbf{Parameter values changed}  \\

\rowcolor[HTML]{FFFFFF} 

1 & 
Main simulation scenario & 
N/A \\

\rowcolor[HTML]{F2F2F2}

2 & 
Stronger effects of past variables on monitoring $N_k$ & 
$\beta_N^0$, $\beta_N^L$, $\beta_N^A$, $\beta_N^N$ \\

\rowcolor[HTML]{FFFFFF}

3 & 
No effect of past variables on monitoring $N_k$ & 
$\beta_N^0$, $\beta_N^L=0$, $\beta_N^A=0$, $\beta_N^N=0$ \\

\rowcolor[HTML]{F2F2F2}

4 & 
Stronger effect of $N_{k-1}$ on treatment $A_k$ & 
$\beta_A^N$
\\

\rowcolor[HTML]{FFFFFF}

5 & 
No effect of $N_{k-1}$ on treatment $A_k$ &
$\beta_A^N=0$

\end{tabular}
\end{adjustbox}
\caption{Simulation scenario descriptions and overview of parameter changes relative to main simulation scenario}
\label{tab:sim_scenario_desc}
\end{table}

% --------------------------------------------------------------------
% --------------------------------------------------------------------
% END OF TABLE
% --------------------------------------------------------------------
% --------------------------------------------------------------------

\subsubsection{Treatment strategies and estimands}

\begin{figure}[H]
\begin{center}
\begin{tikzpicture}[
  >=Stealth,
  thick,
  every node/.style={font=\small},
  timeline/.style={->, thick},
  ventilate/.style={thick, solid},
  noVent/.style={thick, dotted},
  natural/.style={thick, dashed},
  labelStyle/.style={align=left, text width=5cm}
]

% length of a 1 day interval
\def\oneday{3}

% draw the x-axis and label it
\draw[-{Stealth}, thick] (0,0) -- (15.5,0);
\node at ({0.5*15}, -0.8) {Days from ICU admission};

% add ICU circle
%\draw[fill=gray!10] (0,-1.5) circle (0.5);
%\node at (0,-1.5) {ICU};

% tick marks
\foreach \x in {0,...,5}
  \draw ({\x*\oneday},0.1) -- ({\x*\oneday},-0.1) node[below] {\x};

% strategy positions
\def\rowA{3}
\def\rowB{1}

% ---------------
% Ventilate early
% ---------------

% treatment strategy name
\node[draw=black, text=black, thick, align=left, text width=1.5cm] at (-1,\rowA) {\textbf{Ventilate\\early}};

% arrows
\draw[{Stealth}-{Stealth}, thick, black] (0,\rowA) -- (\oneday*3,\rowA);
\draw[natural, black, thick, {Stealth}-{Stealth}] (\oneday*3,\rowA) -- (\oneday*5,\rowA);

% labels
\node[above, text=black, labelStyle, align=center, text width=\oneday*3cm] at (\oneday*3*0.5,\rowA+0.35) {Initiate ventilation at time 0 and remain ventilated for at least 3 days};
\node[above, text=black, labelStyle, align=center] at (\oneday*4,\rowA+0.35) {Allow ventilation to follow natural course};

% -----------------
% Wait to ventilate
% -----------------

% treatment strategy name
\node[draw=black, text=black, thick, align=left, text width=1.5cm] at (-1,\rowB) {\textbf{\textcolor{black}{Wait to\\ventilate}}};

% arrows
\draw[{Stealth}-{Stealth}, thick, black] (0,\rowB) -- (\oneday*2,\rowB);
\draw[{Stealth}-, thick, black, dashed] (\oneday*2,\rowB) -- (\oneday*3,\rowB);
\draw[thick, black] (\oneday*3,\rowB) -- (\oneday*4,\rowB);
\draw[-{Stealth}, thick, black, dashed] (\oneday*4,\rowB) -- (\oneday*5,\rowB);

% labels
\node[above, text=black, labelStyle, align=center, text width=\oneday*2cm] at (\oneday*2*0.5,\rowB+0.35) {No ventilation};

\node[above, text=black, labelStyle, align=center, text width=\oneday*3cm] at (\oneday*3.5,\rowB+0.35) {Initiate ventilation on either day 2 or day 3 and remain on ventilation for at least 2 days};

\end{tikzpicture}
\end{center}
\caption{Ventilation status under treatment strategies \textit{ventilate early} and \textit{wait to ventilate} for 5 potential observation times. In the \textit{ventilate early} strategy the solid line indicates that the patient must be ventilated ($A_0=A_1=A_2=1$), and the dashed line indicates that ventilation status is allowed to take its natural value. In the \textit{wait to ventilate} strategy the first solid line indicates that the patient must remain unventilated ($A_0=A_1=0$), and the mixed dashed and solid line indicates that the patient must be ventilated for at least two consecutive time periods within this three period interval.}
\label{fig:treatment_strategies_5tps}
\end{figure}
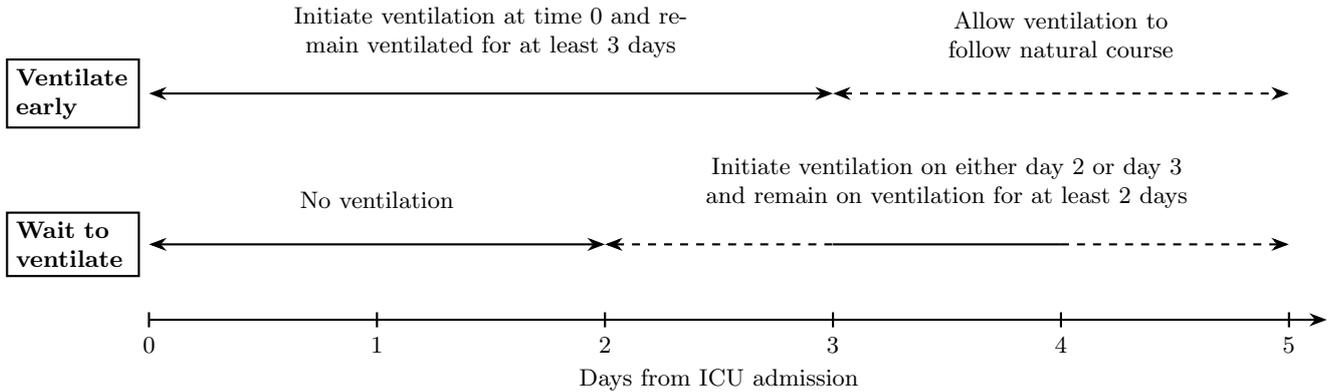

The estimand of interest is the marginal survival probability $\Pr(Y^{\Bar{s}_k}_k=1)$ ($k=1,\ldots,5$) under the following four treatment strategies $\Bar{s}_k$: \textit{always ventilate}, \textit{never ventilate}, \textit{ventilate early} and \textit{wait to ventilate}. The \textit{always ventilate} and \textit{never ventilate} strategies involve setting $A_k=1$ and $A_k=0$, respectively, for the duration of follow-up. The \textit{ventilate early} and \textit{wait to ventilate} strategies, which aim to better reflect more realistic clinical decision making in the context of our motivating example of invasive ventilation in an \ac{ICU} setting, are described in more detail below and illustrated in figure \ref{fig:treatment_strategies_5tps}. 

The \textit{ventilate early} and \textit{wait to ventilate} strategies are examples of \textit{natural grace-period treatment strategies} \cite{wanis_grace_2024}, with both treatment strategies allowing for \textit{grace-periods} when ventilation is allowed to follow its natural course, and additionally with the \textit{wait to ventilate} strategy allowing a window of time (rather than a specific time point) when ventilation must be initiated. In our simulation example with five potential observation times, the \textit{ventilate early} strategy involves setting $A_0=1$, $A_1=1$ and $A_2=1$, and letting $A_3$ and $A_4$ take their natural value. This reflects a ventilation strategy that requires \ac{IMV} to be initiated immediately and sustained for a set duration, but subsequently allows the treating clinicians to wean the patient from the ventilator when they see necessary according to the specific patient's needs. In the observed data, individuals following this strategy until the end of follow-up would have treatment trajectories $\Bar{A}=\{ 1,1,1,0,0 \}$, $\Bar{A}=\{ 1,1,1,0,1 \}$, $\Bar{A}=\{ 1,1,1,1,0 \}$ or $\Bar{A}=\{ 1,1,1,1,1 \}$. The \textit{wait to ventilate} strategy involves delaying ventilation by two time points, i.e. setting $A_0=0$ and $A_1=0$, and then initiating ventilation in one of the subsequent two periods ($k=2$ or $k=3$), then remaining ventilated for at least two consecutive periods. In the observed data, individuals following this strategy until the end of follow-up would have treatment trajectories $\Bar{A}=\{ 0,0,0,1,1 \}$, $\Bar{A}=\{ 0,0,1,1,0 \}$ or $\Bar{A}=\{ 0,0,1,1,1 \}$.

% -----------------------------------------------------------------
% -----------------------------------------------------------------
% START OF FIGURE
% -----------------------------------------------------------------
% -----------------------------------------------------------------

% -----------------------------------------------------------------
% -----------------------------------------------------------------
% END OF FIGURE
% -----------------------------------------------------------------
% -----------------------------------------------------------------

\subsubsection{Methods accommodating monitoring}

The estimands of interest were estimated using \ac{IPW}, G-computation, and \ac{TMLE}, as described in section \ref{sec:methods}. For \ac{IPW} we fit propensity score models of the same form as used in the data-generating process using logistic regression. For G-computation we need to specify parametric models for treatment $A_k$, monitoring $N_k$, observed covariate $L_k$ and outcome $Y_k$, conditional on past variables, using only the variables observed (i.e. not $L^*_k$). We specify models for $A_k$ and $N_k$ of the same form as those used in the data-generating process using logistic regression. However, given that we do not generate $L_k$ directly from a model in the data-generating process (rather $L_k$ is assigned deterministically using the values of $L^*_k$, $L_{k-1}$ and $N_{k-1}$), we must choose a suitable parametric regression model for $L_k$ conditional on only the variables observed in the real-world data. We used a model of the form

\begin{equation}
\begin{split}
    \mathrm{E}(L_k|\overline{A}_{k-1},\overline{L}_{k-1},\overline{N}_{k-1}) =&
    L_{k-1} + \alpha_1 N_{k-1} + \alpha_2 L_{k-1} N_{k-1} + \\
    &\alpha_3 L_{k-1} N_{k-1} N_{k-2} + \alpha_4 A_{k-1} N_{k-1} + \alpha_5 A_{k-1} N_{k-1} N_{k-2}.
\end{split}
\label{eq:model_L}
\end{equation}

We fit this model with no intercept and with $L_{k-1}$ as an offset term so that when $N_{k-1}=0$ (\ref{eq:model_L}) reduces to $\mathrm{E}(L_k|\overline{A}_{k-1},\overline{L}_{k-1},\overline{N}_{k-1}) = L_{k-1}$, and when $N_{k-1}=1$ and $N_{k-2}=1$ (and hence $L_{k}=L^*_{k}$ and $L_{k-1}=L^*_{k-1}$) (\ref{eq:model_L}) simplifies to the correctly-specified model for $L^*_k$ from the data-generating process. We follow similar logic to specify a suitable model for $Y_k$ in terms of $L_k$, $N_k$ and $A_k$, including appropriate interactions terms between $L_k$ and $N_k$ to allow for the fact that $L_k$ is only a confounder when it is observed. We used a model of the form
\begin{equation}
\begin{split}
    logit ( \Pr (Y_k=1|\overline{A}_{k-1},\overline{L}_{k-1},\overline{N}_{k-2}) ) = &\beta_0 + \beta_1 A_{k-1} + \beta_2 A_{k-2} + \beta_3 A_{k-3} + \\
    &\beta_4 N_{k-2} L_{k-1} + \beta_5 N_{k-3} L_{k-2} + \beta_6 N_{k-4} L_{k-3}.
\end{split}
\label{eq:model_Y}
\end{equation}

We implemented G-computation using the \texttt{gfoRmula} package in R, which can incorporate user-defined functions for the flexible treatment strategies \textit{ventilate early} and \textit{wait to ventilate}.

As with G-computation, \ac{TMLE} requires estimation of the observed outcome $Y_k$ using only those variables observed in the real-world data ($\Bar{L}_k$, $\Bar{A}_k$ and $\Bar{N}_k$). We fit a parametric model of the same form as \ref{eq:model_Y}. As with \ac{IPW}, for the propensity score model we fit a parametric model of the same form as used in the data-generating process using logistic regression. 

To estimate both the observed outcome and propensity score we could also make use of the SuperLearner which uses a range of machine learning methods to build the best weighted combination of algorithms. A detailed guide to the SuperLearner can be found here \cite{phillips_practical_2023}. We make use of the SuperLearner in our real-world ventilation application where we have a larger number of potential confounders and an unknown data-generating process.

We implemented \ac{TMLE} using the \texttt{lmtp} package, which, like \texttt{gfoRmula}, can incorporate user-defined functions for the flexible treatment strategies \textit{ventilate early} and \textit{wait to ventilate}.

\subsubsection{Naive versions of the methods ignoring monitoring}

We also implemented \textit{naive} versions of the three methods described above. The idea was to remove the variable $N_k$ from our observed dataset, and to treat the observed covarariate $L_k$ as if it were the true, underlying covariate $L^*_k$, as is often done in practical implementation of causal inference methods for time-dependent confounding, when covariates are monitored irregularly. For naive \ac{IPW} this involved amending the propensity score model to the form $logit(A_k|L_{k},A_{k-1}) = \alpha_0 + \alpha_1 A_{k-1} + \alpha_2 L_{k}$ (i.e. omitting $N_k$). For naive G-computation, we omitted the model for $N_k$ and fitted models for $L_k$ and $Y_k$ of the forms
$\mathrm{E}(L_k|L_{k-1},A_{k-1}) = \gamma_0 + \gamma_1 L_{k-1} + \gamma_2 A_{k-1} $ and $\mathrm{E}(Y_k|\Bar{L}_{k},\Bar{A}_{k}) = \gamma_0 + \gamma_1 (L_{k} + L_{k-1} + L_{k-2}) + \gamma_2 (A_{k} + A_{k-1} + A_{k-2})$. For naive \ac{TMLE} we amended the propensity score model to the form $logit(A_k|L_{k},A_{k-1}) = \alpha_0 + \alpha_1 A_{k-1} + \alpha_2 L_{k}$ and the model for the observed outcome to the form $\mathrm{E}(Y_k|\Bar{L}_{k},\Bar{A}_{k}) = \gamma_0 + \gamma_1 (L_{k} + L_{k-1} + L_{k-2}) + \gamma_2 (A_{k} + A_{k-1} + A_{k-2})$.

\subsubsection{Obtaining true values}

To obtain true values of the estimands we used a simulation-based approach \cite{keogh_simulating_2021}. This involves generating longitudinal data in a similar way to that described in the simulation algorithm, where the relationships between the variables are the same as in the observational study, with the exception that, for time-periods where the treatment intervention involves setting $A_k=a_k$, $L_k$ and $N_k$ do not affect the value of $A_k$. Instead, $A_k$ is set by intervention to the fixed value determined by the treatment regime. For the \textit{always ventilate} and \textit{never ventilate} strategies this involves setting $A_k=1$ and $A_k=0$, respectively, for all $k$. For the \textit{ventilate early} strategy this involves setting $A_0=A_1=A_2=1$ and then allowing $A_3$ and $A_4$ to follow their natural (simulated) values as in the observational data generation process. For the \textit{wait to ventilate} strategy the process is as follows:
\begin{enumerate}
    \item Set $A_0=0$ and $A_1=0$
    \item Allow $A_2$ to follow its natural (simulated) value
    \item Set $A_3=1$ (as individuals must initiate ventilation by this point)
    \item Set $A_4=1$ for individuals with $A_2=0$ (because we require two consecutive periods of ventilation), otherwise, for individuals with $A_2=1$, allow $A_4$ to follow its natural (simulated) value.
\end{enumerate} 

We generated data in this way for 1,000,000 individuals under each of the four treatment strategies. For each treatment strategy estimates of the survival probability at each time-point were obtained using simple proportions, since there is no time-dependent confounding and only administrative censoring in these data.

\subsubsection{Performance measures}

For each estimand we present the mean value of the estimates across all simulations, and the corresponding bias. We also obtain empirical standard errors of the estimates as the standard deviation of the estimates across all simulations. For bias we obtain Monte Carlo standard errors \cite{morris_using_2019}.

\subsection{Results}

\begin{figure}[H]
\centering
\hspace*{-1cm}
\includegraphics[width=1\columnwidth]{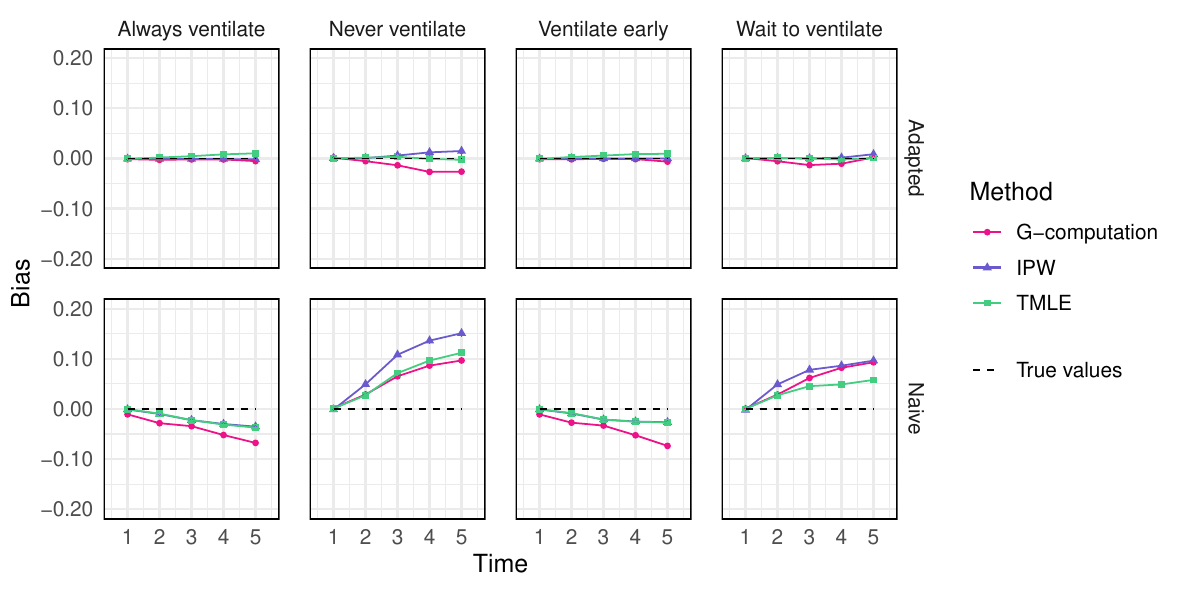}
\caption{Bias (shown as a percentage of true survival probability) of the estimated survival probabilities for the treatment strategies \textit{always ventilate}, \textit{never ventilate}, \textit{ventilate early} and \textit{wait to ventilate} at observation times 1-5 obtained using adapted (top panel) and naive (bottom panel) versions of IPW, G-computation and TMLE from 1000 simulations.}
\label{fig:bias_mc_se}
\end{figure}

Results from the simulation study are summarised in tables \ref{tab:sim_results_naive} and \ref{tab:sim_results_adapted} and figure \ref{fig:bias_mc_se}. \textit{Naive} methods that ignore informative monitoring (by removing $N_k$ from the observed data and treating $L_k$ as if it were the true, underlying covariate $L^*_k$) result in biased estimates of the survival probabilities. This bias increases over time as estimated cumulative \ac{IPW} weights deviate further from their true values and the outcome model becomes increasingly misspecified. The amount of bias varies by method type. For \ac{IPW}, bias occurs because we misspecify the propensity score model. For G-computation, bias occurs because we misspecify both the outcome model and the confounder models. For \ac{TMLE} bias occurs because we misspecify both the outcome and propensity score models. The bias resulting from \ac{TMLE} is generally only as large as the smaller of the biases resulting from \ac{IPW} or G-computation. 

Results from the alternative simulation scenarios, in which we vary the informativeness of the monitoring (i.e. on the strength of the effects of $\Bar{L}_k$, $\Bar{A}_k$ and $\Bar{N}_{k-1}$ on $N_k$) and the extent to which monitoring affects treatment decisions (i.e. on the strength of the effect of $\Bar{N}_{k-1}$ on $A_k$), are presented in the supplementary material (tables S2-S5 and figures S1-S4). We see that increasing the strength of the effects of $\Bar{L}_k$, $\Bar{A}_k$ and $\Bar{N}_{k-1}$ on $N_k$ primarily affects the estimates from naive G-computation, with increased bias due to the fact that the outcome and confounder models are increasingly misspecified. Removing the effects of $\Bar{L}_k$, $\Bar{A}_k$ and $\Bar{N}_{k-1}$ on $N_k$ (i.e. simulating $N_k \sim Bernoulli(0.5)$) does not remove bias from the naive estimates. This is because the backdoor pathways from $A_k$ to $Y_{k+1}$ via $N_{k-1}$ highlighted in the DAG in figure \ref{fig:dag_im} are still present even when $N$ is random. Conversely, increasing the strength of the effect of $\Bar{N}_{k-1}$ on $A_k$ primarily affects the estimates from naive \ac{IPW} and naive \ac{TMLE}, with increased bias due to the fact that the propensity score model is increasingly misspecified. Removing the effect of $\Bar{N}_{k-1}$ on $A_k$ (i.e. setting the parameter $\beta_{a_N}=0$) removes the bias from the naive \ac{IPW} and naive \ac{TMLE} estimates, but not from the naive G-computation estimates, because these still rely on correct specification of models for $L_k$ and $Y_k$ in terms of $\Bar{L}_k$, $\Bar{A}_k$ and $\Bar{N}_{k}$.

All three methods that account for informative monitoring gave approximately unbiased estimates. The small bias in some of the survival probabilities might be due to finite sample bias, especially as this tends to occur more at later observation times as the numbers of individuals still alive and following each strategy of interest decreases. The empirical standard errors are larger for \ac{IPW} compared with G-computation. This is expected, as inverse probability weights can become large and unstable, especially in a longitudinal setting, but the more precise estimates typically come at the cost of more opportunity for model mis-specification. Survival probabilities estimated using \ac{TMLE} result in the smallest bias overall across the range of different treatment strategies, with empirical standard errors typically falling somewhere between those from \ac{IPW} and G-computation.

In the alternative simulation scenarios, all three methods that account for informative monitoring give approximately unbiased estimates, with the exception of G-computation in scenario 3, where we simulate $N_k \sim Bernoulli(0.5)$, as bias is introduced by overspecifying the model for $N_k$ in terms of $\Bar{L}_k$, $\Bar{A}_k$ and $\Bar{N}_{k-1}$.

% *******************************************************
% *******************************************************
% Results table: NAIVE METHODS
% *******************************************************
% *******************************************************

\begin{table}[H]
\centering
\large
\begin{adjustbox}{width=1\textwidth}
\begin{tabular}{p{1.5cm} p{2cm} p{2.5cm} p{2.5cm} p{2.5cm} p{2.5cm} p{2.5cm} p{2.5cm}}

% ------------------
% row 1
% ------------------
\rowcolor[HTML]{D9D9D9} 
 &  & \multicolumn{2}{c}{\textbf{Naïve IPW}}      & \multicolumn{2}{c}{\textbf{Naïve G-computation}} & \multicolumn{2}{c}{\textbf{Naïve TMLE}} \\ 
\cline{3-8}

% ------------------
% row 2
% ------------------
\rowcolor[HTML]{D9D9D9} 
\textbf{Time} & \textbf{True value} & \textbf{Mean estimate (Empirical SE)} & \textbf{Bias (Monte Carlo SE)}         & \textbf{Mean estimate (Empirical SE)}  & \textbf{Bias (Monte Carlo SE)}  & \textbf{Mean estimate (Empirical SE)}   & \textbf{Bias (Monte Carlo SE)} \\

% ------------------
% row 3
% ------------------
\rowcolor[HTML]{F2F2F2} 
\multicolumn{2}{l}{\textbf{Ventilate early}} & & & & & & \\

% ------------------
% rows 4-8
% ------------------
\rowcolor[HTML]{FFFFFF} 
1 & 0.717 & 0.716 (0.010) & -0.001 (0.000) & 0.709 (0.007) & -0.008 (0.000) & 0.716 (0.009) & -0.000 (0.000) \\
\rowcolor[HTML]{F2F2F2} 
2 & 0.598 & 0.593 (0.011) & -0.005 (0.000) & 0.582 (0.009) & -0.016 (0.000) & 0.593 (0.011) & -0.005 (0.000) \\
\rowcolor[HTML]{FFFFFF} 
3 & 0.559 & 0.547 (0.012) & -0.012 (0.000) & 0.540 (0.009) & -0.019 (0.000) & 0.547 (0.012) & -0.012 (0.000) \\
\rowcolor[HTML]{F2F2F2} 
4 & 0.546 & 0.532 (0.013) & -0.014 (0.000) & 0.517 (0.009) & -0.029 (0.000) & 0.532 (0.012) & -0.014 (0.000) \\
\rowcolor[HTML]{FFFFFF} 
5 & 0.539 & 0.525 (0.013) & -0.014 (0.000) & 0.499 (0.010) & -0.040 (0.000) & 0.524 (0.012) & -0.015 (0.000) \\

% ------------------
% row 9
% ------------------
\rowcolor[HTML]{F2F2F2} 
\multicolumn{2}{l}{\textbf{Wait to ventilate}} & & & & & & \\

% ------------------
% rows 10-14
% ------------------
\rowcolor[HTML]{FFFFFF} 
1 & 0.656 & 0.655 (0.009) & -0.001 (0.000) & 0.657 (0.008) & 0.000 (0.000)  & 0.656 (0.009) & 0.000 (0.000)  \\
\rowcolor[HTML]{F2F2F2} 
2 & 0.432 & 0.453 (0.012) & 0.021 (0.000)  & 0.444 (0.009) & 0.012 (0.000)  & 0.443 (0.012) & 0.012 (0.000)  \\
\rowcolor[HTML]{FFFFFF} 
3 & 0.321 & 0.347 (0.011) & 0.025 (0.000)  & 0.341 (0.008) & 0.020 (0.000)  & 0.336 (0.011) & 0.015 (0.000)  \\
\rowcolor[HTML]{F2F2F2} 
4 & 0.274 & 0.297 (0.012) & 0.024 (0.000)  & 0.296 (0.007) & 0.022 (0.000)  & 0.287 (0.011) & 0.013 (0.000)  \\
\rowcolor[HTML]{FFFFFF} 
5 & 0.254 & 0.278 (0.013) & 0.025 (0.000)  & 0.278 (0.007) & 0.024 (0.000)  & 0.269 (0.012) & 0.015 (0.000)  \\

% ------------------
% row 15
% ------------------
\rowcolor[HTML]{F2F2F2} 
\multicolumn{2}{l}{\textbf{Always ventilate}} & & & & & & \\

% ------------------
% rows 16-20
% ------------------
\rowcolor[HTML]{FFFFFF} 
1 & 0.717 & 0.716 (0.010) & -0.001 (0.000) & 0.709 (0.007) & -0.008 (0.000) & 0.716 (0.009) & -0.000 (0.000) \\
\rowcolor[HTML]{F2F2F2} 
2 & 0.599 & 0.593 (0.011) & -0.006 (0.000) & 0.582 (0.009) & -0.017 (0.000) & 0.593 (0.011) & -0.006 (0.000) \\
\rowcolor[HTML]{FFFFFF} 
3 & 0.559 & 0.547 (0.012) & -0.013 (0.000) & 0.540 (0.009) & -0.019 (0.000) & 0.547 (0.012) & -0.012 (0.000) \\
\rowcolor[HTML]{F2F2F2} 
4 & 0.547 & 0.530 (0.013) & -0.016 (0.000) & 0.518 (0.010) & -0.028 (0.000) & 0.530 (0.012) & -0.017 (0.000) \\
\rowcolor[HTML]{FFFFFF} 
5 & 0.542 & 0.523 (0.014) & -0.019 (0.000) & 0.505 (0.010) & -0.037 (0.000) & 0.522 (0.013) & -0.020 (0.000) \\

% ------------------
% row 21
% ------------------
\rowcolor[HTML]{F2F2F2} 
\multicolumn{2}{l}{\textbf{Never ventilate}} & & & & & & \\

% ------------------
% rows 22-26
% ------------------
\rowcolor[HTML]{FFFFFF} 
1 & 0.656 & 0.655 (0.009) & -0.001 (0.000) & 0.657 (0.008) & 0.001 (0.000)  & 0.656 (0.009) & 0.000 (0.000)  \\
\rowcolor[HTML]{F2F2F2} 
2 & 0.432 & 0.453 (0.012) & 0.021 (0.000)  & 0.444 (0.009) & 0.012 (0.000)  & 0.443 (0.012) & 0.012 (0.000)  \\
\rowcolor[HTML]{FFFFFF} 
3 & 0.314 & 0.348 (0.014) & 0.034 (0.000)  & 0.335 (0.008) & 0.020 (0.000)  & 0.337 (0.013) & 0.022 (0.000)  \\
\rowcolor[HTML]{F2F2F2} 
4 & 0.252 & 0.287 (0.015) & 0.034 (0.000)  & 0.274 (0.008) & 0.022 (0.000)  & 0.277 (0.014) & 0.024 (0.000)  \\
\rowcolor[HTML]{FFFFFF} 
5 & 0.213 & 0.246 (0.016) & 0.032 (0.000)  & 0.234 (0.008) & 0.021 (0.000)  & 0.237 (0.015) & 0.024 (0.000) 
                 
\end{tabular}
\end{adjustbox}
\caption{Survival probabilities for the treatment strategies \textit{ventilate early} and \textit{wait to ventilate} at observation times 1-5: true values, mean estimates (and empirical standard errors) and bias in the estimates (and Monte Carlo standard errors) obtained using naive versions of IPW, G-computation and TMLE from 1000 simulations.}
\label{tab:sim_results_naive}
\end{table}

% *******************************************************
% *******************************************************
% Results table: ADAPTED METHODS
% *******************************************************
% *******************************************************

\begin{table}[H]
\centering
\large
\begin{adjustbox}{width=1\textwidth}
\begin{tabular}{p{1.5cm} p{2cm} p{2.5cm} p{2.5cm} p{2.5cm} p{2.5cm} p{2.5cm} p{2.5cm}}

% ------------------
% row 1
% ------------------
\rowcolor[HTML]{D9D9D9} 
 &  & \multicolumn{2}{c}{\textbf{Adapted IPW}}      & \multicolumn{2}{c}{\textbf{Adapted G-computation}} & \multicolumn{2}{c}{\textbf{Adapted TMLE}} \\ 
\cline{3-8}

% ------------------
% row 2
% ------------------
\rowcolor[HTML]{D9D9D9} 
\textbf{Time} & \textbf{True value} & \textbf{Mean estimate (Empirical SE)} & \textbf{Bias (Monte Carlo SE)} & \textbf{Mean estimate (Empirical SE)}  & \textbf{Bias (Monte Carlo SE)}  & \textbf{Mean estimate (Empirical SE)}   & \textbf{Bias (Monte Carlo SE)} \\

% ------------------
% row 3
% ------------------
\rowcolor[HTML]{F2F2F2} 
\multicolumn{2}{l}{\textbf{Ventilate early}} & & & & & & \\

% ------------------
% rows 4-8
% ------------------
\rowcolor[HTML]{FFFFFF}
1 & 0.717 & 0.716 (0.009) & -0.000 (0.000) & 0.716 (0.009) & -0.001 (0.000) & 0.716 (0.009) & -0.000 (0.000) \\
\rowcolor[HTML]{F2F2F2} 
2 & 0.598    & 0.598 (0.011) & 0.000 (0.000)  & 0.597 (0.010) & -0.001 (0.000) & 0.600 (0.011) & 0.002 (0.000)  \\
\rowcolor[HTML]{FFFFFF}
3 & 0.559    & 0.559 (0.012) & 0.000 (0.000)  & 0.558 (0.010) & -0.000 (0.000) & 0.562 (0.012) & 0.003 (0.000)  \\
\rowcolor[HTML]{F2F2F2} 
4 & 0.546    & 0.546 (0.012) & -0.000 (0.000) & 0.545 (0.010) & -0.001 (0.000) & 0.550 (0.012) & 0.005 (0.000)  \\
\rowcolor[HTML]{FFFFFF}
5 & 0.539    & 0.539 (0.012) & 0.000 (0.000)  & 0.535 (0.010) & -0.003 (0.000) & 0.544 (0.012) & 0.005 (0.000)  \\

% ------------------
% row 9
% ------------------
\rowcolor[HTML]{F2F2F2} 
\multicolumn{2}{l}{\textbf{Wait to ventilate}} & & & & & & \\

% ------------------
% rows 10-14
% ------------------
\rowcolor[HTML]{FFFFFF} 
1 & 0.656    & 0.656 (0.009) & 0.000 (0.000)  & 0.657 (0.009) & 0.001 (0.000)  & 0.656 (0.009) & -0.000 (0.000) \\
\rowcolor[HTML]{F2F2F2} 
2 & 0.432    & 0.432 (0.015) & 0.000 (0.000)  & 0.429 (0.010) & -0.002 (0.000) & 0.432 (0.013) & 0.001 (0.000)  \\
\rowcolor[HTML]{FFFFFF} 
3 & 0.321    & 0.321 (0.013) & 0.000 (0.000)  & 0.317 (0.009) & -0.004 (0.000) & 0.321 (0.010) & 0.000 (0.000)  \\
\rowcolor[HTML]{F2F2F2} 
4 & 0.274    & 0.274 (0.012) & 0.000 (0.000)  & 0.271 (0.009) & -0.003 (0.000) & 0.273 (0.011) & -0.001 (0.000) \\
\rowcolor[HTML]{FFFFFF} 
5 & 0.254    & 0.256 (0.013) & 0.002 (0.000)  & 0.254 (0.009) & 0.001 (0.000)  & 0.254 (0.012) & 0.000 (0.000)  \\ 

% ------------------
% row 15
% ------------------
\rowcolor[HTML]{F2F2F2} 
\multicolumn{2}{l}{\textbf{Always ventilate}} & & & & & & \\

% ------------------
% rows 16-20
% ------------------
\rowcolor[HTML]{FFFFFF} 
1 & 0.717    & 0.716 (0.009) & -0.000 (0.000) & 0.716 (0.009) & -0.001 (0.000) & 0.716 (0.009) & -0.000 (0.000) \\
\rowcolor[HTML]{F2F2F2}
2 & 0.599    & 0.598 (0.011) & -0.001 (0.000) & 0.597 (0.010) & -0.002 (0.000) & 0.600 (0.011) & 0.001 (0.000)  \\
\rowcolor[HTML]{FFFFFF} 
3 & 0.559    & 0.559 (0.012) & -0.001 (0.000) & 0.558 (0.010) & -0.001 (0.000) & 0.562 (0.012) & 0.003 (0.000)  \\
\rowcolor[HTML]{F2F2F2}
4 & 0.547    & 0.546 (0.012) & -0.001 (0.000) & 0.546 (0.010) & -0.001 (0.000) & 0.551 (0.012) & 0.005 (0.000)  \\
\rowcolor[HTML]{FFFFFF} 
5 & 0.542    & 0.541 (0.013) & -0.001 (0.000) & 0.539 (0.010) & -0.003 (0.000) & 0.548 (0.013) & 0.006 (0.000)  \\

% ------------------
% row 21
% ------------------
\rowcolor[HTML]{F2F2F2} 
\multicolumn{2}{l}{\textbf{Never ventilate}} & & & & & & \\

% ------------------
% rows 22-26
% ------------------
\rowcolor[HTML]{FFFFFF} 
1 & 0.656    & 0.656 (0.009) & 0.000 (0.000)  & 0.657 (0.009) & 0.001 (0.000)  & 0.656 (0.009) & 0.000 (0.000)  \\
\rowcolor[HTML]{F2F2F2} 
2 & 0.432    & 0.432 (0.015) & 0.000 (0.000)  & 0.429 (0.010) & -0.002 (0.000) & 0.432 (0.013) & 0.001 (0.000)  \\
\rowcolor[HTML]{FFFFFF} 
3 & 0.314    & 0.316 (0.022) & 0.002 (0.001)  & 0.310 (0.009) & -0.004 (0.000) & 0.315 (0.013) & 0.001 (0.000)  \\
\rowcolor[HTML]{F2F2F2} 
4 & 0.252    & 0.255 (0.023) & 0.003 (0.001)  & 0.245 (0.009) & -0.007 (0.000) & 0.252 (0.014) & -0.000 (0.000) \\
\rowcolor[HTML]{FFFFFF} 
5 & 0.213    & 0.217 (0.023) & 0.003 (0.001)  & 0.208 (0.009) & -0.006 (0.000) & 0.213 (0.015) & -0.001 (0.000) \\

\end{tabular}
\end{adjustbox}
\caption{Survival probabilities for the treatment strategies \textit{ventilate early} and \textit{wait to ventilate} at observation times 1-5: true values, mean estimates (and empirical standard errors) and bias in the estimates (and Monte Carlo standard errors) obtained using versions of IPW, G-computation and TMLE adapted to incorporate monitoring nodes from 1000 simulations.}
\label{tab:sim_results_adapted}
\end{table}

% *******************************
% *******************************
% Figure 
% *******************************
% *******************************

\newpage

\section{An application to mechanical ventilation in COVID-19 patients using UCLH intensive care data}
\label{sec:application}

\subsection{Study design and methods}

We use the target trial framework \cite{hernan_using_2016} to design and emulate an idealised trial to investigate the effect of different ventilation strategies on mortality of patients in the \ac{ICU}, using routinely collected hospital data from \ac{UCLH}. Below we describe the target trial design and emulation using the \ac{CC-HIC} data.

\textbf{Eligibility criteria:} The target trial would enrol adult patients admitted to the \ac{ICU}. To ensure we could reasonably consider equipoise to exist over treatment strategies, patients would only be eligible for the trial if they were receiving supplemental oxygen at or upon admission, had a sufficiently stable respiratory rate defined as between 20 and 40 breaths per minute, and did not have a treatment limitation order precluding treatment in a critical care (level 3) environment with invasive or non-invasive ventilation. We emulated the target trial using routinely-collected hospital data containing all patients admitted to \ac{UCLH} \ac{ICU} between January 2020 and July 2023. We replicated the target trial eligibility criteria as follows. We included only patients who were 18 years or older at the time of \ac{ICU} admission. We included patients receiving at least supplemental oxygen at baseline (as determined by a measurement of $F_I O_2 >$ 21\% or a measurement of $O_2 >$ 0 litres per minute, in addition to a measurement of $SpO_2 \leq$ 98\%, observed in the 24-hour period from \ac{ICU} admission). We excluded patients whose average respiratory rate in the 24-hour period from \ac{ICU} admission was outside the range [20,40] breaths per minute. As we do not have measurements of $F_I O_2 >$, $0_2$, $SpO_2$ or respiratory rate available prior to ICU admission, eligibility is ascertained based on measurements observed in the first 24-hour period from ICU admission. We make the assumption that measurements of variables relating to eligibility occur before ventilation is initiated. We also excluded patients with a ``not for (cardiopulmonary) resuscitation'' order issued within 24 hours of admission to the \ac{ICU} (with this 24-hour limit aiming to prevent exclusion of patients who receive treatment limitation orders very close to death because further treatment is deemed to be futile). We included only the first admission to the \ac{ICU} within our study period. Patients were excluded if they had missing values of the variables required to identify eligibility.

\begin{figure}[htbp]
\begin{center}
\begin{tikzpicture}[
  >=Stealth,
  thick,
  every node/.style={font=\small},
  timeline/.style={->, thick},
  ventilate/.style={thick, solid},
  noVent/.style={thick, dotted},
  natural/.style={thick, dashed},
  labelStyle/.style={align=left, text width=5cm}
]

% length of a 1 day interval
\def\oneday{0.75}

% draw the x-axis and label it
\draw[-{Stealth}, thick] (0,0) -- (15,0);
\node at ({0.5*15}, -0.8) {Days from ICU admission};

% add ICU circle
%\draw[fill=gray!10] (0,-1.5) circle (0.5);
%\node at (0,-1.5) {ICU};

% tick marks
\foreach \x in {0,...,16}
  \draw ({\x*\oneday},0.1) -- ({\x*\oneday},-0.1) node[below] {\x};
\foreach \x in {29,30}
  \draw ({(\x-11)*\oneday},0.1) -- ({(\x-11)*\oneday},-0.1) node[below] {\x};

% add slashes indicating continuation
\draw[thick] (12.6,-0.2) -- (12.7,0.2);
\draw[thick] (12.7,-0.2) -- (12.8,0.2);

% strategy positions
\def\rowA{3}
\def\rowB{1}

% ---------------
% Ventilate early
% ---------------

% treatment strategy name
\node[draw=black, text=black, thick, align=left, text width=1.5cm] at (-1,\rowA) {\textbf{Ventilate\\early}};

% arrows
\draw[{Stealth}-{Stealth}, thick, black] (0,\rowA) -- (\oneday*3,\rowA);
\draw[natural, black, thick, {Stealth}-{Stealth}] (\oneday*3,\rowA) -- (\oneday*19,\rowA);

% labels
\node[above, text=black, labelStyle, align=center, text width=\oneday*5cm] at (\oneday*3*0.5,\rowA+0.35) {Initiate ventilation on\\day 0 and remain\\ventilated for at least\\3 days};
\node[above, text=black, labelStyle, align=center] at (\oneday*11,\rowA+0.35) {Allow ventilation to follow natural course};

% -----------------
% Wait to ventilate
% -----------------

% treatment strategy name
\node[draw=black, text=black, thick, align=left, text width=1.5cm] at (-1,\rowB) {\textbf{\textcolor{black}{Wait to\\ventilate}}};

% arrows
\draw[{Stealth}-{Stealth}, thick, black] (0,\rowB) -- (\oneday*3,\rowB);
\draw[{Stealth}-{Stealth}, thick, black] (\oneday*3,\rowB) -- (\oneday*14,\rowB);
\draw[natural, black, thick, {Stealth}-{Stealth}] (\oneday*14,\rowB) -- (\oneday*19,\rowB);

% labels
\node[above, text=black, labelStyle, align=center, text width=\oneday*5cm] at (\oneday*3*0.5,\rowB+0.35) {No ventilation};

\node[above, text=black, labelStyle, align=center, text width=\oneday*10cm] at (\oneday*8.5,\rowB+0.35) {Initiate ventilation between days 3 to 11 and remain on ventilation for at least 3 days};

\node[above, text=black, labelStyle, align=center, text width=\oneday*5cm] at (\oneday*16.5,\rowB+0.35) {Allow ventilation to follow natural course};

\end{tikzpicture}
\end{center}
\caption{Ventilation status under treatment strategies \textit{ventilate early} and \textit{wait to ventilate}}
\label{fig:treatment_strategies}
\end{figure}
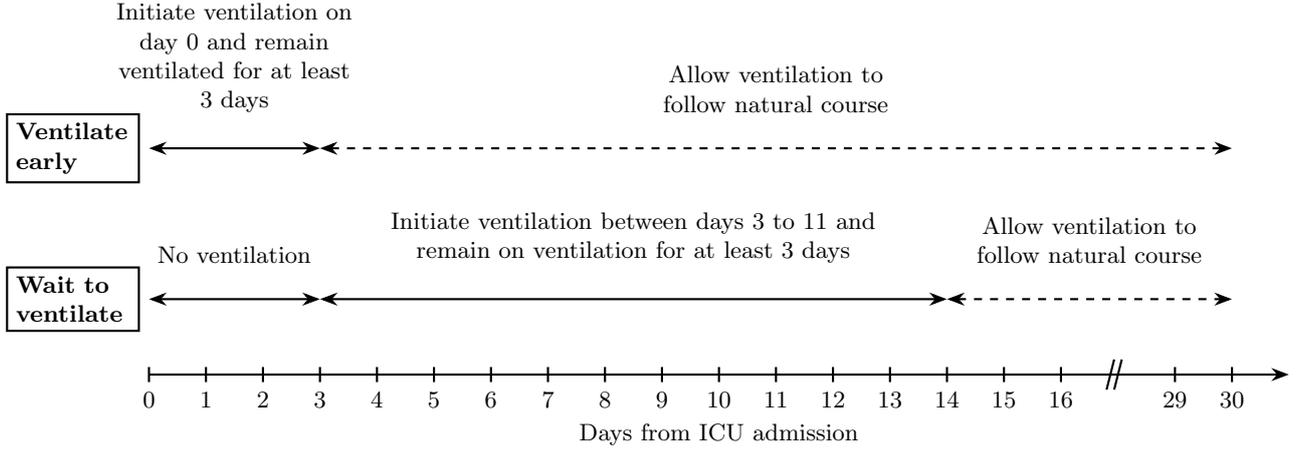

\textbf{Treatment strategies:} We considered two ventilation strategies based on early and delayed initiation of \ac{IMV}. Treatment strategies were designed with the aim of gaining insight into the effects of delaying initiation of \ac{IMV}, while still reflecting realistic clinical decision making (by allowing flexibility in terms of the exact timing of initiation of delayed ventilation and also in terms of when the patient is weaned from the ventilator). The \textit{ventilate early} strategy involves initiating \ac{IMV} within the first 24 hours of \ac{ICU} admission (day zero), and remaining ventilated for at least three consecutive days (i.e. days 0,1,2). From day three onwards, the treating clinician should use their clinical judgement to determine whether a patient should remain on ventilation or not. The \textit{wait to ventilate} strategy requires that the patient remains unventilated for the first three days of \ac{ICU} admission. For a patient who is still alive and still in the \ac{ICU}, \ac{IMV} must be initiated between days three and 10, inclusive, and, once initiated, the patient must remain ventilated for at least three consecutive days. The exact day of initiation of delayed \ac{IMV} will depend on the individual patient's condition and will be determined by the treating clinicians. After the three consecutive days of ventilation, the treating clinician should again use their clinical judgement to determine whether a patient should remain on ventilation or not. Figure \ref{fig:treatment_strategies} presents a schematic for these treatment strategies. The treatment strategies are the same in the target trial and emulated trial.

% --------------------------------------------------------------------------
% treatment strategy schematic
% --------------------------------------------------------------------------

% --------------------------------------------------------------------------
% --------------------------------------------------------------------------

\textbf{Treatment assignment:} In the target trial patients would be randomised to receive either the \textit{ventilate early} or the \textit{wait to ventilate} treatment strategies. In the emulated trial we assume that treatment assignment is random conditional on a set of baseline and time-dependent variables. Baseline variables are age, sex, ethnicity, \ac{IMD} (quintile), \ac{BMI}, and a set of comorbidities (chronic kidney disease, hypertension, diabetes, asthma, heart failure, chronic respiratory disease and pneumonia). Time-dependent variables are Glasgow Coma Score (GCS), respiratory rate (RR), $F_I O_2$, $SpO_2$, pH, a binary variable indicating steroid use, a binary variable indicating use of \ac{NIV}, a binary variable indicating whether the patient has been discharged from the \ac{ICU} or not, and, for GCS, RR, $F_I O_2$, $SpO_2$ and pH, the corresponding set of monitoring indicator variables.

\textbf{Follow-up and outcome}: In the target trial patients meeting the eligibility criteria would be randomised at \ac{ICU} admission to one of the two treatment strategies, with patients and their care teams aware of their treatment assignment. Patients would be followed up from treatment assignment to the time of death or the end of 30-day follow-up. The primary outcome would be the risk of 30-day mortality under each treatment strategy and the risk difference. The follow-up and outcome in the emulated trial are as per the target trial. In both the target trial and the emulated trial the 30 days of follow up might include a period of time post-discharge as some patients will be discharged from the \ac{ICU} before 30 days.

\textbf{Data set construction}: For each patient the \ac{CC-HIC} database records demographics (e.g. age, sex, ethnicity), measurements of clinical biomarkers (e.g. vital signs, blood tests, microbiological assays, COVID PCR), all medicine administrations, highly granular information on non-pharmaceutical interventions (e.g. ventilator settings), comorbidities and new diagnoses. Dates of death are populated at the time of data extraction using \ac{NHS} Spine. For baseline comorbidities we determine whether the condition was present before the start of the admission to the \ac{ICU}. For time-varying confounders and ventilation status, we obtain the complete set of measurements for each patient and their corresponding measurement datetime. Ventilation status is derived using a set of measurements describing breathing support, airway and oxygen delivery devices, use and amount of oxygen supplied to the patient, and the presence of certain ventilator settings relating to the volume and pressure of gas supplied. The set of available measurements are used to determine whether the patient is receiving no ventilatory support, oxygen support only, non-invasive ventilation or IMV. More details are provided in the supplementary material. Data are converted into a grid structure by discretising time into 24-hour periods, starting from the day of the patient's admission to the \ac{ICU}, and ending with the earliest of death or 30-day follow-up. When the end of follow-up is due to death, the final period may be less than 24 hours. For time-dependent confounders and ventilation status, measurements are grouped and summarised over these 24-hour time periods to create a single record for each time-dependent confounder ($\mathbf{L}_k$) and ventilation status ($A_k$) in each 24-hour period, and we assume that measurements of $L_k$ precede measurements of $A_k$. The summary measure is either the minimum, maximum or mean for continuous variables, or the mode or most extreme category for categorical variables (with the selection being based on clinical judgement). More details are provided in supplementary table S6. For ventilation status, if a measurement of \ac{IMV} exists at any point during the 24-hour period, we consider the patient to be ventilated during this period ($A_k=1$); otherwise, we set $A_k=0$. A similar variable is created for \ac{NIV}, which takes the value of one when the patient has any record of receiving non-invasive ventilation during the 24-hour period, and zero otherwise. The binary outcome $Y_{k+1}$ identifies whether death has occurred by the end of the 24-hour period $k$. We also generate a binary variable for \ac{ICU} status ($I_k$) indicating whether the patient is admitted to the \ac{ICU} or has been discharged at the start of period $k$.

\textbf{Analysis plan}: We estimated the risk of death up to day $k$ for $k=1,\ldots,30$ days under each treatment strategy by applying the three methods identified and described in section \ref{sec:methods}: \ac{IPW}, G-computation and \ac{TMLE}. We adjusted for the set of baseline and time-varying confounders described above.

For the \ac{IPW} method, models for the weights were fitted using logistic regression including the set of baseline and time-varying confounders described above, including the set of monitoring indicator variables $\mathbf{N}_k$ (indicators of whether at least one measurement of the corresponding time-dependent confounder was observed during the 24-hour period). We fitted an \ac{MSM} with stabilised \ac{IPW} weights using pooled logistic regression. The covariates in the MSM were current treatment $A_k$, day from admission $k$, and all time-fixed confounders and baseline values of the time-dependent confounders listed above. All continuous variables (day from admission, age, logarithm of BMI, respiratory rate, $F_I O_2$, $SpO_2$ and pH were modelled using restricted cubic splines with three knots. For G-computation we fitted models for the time-varying confounders (as described above) $\mathbf{L}_k$, the monitoring variables $\mathbf{N}_k$, ventilation status $A_k$ and outcome status $Y_k$, using linear and logistic regression. For \ac{TMLE} we used a SuperLearner ensemble to estimate both the conditional outcome and treatment models.

All analyses were conducted in R version 4.4.1. G-computation was implemented using the \texttt{gfoRmula} package and \ac{TMLE} was implemented using the \texttt{lmtp} package (with built in \texttt{SuperLearner}).

\subsection{Results}

\begin{figure}[H]
    \centering
    \includegraphics[width=0.5\linewidth]{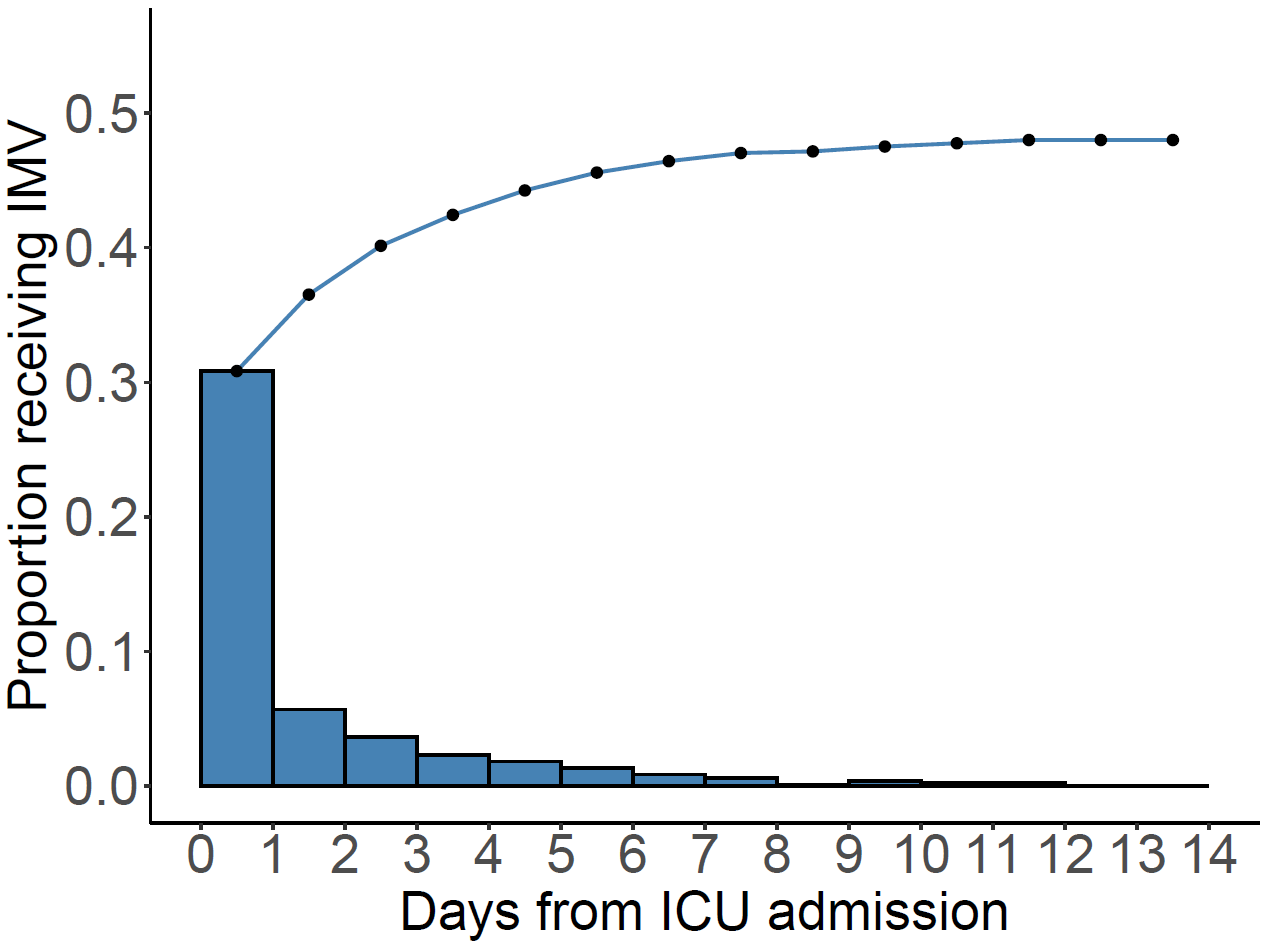}
    \caption{Proportion of individuals initiating IMV by day since ICU admission (bar plot) and the cumulative incidence of IMV initiation by day since ICU admission (line) among eligible individuals}
    \label{fig:desc_imv}
\end{figure}

\textbf{Descriptive statistics}: From a total of 8676 patients admitted to the \ac{ICU} between January 2020 and July 2023, a total of 827 patients met the eligibility criteria (for details see supplementary table S7). The baseline characteristics (including baseline values of time-varying covariates) for eligible patients are tabulated in supplementary table S8. The proportion of eligible patients receiving \ac{IMV} on each day from \ac{ICU} admission is illustrated in figure \ref{fig:desc_imv}. The majority of people who received \ac{IMV} initiated it on their fist day of admission. The outcome was observed in 18.7\% of eligible patients overall, in 32.1\% of those who ever received \ac{IMV}, and in 28.9\% of those who initiated \ac{IMV} between days 1 and 3 of \ac{ICU} admission (i.e. within the first 72-hours), and in 43.1\% of those who initiated \ac{IMV} between days 4 and 12 of \ac{ICU} admission. Complete follow-up, with respect to 30-day mortality, was available for all individuals.

% -----------------------------------------------
% figure: initiation of ventilation (descriptive)
% -----------------------------------------------

% -----------------------------------------------
% -----------------------------------------------

\begin{figure}[H]
\centering
\hspace*{-0.5cm}
\subfloat[IPW]{\includegraphics[width=0.33\columnwidth]{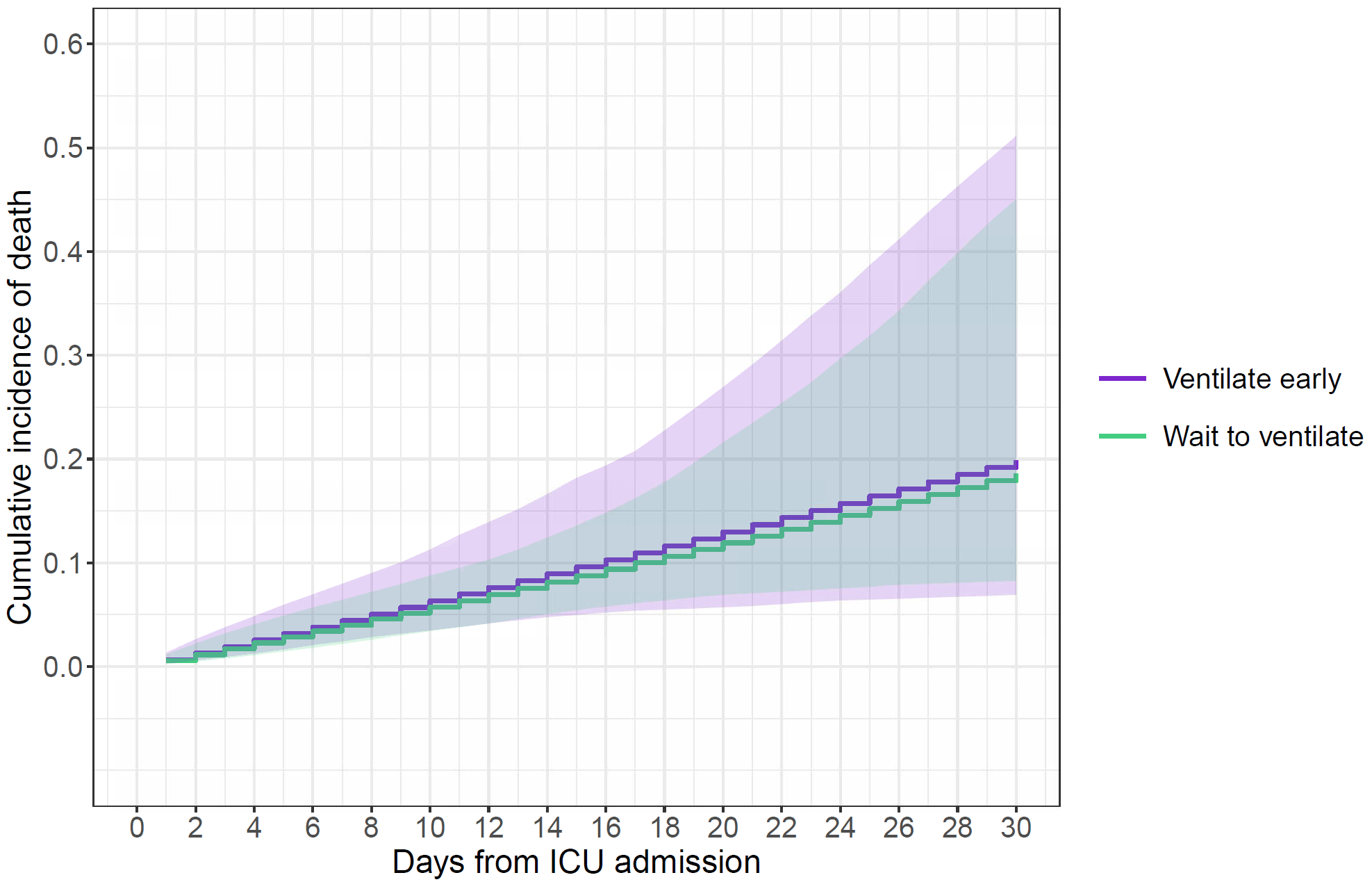}}
\subfloat[G-computation]{\includegraphics[width=0.33\columnwidth]{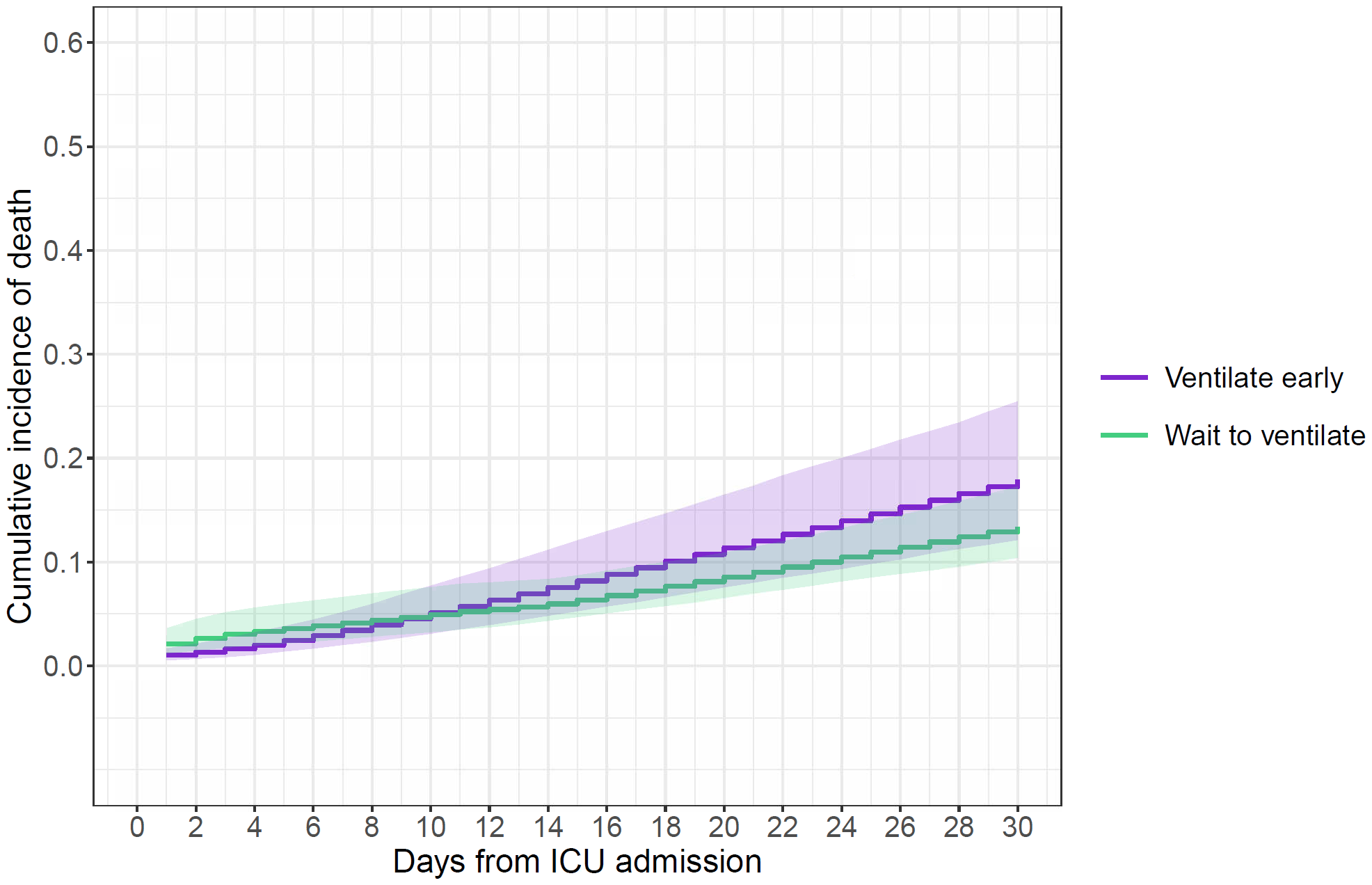}}
\subfloat[TMLE]{\includegraphics[width=0.33\columnwidth]{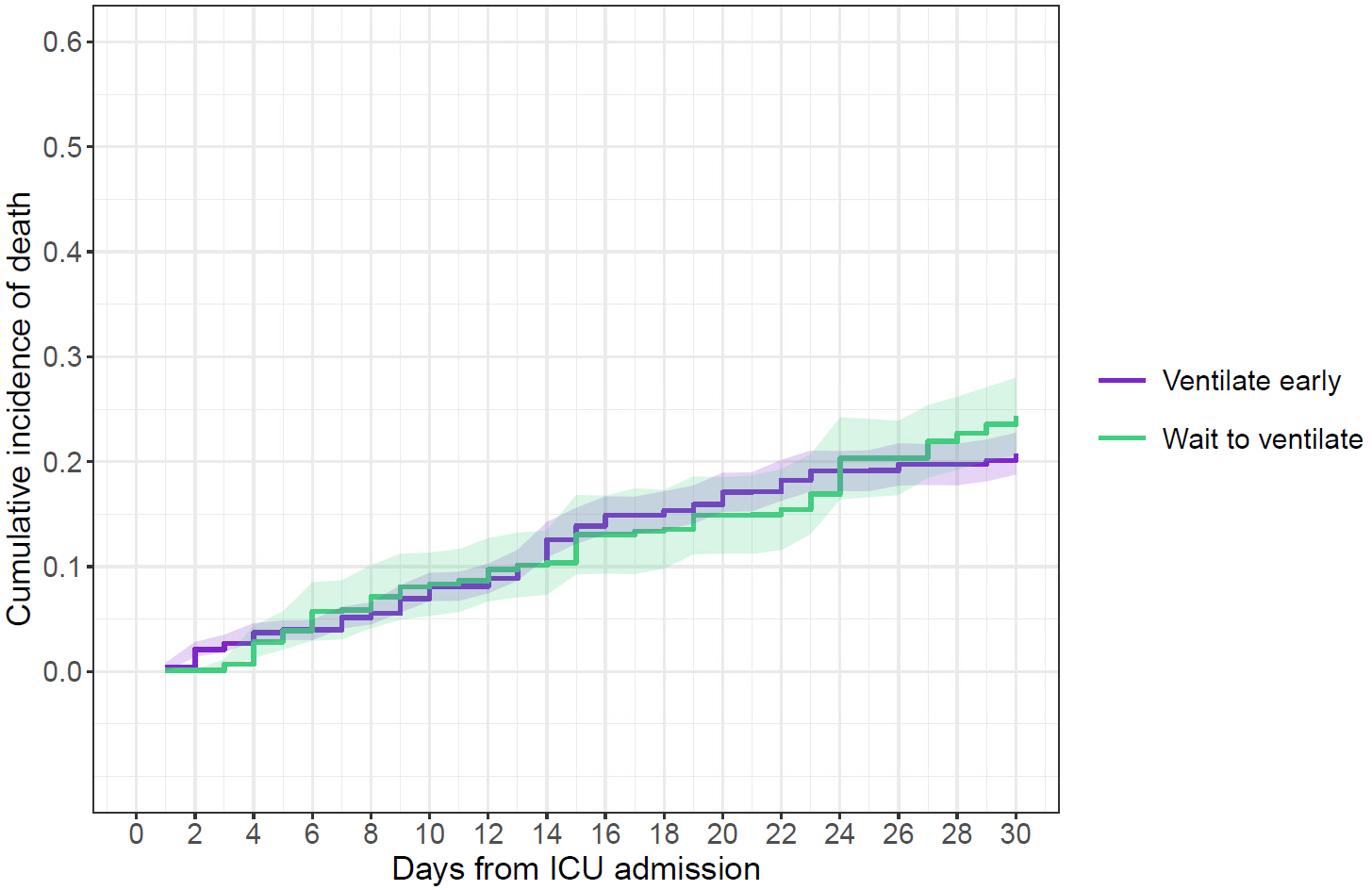}}
\caption{Estimated daily cumulative incidences of death up to 30 days (and 95\% confidence interval) under the treatment strategies \textit{ventilate early} (purple) and \textit{wait to ventilate} (green) for eligible patients in the CC-HIC cohort, using three estimation approaches to account for time-varying confounding and informative monitoring}
\label{fig:results_uclh}
\end{figure}

\textbf{Causal estimates}: The estimated cumulative incidence of death up to 30-days using \ac{IPW}-\ac{MSM} was 0.199 \% (95\% CI 0.069\% - 0.512\%) under the \textit{ventilate early} strategy and 0.186 \% (95\% CI 0.083\% - 0.451\%) under the \textit{wait to ventilate} strategy. These results, and the corresponding results for G-computation and \ac{TMLE}, are presented in table \ref{tab:results_uclh}. The daily cumulative incidences up to 30 days obtained using each method are presented in figure \ref{fig:results_uclh}. 

The confidence intervals for the IPW-MSM method are much wider than those for both G-computation and TMLE. As noted earlier in section \ref{sec:simulation_study}, this is expected. The IPW-MSM 95\% confidence intervals for both ventilate early and wait to ventilate include the point estimates from both G-computation and TMLE. Conversely, however, the 95\% confidence intervals for G-computation and TMLE do not include the point estimates from the alternative methods. Most notably, for the wait to ventilate strategy, the 95\% confidence interval for the G-computation estimate (0.104\% - 0.173\%) does not include the point estimate from the TMLE method (0.243\%). This highlights that different modelling assumptions can result in substantially different results. However, when we look at the estimated cumulative incidences up to 30 days in figure \ref{fig:results_uclh} the differences between methods appear less meaningful; the plots show overlapping cumulative incidence curves for ventilate early and wait to ventilate within each method. 

% ----------------------------------------
% table of results : uclh ventilation
% ----------------------------------------
\begin{table}[H]
\centering
\begin{adjustbox}{width=0.7\textwidth}
\large
\begin{tabular}{lccc}
\rowcolor[HTML]{D9D9D9} 
& \textbf{IPW-MSM} & \textbf{G-computation} & \textbf{Longitudinal TMLE} \\
\rowcolor[HTML]{FFFFFF} 
Ventilate early   & 0.199 (0.069 to 0.512) & 0.179 (0.121 to 0.255) & 0.208 (0.188 to 0.228) \\
\rowcolor[HTML]{F2F2F2} 
Wait to ventilate & 0.186 (0.083 to 0.451) & 0.134 (0.104 to 0.173) & 0.243 (0.207 to 0.280) \\
\end{tabular}
\end{adjustbox}
\caption{Estimated cumulative incidence of death at 30-days (and 95\% confidence intervals) under the treatment strategies \textit{ventilate early} and \textit{wait to ventilate} for eligible patients in the CC-HIC cohort, using three estimation approaches to account for time-varying confounding and informative monitoring}
\label{tab:results_uclh}
\end{table}

% ----------------------------------------
% figure of results : uclh ventilation
% ----------------------------------------

\section{Discussion}
\label{sec:discussion}

This paper has discussed three methods for estimating the effects of flexible static treatment strategies on time-to-event outcomes using routinely collected data in a setting where time-dependent confounders are monitored irregularly and potentially informatively. Previous work has described the observational data structure in the presence of informative monitoring, and derived estimators to estimate the effects of longitudinal treatment strategies under this assumed data structure, however, none of these papers describe methods other than \ac{IPW}, nor do they evaluate methods using simulations. We described methods based on \ac{IPW}, G-computation and \ac{TMLE}, generalising methods to handle \textit{natural grace-period} treatment strategies. We showed using a simulation study that using \ac{LOCF} to populate missing values of time-dependent confounders but ignoring the irregular nature of monitoring of these confounders introduces bias in the estimated survival probabilities, but that this bias can be avoided with a simple adaptation based on including monitoring indicator variables as additional time-varying confounders. Routinely-collected observational data originating from \ac{EHR} are increasingly used for research purposes to investigate causal effects of treatments. In these data measurements of time-dependent variables are almost always observed in an irregular and potentially informative way, either because they relate to when a patient attended their GP practice or was monitored in a hospital setting. Our description of methods to account for irregular and informative monitoring will help other researchers using routinely collected \ac{EHR} data to estimate the effects of longitudinal static treatment strategies on time-to-event outcomes.

We present methods based on \ac{IPW}, G-computation and \ac{TMLE}. There are advantages and disadvantages of all three methods. Both \ac{IPW} and G-computation are relatively straightforward to implement, with the option to keep data in long format containing one row per person time-interval. \Ac{IPW} relies on correct specification of propensity score model whereas G-computation (the \ac{NICE} version, section \ref{sec:method_g_comp}) requires correct specification of models for the time-dependent confounders, monitoring nodes and observed outcome. The potential for model misspecification is relatively higher using G-computation, especially in the setting with informative monitoring where theoretically plausible models for $L_k$ and $Y_k$ likely involve complex interactions between $N_k$ and other conditioning variables. Disadvantages of \ac{IPW} are that (near) violations in the positivity assumption are common, leading to large and unstable weights and wide confidence intervals, although this is mitigated somewhat by using stabilised weights. This may be a particular issue in our adapted version of \ac{IPW}, allowing for informative monitoring, as, for example, patients may be unlikely to initiate treatment in time-periods where no time-varying confounders are monitored. The double-robustness property of \ac{TMLE}, and the fact that it is well-suited to use of machine learning methods to estimate both conditional outcome and propensity score, mean that it is less susceptible to bias due to model misspecification. Implementation of \ac{TMLE} is more involved although R packages such as \texttt{lmtp} negate this challenge. However, this approach is computationally intensive, especially when using machine learning methods to estimate the nuisance models.

We focused on a simplified setting with no loss-to-follow-up except through administrative censoring. The general results apply when there are other types of right-censoring, including when right-censoring depends on treatment, covariates, or monitoring decisions. We can consider censoring $C_k$ as an additional variable in the causal \ac{DAG}, where $C_k=1$ indicates that an individual has been censored by the end of period $k$. If an individual is censored by period $k$ then treatments, confounders and outcomes measured after time $k$ are unobserved. Censoring can be thought of as an additional treatment node and the goal is to evaluate our treatment strategy in a world where all individuals remain uncensored. We can write the estimand as $\Pr[Y_{k}^{\Bar{s}_k,\Bar{c}=0}=1]$. Under the identifiability assumptions described in section \ref{sec:assumps} but with treatment $A_k$ replaced with $(A_k,C_k)$, the g-formula in expression \ref{eq:gform_nice} can be rewritten such that all the terms condition on $\Bar{C}_k=0$. The G-computation procedure remains unchanged as we simulate a $\tilde{Y}_{i,k}$ for each individual at each time-period in a hypothetical world in which all individuals remain uncensored. For \ac{IPW} and \ac{TMLE}, censoring that depends on past treatment, covariates, or monitoring decisions can be handled using inverse probability of censoring weights
\begin{equation}
    {W}^{\Bar{c}}_k = \prod_{j=0}^{k} \frac{1}{\Pr(C_j=0|\overline{N}_{j-1},\overline{A}_{j-1},\overline{L}_{j},Y_{j-1}=0,C_{j-1}=0)}
\end{equation}
which are multiplied together with the inverse probability of treatment weights. We propose conditioning on $N_k$ in the censoring model since it is plausible that previous monitoring decisions may affect loss-to-follow-up. We focused on a situation with a single event of interest, such as death, but methods could be extended to handle competing events using the framework set out by Young et al. \cite{young_causal_2020}.

We use routinely collected hospital data from the \ac{CC-HIC} database to evaluate the effect of early versus delayed ventilation strategies on mortality of intensive care patients, in a setting where several time-dependent confounders are monitored informatively. Our results suggest little difference in the effectiveness of interventions that either initiate early or delay initiation of mechanical ventilation in patients in the \ac{ICU} meeting our eligibility criteria. Results are consistent across the three estimation approaches, in that all three methods suggest no treatment effect, although some differences exist in terms of estimated survival probabilities, which can be attributed to differences in modelling assumptions. In particular, the confidence intervals are much wider for \ac{IPW}, especially at later observation times, as inverse-probability weights can become large and unstable. The cumulative incidence curves for \ac{IPW} and G-computation are smooth, as models for the (potential) outcome are fitted using pooled regression, whereas for \ac{TMLE} the outcome is estimated separately for each observation time, resulting in stepwise curves.

Our study has limitations, both in terms of our methods and our application to ventilation. In terms of the methods, although we considered \textit{grace-period} treatment strategies, we did this in the context of static treatment strategies. In further work it is of interest to study the effects of dynamic treatment strategies, which involve rules that adapt the treatment to the current status of the patient, for example, ``\textit{initiate ventilation when respiratory rate exceeds 30 breaths per minute}''. Dynamic treatment strategies can also incorporate grace-periods. Such strategies may better reflect realistic clinical decision making. When examining the impact of dynamic treatment strategies, in addition to acting as a confounder, monitoring poses a further challenge. The ``optimal'' dynamic treatment strategy will depend on the frequency of monitoring, as clinicians are able to intervene on treatment sooner when recent, abnormal values of certain variables have been observed. For example, in a population where respiratory rate is monitored every 12 hours, the optimal respiratory rate threshold for initiating invasive mechanical ventilation might differ from that in a population where respiratory rate is monitored every hour \cite{kreif_exploiting_2021}. This adds complexity to the problem of informative monitoring in routinely-collected hospital data, which we plan to explore in future work.

Our methods assume that there is an underlying grid of times at which, for each individual, measurements of covariates can potentially be observed and changes in treatment status can occur. This could be, say, minutes, hours, days, etc. depending on the time grid representing the true underlying data generating process. The resulting dataset may have a very large number of observation times, for example, if the grid times are hours and we wish to follow individuals for 30 days, the number of potential observation times would be $K=720$. This might result in problems in terms of computational feasibility and data support, and therefore it may be necessary to coarsen the data to reduce the number of time-points for analysis purposes. We explore this problem in ongoing work.

In terms of our data application, despite our efforts to control for confounding using a variety of methods, the potential for unmeasured confounding exists. Despite restricting the study population using strict eligibility criteria, this population is still broader than the populations other studies have considered (e.g. only including individuals with \ac{ARDS} resulting from COVID-19). This is a challenge when studying \ac{ICU} populations which are heterogenous in nature. Results may not be transportable to other specific \ac{ICU} populations, and also, effects may vary for specific subsets of the population not identified in this study.

\newpage

\section{Supplementary material}

\setcounter{table}{0}
\renewcommand{\thetable}{S\arabic{table}}

\setcounter{figure}{0}
\renewcommand{\thefigure}{S\arabic{figure}}

\subsection{Simulation study}

% --------------------------------
% TABLE: PARAMETER VALUES
% --------------------------------

\renewcommand{\arraystretch}{1}
\begin{table}[H]
\centering
\setlength{\tabcolsep}{4pt}

\begin{adjustbox}{max width=0.7\textwidth}

\begin{tabular}{lccccc}
\rowcolor[HTML]{BFBFBF} 
  & \multicolumn{5}{c}{\cellcolor[HTML]{BFBFBF}\textbf{Parameter value}} \\
  
\rowcolor[HTML]{BFBFBF} 
\textbf{Parameter} & \textbf{Scenario 1} & \textbf{Scenario 2} & \textbf{Scenario 3} & \textbf{Scenario 4}  & \textbf{Scenario 5} \\

% MODEL FOR L*
\rowcolor[HTML]{F2F2F2} 
\textbf{Model for $L^*_k$} & & & & & \\

\rowcolor[HTML]{FFFFFF} 
$\beta_{l_0}$ & 0.1 & 0.1 & 0.1 & 0.1 & 0.1  \\
\rowcolor[HTML]{FFFFFF} 
$\beta_{l_{L^*}}$ & 1.2 & 1.2 & 1.2 & 1.2 & 1.2  \\
\rowcolor[HTML]{FFFFFF} 
$\beta_{l_A}$ & -1.2 & -1.2 & -1.2 & -1.2 & -1.2 \\
\rowcolor[HTML]{FFFFFF} 
$\beta_{l_U}$& 1.0& 1.0& 1.0& 1.0& 1.0 \\
\rowcolor[HTML]{FFFFFF} 
$\sigma^2$   & 1.0& 1.0& 1.0& 1.0& 1.0 \\

% MODEL FOR A
\rowcolor[HTML]{F2F2F2}
\textbf{Model for $A_k$}   & & & & &  \\

\rowcolor[HTML]{FFFFFF} 
$\beta_{a_0}$& -0.3 & -0.3 & -0.3 & -0.3 & -0.3  \\

\rowcolor[HTML]{FFFFFF} 
$\beta_{a_L}$& 0.5& 0.5& 0.5& 0.5& 0.5 \\

\rowcolor[HTML]{FFFFFF} 
$\beta_{a_A}$& 0.7& 0.7& 0.7& 0.7& 0.7 \\

\rowcolor[HTML]{FFFFFF} 
$\beta_{a_N}$& 1.2& 1.2& 1.2 & \textbf{2.0} & \textbf{0.0}\\

% MODEL FOR N
\rowcolor[HTML]{F2F2F2} 
\textbf{Model for $N_k$}   &  &  &  &  &  \\

\rowcolor[HTML]{FFFFFF} 
$\beta_{n_0}$& -2.0 & \textbf{-5.0}& \textbf{-0.7} & -2.0  & -2.0 \\

\rowcolor[HTML]{FFFFFF} 
$\beta_{n_L}$& 2.0 & \textbf{3.0} & \textbf{0.0} & 2.0 & 2.0 \\

\rowcolor[HTML]{FFFFFF} 
$\beta_{n_A}$& 2.0& \textbf{3.0} & \textbf{0.0} & 2.0 & 2.0 \\

\rowcolor[HTML]{FFFFFF} 
$\beta_{n_N}$& 0.5& \textbf{1.0} & \textbf{0.0} & 0.5 & 0.5\\

% MODEL FOR Y
\rowcolor[HTML]{F2F2F2} 
\textbf{Model for $Y_k$} & & & & & \\

\rowcolor[HTML]{FFFFFF} 
$\beta_{y_0}$& -0.7 & -0.7 & -0.7 & -0.7 & -0.7  \\

\rowcolor[HTML]{FFFFFF} 
$\beta_{y_{L^*}}$ & 0.6& 0.6& 0.6& 0.6 & 0.6 \\

\rowcolor[HTML]{FFFFFF} 
$\beta_{y_A}$& -0.3 & -0.3 & -0.3 & -0.3 & -0.3 \\

\rowcolor[HTML]{FFFFFF} 
$\beta_{y_U}$& 0.05 & 0.05 & 0.05 & 0.05 & 0.05             
\end{tabular}

\renewcommand{\arraystretch}{1}
\end{adjustbox}
\caption{Parameter values for each simulation scenario. For scenarios 2-6 the changed parameter values are highlighted in bold.}
\label{tab:parameter_values}
\end{table}

% *******************************************************
% *******************************************************

\newpage

% *******************************************************
% *******************************************************
% Results table - SCENARIO 2
% *******************************************************
% *******************************************************

\renewcommand{\arraystretch}{1.4}
\begin{table}[H]
\centering
\large
\setlength{\tabcolsep}{4pt}

\begin{adjustbox}{max width=\textwidth}
\begin{tabular}{llccccccc}

\rowcolor[HTML]{BFBFBF} 
\textbf{} & \textbf{} & \textbf{} & \multicolumn{3}{c}{\cellcolor[HTML]{BFBFBF}\textbf{Adapted}} & \multicolumn{3}{c}{\cellcolor[HTML]{BFBFBF}\textbf{Naïve}} \\

\cline{4-9}
\rowcolor[HTML]{BFBFBF} 

\textbf{Treatment strategy} & \textbf{Time} & \textbf{True value} & \textbf{IPW}     & \textbf{G-computation}  & \textbf{TMLE}   & \textbf{IPW}    & \textbf{G-computation}  & \textbf{TMLE}  \\

\rowcolor[HTML]{FFFFFF} 

% ---------------------------------
% section to replace for each table
% ---------------------------------

\rowcolor[HTML]{FFFFFF} 
Always   ventilate & 1 & 0.716683 & -0.000 (0.000) & -0.001 (0.000) & -0.000 (0.000) & -0.000 (0.000) & -0.014 (0.000) & -0.000 (0.000) \\
\rowcolor[HTML]{FFFFFF} 
 & 2 & 0.598304 & -0.000 (0.000) & -0.001 (0.000) & 0.001 (0.000) & -0.008 (0.000) & -0.025 (0.000) & -0.008 (0.000) \\
\rowcolor[HTML]{FFFFFF} 
 & 3 & 0.558907 & -0.000 (0.000) & -0.000 (0.000) & 0.002 (0.000) & -0.015 (0.000) & -0.026 (0.000) & -0.015 (0.000) \\
\rowcolor[HTML]{FFFFFF} 
 & 4 & 0.546649 & -0.000 (0.000) & -0.001 (0.000) & 0.003 (0.000) & -0.018 (0.000) & -0.037 (0.000) & -0.019 (0.000) \\
\rowcolor[HTML]{FFFFFF} 
 & 5 & 0.541889 & -0.000 (0.000) & -0.004 (0.000) & 0.003 (0.000) & -0.020 (0.000) & -0.048 (0.000) & -0.020 (0.000) \\
\rowcolor[HTML]{F2F2F2} 
Never ventilate & 1 & 0.656192 & 0.000 (0.000) & 0.001 (0.000) & 0.000 (0.000) & -0.000 (0.000) & -0.004 (0.000) & 0.000 (0.000) \\
\rowcolor[HTML]{F2F2F2} 
 & 2 & 0.431255 & 0.000 (0.000) & -0.001 (0.000) & 0.001 (0.000) & 0.012 (0.000) & 0.015 (0.000) & 0.007 (0.000) \\
\rowcolor[HTML]{F2F2F2} 
 & 3 & 0.314202 & 0.002 (0.001) & -0.002 (0.000) & 0.000 (0.000) & 0.015 (0.000) & 0.028 (0.000) & 0.009 (0.000) \\
\rowcolor[HTML]{F2F2F2} 
 & 4 & 0.251886 & 0.002 (0.001) & -0.005 (0.000) & -0.000 (0.000) & 0.014 (0.000) & 0.031 (0.000) & 0.009 (0.000) \\
\rowcolor[HTML]{F2F2F2} 
 & 5 & 0.213336 & 0.002 (0.001) & -0.007 (0.000) & -0.001 (0.000) & 0.012 (0.000) & 0.030 (0.000) & 0.008 (0.000) \\
\rowcolor[HTML]{FFFFFF} 
Ventilate early & 1 & 0.71684 & -0.000 (0.000) & -0.001 (0.000) & -0.000 (0.000) & -0.000 (0.000) & -0.014 (0.000) & -0.000 (0.000) \\
\rowcolor[HTML]{FFFFFF} 
 & 2 & 0.598069 & 0.000 (0.000) & -0.001 (0.000) & 0.001 (0.000) & -0.008 (0.000) & -0.025 (0.000) & -0.008 (0.000) \\
\rowcolor[HTML]{FFFFFF} 
 & 3 & 0.558623 & 0.000 (0.000) & 0.000 (0.000) & 0.002 (0.000) & -0.015 (0.000) & -0.025 (0.000) & -0.015 (0.000) \\
\rowcolor[HTML]{FFFFFF} 
 & 4 & 0.545497 & 0.000 (0.000) & -0.001 (0.000) & 0.003 (0.000) & -0.017 (0.000) & -0.038 (0.000) & -0.017 (0.000) \\
\rowcolor[HTML]{FFFFFF} 
 & 5 & 0.538519 & 0.000 (0.000) & -0.004 (0.000) & 0.003 (0.000) & -0.017 (0.000) & -0.053 (0.000) & -0.018 (0.000) \\
\rowcolor[HTML]{F2F2F2} 
Wait to ventilate & 1 & 0.656582 & -0.000 (0.000) & 0.000 (0.000) & -0.000 (0.000) & -0.001 (0.000) & -0.005 (0.000) & -0.000 (0.000) \\
\rowcolor[HTML]{F2F2F2} 
 & 2 & 0.432248 & -0.001 (0.000) & -0.002 (0.000) & -0.000 (0.000) & 0.011 (0.000) & 0.014 (0.000) & 0.006 (0.000) \\
\rowcolor[HTML]{F2F2F2} 
 & 3 & 0.321668 & -0.001 (0.000) & -0.004 (0.000) & -0.001 (0.000) & 0.012 (0.000) & 0.027 (0.000) & 0.006 (0.000) \\
\rowcolor[HTML]{F2F2F2} 
 & 4 & 0.273636 & 0.000 (0.000) & -0.004 (0.000) & -0.002 (0.000) & 0.011 (0.000) & 0.031 (0.000) & 0.005 (0.000) \\
\rowcolor[HTML]{F2F2F2} 
 & 5 & 0.253851 & 0.002 (0.000) & -0.003 (0.000) & -0.002 (0.000) & 0.012 (0.000) & 0.033 (0.000) & 0.007 (0.000)

% ---------------------------------
% end of section to replace
% ---------------------------------

\end{tabular}
\end{adjustbox}
\caption{Bias (with corresponding Monte Carlo standard errors) from 1000 simulations for simulation scenario 2: stronger effect of past treatment, covariate and monitoring on current monitoring.}
\label{tab:sim_results_scenario2}
\end{table}

% *******************************************************
% *******************************************************
% Results table - SCENARIO 3
% *******************************************************
% *******************************************************

\renewcommand{\arraystretch}{1.4}
\begin{table}[H]
\centering
\large
\setlength{\tabcolsep}{4pt}

\begin{adjustbox}{max width=\textwidth}
\begin{tabular}{llccccccc}

\rowcolor[HTML]{BFBFBF} 
\textbf{} & \textbf{} & \textbf{} & \multicolumn{3}{c}{\cellcolor[HTML]{BFBFBF}\textbf{Adapted}} & \multicolumn{3}{c}{\cellcolor[HTML]{BFBFBF}\textbf{Naïve}} \\

\cline{4-9}
\rowcolor[HTML]{BFBFBF} 

\textbf{Treatment strategy} & \textbf{Time} & \textbf{True value} & \textbf{IPW}     & \textbf{G-computation}  & \textbf{TMLE}   & \textbf{IPW}    & \textbf{G-computation}  & \textbf{TMLE}  \\

\rowcolor[HTML]{FFFFFF} 

% ---------------------------------
% section to replace for each table
% ---------------------------------

\rowcolor[HTML]{FFFFFF} 
Always   ventilate & 1 & 0.716798 & -0.000 (0.000) & -0.001 (0.000) & -0.000 (0.000) & -0.009 (0.000) & -0.012 (0.000) & -0.000 (0.000) \\
\rowcolor[HTML]{FFFFFF} 
 & 2 & 0.598707 & -0.001 (0.000) & -0.005 (0.000) & 0.001 (0.000) & -0.016 (0.000) & -0.012 (0.000) & -0.003 (0.000) \\
\rowcolor[HTML]{FFFFFF} 
 & 3 & 0.559296 & -0.001 (0.000) & -0.008 (0.000) & 0.002 (0.000) & -0.020 (0.000) & -0.010 (0.000) & -0.007 (0.000) \\
\rowcolor[HTML]{FFFFFF} 
 & 4 & 0.546893 & -0.001 (0.000) & -0.014 (0.000) & 0.004 (0.000) & -0.023 (0.000) & -0.021 (0.000) & -0.009 (0.000) \\
\rowcolor[HTML]{FFFFFF} 
 & 5 & 0.542192 & -0.001 (0.000) & -0.018 (0.000) & 0.005 (0.000) & -0.025 (0.000) & -0.032 (0.000) & -0.012 (0.000) \\
\rowcolor[HTML]{F2F2F2} 
Never ventilate & 1 & 0.655996 & 0.000 (0.000) & 0.001 (0.000) & 0.000 (0.000) & 0.005 (0.000) & -0.001 (0.000) & 0.000 (0.000) \\
\rowcolor[HTML]{F2F2F2} 
 & 2 & 0.431636 & 0.000 (0.000) & -0.003 (0.000) & 0.000 (0.000) & 0.019 (0.000) & 0.021 (0.000) & 0.007 (0.000) \\
\rowcolor[HTML]{F2F2F2} 
 & 3 & 0.314312 & 0.001 (0.001) & 0.005 (0.000) & 0.001 (0.000) & 0.032 (0.000) & 0.039 (0.000) & 0.017 (0.000) \\
\rowcolor[HTML]{F2F2F2} 
 & 4 & 0.252236 & 0.002 (0.001) & 0.017 (0.000) & 0.002 (0.000) & 0.039 (0.000) & 0.050 (0.000) & 0.025 (0.000) \\
\rowcolor[HTML]{F2F2F2} 
 & 5 & 0.213494 & 0.003 (0.001) & 0.029 (0.000) & 0.004 (0.001) & 0.045 (0.001) & 0.058 (0.000) & 0.030 (0.001) \\
\rowcolor[HTML]{FFFFFF} 
Ventilate early & 1 & 0.71684 & -0.000 (0.000) & -0.001 (0.000) & -0.000 (0.000) & -0.009 (0.000) & -0.012 (0.000) & -0.000 (0.000) \\
\rowcolor[HTML]{FFFFFF} 
 & 2 & 0.598069 & 0.000 (0.000) & -0.004 (0.000) & 0.002 (0.000) & -0.016 (0.000) & -0.011 (0.000) & -0.002 (0.000) \\
\rowcolor[HTML]{FFFFFF} 
 & 3 & 0.558623 & 0.000 (0.000) & -0.008 (0.000) & 0.003 (0.000) & -0.020 (0.000) & -0.009 (0.000) & -0.006 (0.000) \\
\rowcolor[HTML]{FFFFFF} 
 & 4 & 0.545608 & 0.000 (0.000) & -0.014 (0.000) & 0.004 (0.000) & -0.021 (0.000) & -0.021 (0.000) & -0.007 (0.000) \\
\rowcolor[HTML]{FFFFFF} 
 & 5 & 0.538984 & 0.000 (0.000) & -0.019 (0.000) & 0.005 (0.000) & -0.021 (0.000) & -0.035 (0.000) & -0.007 (0.000) \\
\rowcolor[HTML]{F2F2F2} 
Wait to ventilate & 1 & 0.65644 & 0.000 (0.000) & 0.001 (0.000) & -0.000 (0.000) & 0.005 (0.000) & -0.002 (0.000) & 0.000 (0.000) \\
\rowcolor[HTML]{F2F2F2} 
 & 2 & 0.431714 & 0.000 (0.000) & -0.003 (0.000) & 0.000 (0.000) & 0.019 (0.000) & 0.020 (0.000) & 0.007 (0.000) \\
\rowcolor[HTML]{F2F2F2} 
 & 3 & 0.32236 & 0.000 (0.000) & 0.005 (0.000) & 0.001 (0.000) & 0.023 (0.000) & 0.037 (0.000) & 0.010 (0.000) \\
\rowcolor[HTML]{F2F2F2} 
 & 4 & 0.276383 & -0.000 (0.000) & 0.013 (0.000) & 0.000 (0.000) & 0.022 (0.000) & 0.045 (0.000) & 0.009 (0.000) \\
\rowcolor[HTML]{F2F2F2} 
 & 5 & 0.257577 & 0.001 (0.000) & 0.020 (0.000) & 0.002 (0.000) & 0.022 (0.000) & 0.050 (0.000) & 0.009 (0.000)

% ---------------------------------
% end of section to replace
% ---------------------------------

\end{tabular}
\end{adjustbox}
\caption{Bias (with corresponding Monte Carlo standard errors) from 1000 simulations for simulation scenario 3: no effect of past treatment, covariate and monitoring on current monitoring (i.e. monitoring random).}
\label{tab:sim_results_scenario3}
\end{table}

% *******************************************************
% *******************************************************
% Results table - SCENARIO 4
% *******************************************************
% *******************************************************

\renewcommand{\arraystretch}{1.4}
\begin{table}[H]
\centering
\large
\setlength{\tabcolsep}{4pt}

\begin{adjustbox}{max width=\textwidth}
\begin{tabular}{llccccccc}

\rowcolor[HTML]{BFBFBF} 
\textbf{} & \textbf{} & \textbf{} & \multicolumn{3}{c}{\cellcolor[HTML]{BFBFBF}\textbf{Adapted}} & \multicolumn{3}{c}{\cellcolor[HTML]{BFBFBF}\textbf{Naïve}} \\

\cline{4-9}
\rowcolor[HTML]{BFBFBF} 

\textbf{Treatment strategy} & \textbf{Time} & \textbf{True value} & \textbf{IPW}     & \textbf{G-computation}  & \textbf{TMLE}   & \textbf{IPW}    & \textbf{G-computation}  & \textbf{TMLE}  \\

\rowcolor[HTML]{FFFFFF} 

% ---------------------------------
% section to replace for each table
% ---------------------------------

\rowcolor[HTML]{FFFFFF} 
Always   ventilate & 1 & 0.716798 & -0.000 (0.000) & -0.001 (0.000) & -0.000 (0.000) & -0.002 (0.000) & -0.009 (0.000) & -0.000 (0.000) \\
\rowcolor[HTML]{FFFFFF} 
 & 2 & 0.598707 & -0.001 (0.000) & -0.002 (0.000) & 0.001 (0.000) & -0.010 (0.000) & -0.019 (0.000) & -0.007 (0.000) \\
\rowcolor[HTML]{FFFFFF} 
 & 3 & 0.559296 & -0.001 (0.000) & -0.001 (0.000) & 0.003 (0.000) & -0.019 (0.000) & -0.021 (0.000) & -0.017 (0.000) \\
\rowcolor[HTML]{FFFFFF} 
 & 4 & 0.546893 & -0.001 (0.000) & -0.001 (0.000) & 0.005 (0.000) & -0.024 (0.000) & -0.030 (0.000) & -0.025 (0.000) \\
\rowcolor[HTML]{FFFFFF} 
 & 5 & 0.542192 & -0.001 (0.000) & -0.003 (0.000) & 0.006 (0.000) & -0.028 (0.000) & -0.038 (0.000) & -0.030 (0.000) \\
\rowcolor[HTML]{F2F2F2} 
Never ventilate & 1 & 0.655996 & 0.000 (0.000) & 0.001 (0.000) & 0.000 (0.000) & 0.000 (0.000) & 0.002 (0.000) & 0.000 (0.000) \\
\rowcolor[HTML]{F2F2F2} 
 & 2 & 0.431636 & 0.000 (0.001) & -0.003 (0.000) & 0.001 (0.000) & 0.034 (0.000) & 0.015 (0.000) & 0.019 (0.000) \\
\rowcolor[HTML]{F2F2F2} 
 & 3 & 0.314312 & 0.003 (0.001) & -0.005 (0.000) & 0.001 (0.001) & 0.050 (0.000) & 0.024 (0.000) & 0.033 (0.000) \\
\rowcolor[HTML]{F2F2F2} 
 & 4 & 0.252236 & 0.005 (0.001) & -0.007 (0.000) & -0.000 (0.001) & 0.049 (0.001) & 0.026 (0.000) & 0.035 (0.000) \\
\rowcolor[HTML]{F2F2F2} 
 & 5 & 0.213494 & 0.006 (0.001) & -0.006 (0.000) & -0.000 (0.001) & 0.046 (0.001) & 0.025 (0.000) & 0.034 (0.000) \\
\rowcolor[HTML]{FFFFFF} 
Ventilate early & 1 & 0.71684 & -0.000 (0.000) & -0.001 (0.000) & -0.000 (0.000) & -0.002 (0.000) & -0.009 (0.000) & -0.000 (0.000) \\
\rowcolor[HTML]{FFFFFF} 
 & 2 & 0.598069 & -0.000 (0.000) & -0.001 (0.000) & 0.002 (0.000) & -0.009 (0.000) & -0.018 (0.000) & -0.007 (0.000) \\
\rowcolor[HTML]{FFFFFF} 
 & 3 & 0.558623 & -0.000 (0.000) & -0.000 (0.000) & 0.004 (0.000) & -0.018 (0.000) & -0.020 (0.000) & -0.017 (0.000) \\
\rowcolor[HTML]{FFFFFF} 
 & 4 & 0.545696 & -0.000 (0.000) & -0.001 (0.000) & 0.005 (0.000) & -0.020 (0.000) & -0.030 (0.000) & -0.020 (0.000) \\
\rowcolor[HTML]{FFFFFF} 
 & 5 & 0.539238 & -0.000 (0.000) & -0.003 (0.000) & 0.006 (0.000) & -0.021 (0.000) & -0.040 (0.000) & -0.021 (0.000) \\
\rowcolor[HTML]{F2F2F2} 
Wait to ventilate & 1 & 0.65644 & 0.000 (0.000) & 0.000 (0.000) & -0.000 (0.000) & -0.000 (0.000) & 0.001 (0.000) & 0.000 (0.000) \\
\rowcolor[HTML]{F2F2F2} 
 & 2 & 0.431714 & 0.000 (0.001) & -0.003 (0.000) & 0.001 (0.000) & 0.034 (0.000) & 0.015 (0.000) & 0.019 (0.000) \\
\rowcolor[HTML]{F2F2F2} 
 & 3 & 0.32168 & -0.000 (0.001) & -0.005 (0.000) & 0.001 (0.000) & 0.040 (0.000) & 0.023 (0.000) & 0.024 (0.000) \\
\rowcolor[HTML]{F2F2F2} 
 & 4 & 0.274165 & 0.000 (0.000) & -0.003 (0.000) & -0.000 (0.000) & 0.038 (0.000) & 0.026 (0.000) & 0.021 (0.000) \\
\rowcolor[HTML]{F2F2F2} 
 & 5 & 0.254544 & 0.002 (0.000) & 0.000 (0.000) & 0.001 (0.000) & 0.038 (0.000) & 0.027 (0.000) & 0.022 (0.000)

% ---------------------------------
% end of section to replace
% ---------------------------------

\end{tabular}
\end{adjustbox}
\caption{Bias (with corresponding Monte Carlo standard errors) from 1000 simulations for simulation scenario 4: stronger effect of past monitoring on current treatment.}
\label{tab:sim_results_scenario4}
\end{table}

% *******************************************************
% *******************************************************
% Results table - SCENARIO 5
% *******************************************************
% *******************************************************

\renewcommand{\arraystretch}{1.4}
\begin{table}[H]
\centering
\large
\setlength{\tabcolsep}{4pt}

\begin{adjustbox}{max width=\textwidth}
\begin{tabular}{llccccccc}

\rowcolor[HTML]{BFBFBF} 
\textbf{} & \textbf{} & \textbf{} & \multicolumn{3}{c}{\cellcolor[HTML]{BFBFBF}\textbf{Adapted}} & \multicolumn{3}{c}{\cellcolor[HTML]{BFBFBF}\textbf{Naïve}} \\

\cline{4-9}
\rowcolor[HTML]{BFBFBF} 

\textbf{Treatment strategy} & \textbf{Time} & \textbf{True value} & \textbf{IPW}     & \textbf{G-computation}  & \textbf{TMLE}   & \textbf{IPW}    & \textbf{G-computation}  & \textbf{TMLE}  \\

\rowcolor[HTML]{FFFFFF} 

% ---------------------------------
% section to replace for each table
% ---------------------------------

\rowcolor[HTML]{FFFFFF} 
Always   ventilate & 1 & 0.716798 & -0.000 (0.000) & -0.001 (0.000) & -0.000 (0.000) & -0.001 (0.000) & -0.006 (0.000) & -0.000 (0.000) \\
\rowcolor[HTML]{FFFFFF} 
 & 2 & 0.598707 & -0.001 (0.000) & -0.001 (0.000) & 0.001 (0.000) & -0.001 (0.000) & -0.013 (0.000) & -0.001 (0.000) \\
\rowcolor[HTML]{FFFFFF} 
 & 3 & 0.559296 & -0.001 (0.000) & -0.000 (0.000) & 0.001 (0.000) & -0.001 (0.000) & -0.015 (0.000) & -0.001 (0.000) \\
\rowcolor[HTML]{FFFFFF} 
 & 4 & 0.546893 & -0.001 (0.000) & -0.000 (0.000) & 0.002 (0.000) & -0.001 (0.000) & -0.025 (0.000) & -0.001 (0.000) \\
\rowcolor[HTML]{FFFFFF} 
 & 5 & 0.542192 & -0.002 (0.000) & -0.002 (0.000) & 0.003 (0.000) & -0.002 (0.001) & -0.033 (0.000) & -0.002 (0.000) \\
\rowcolor[HTML]{F2F2F2} 
Never ventilate & 1 & 0.655996 & 0.000 (0.000) & 0.001 (0.000) & 0.000 (0.000) & 0.001 (0.000) & -0.001 (0.000) & 0.000 (0.000) \\
\rowcolor[HTML]{F2F2F2} 
 & 2 & 0.431636 & -0.000 (0.000) & -0.002 (0.000) & 0.000 (0.000) & 0.000 (0.000) & 0.008 (0.000) & -0.000 (0.000) \\
\rowcolor[HTML]{F2F2F2} 
 & 3 & 0.314312 & 0.001 (0.000) & -0.004 (0.000) & -0.000 (0.000) & 0.001 (0.000) & 0.014 (0.000) & 0.001 (0.000) \\
\rowcolor[HTML]{F2F2F2} 
 & 4 & 0.252236 & 0.001 (0.000) & -0.006 (0.000) & -0.001 (0.000) & 0.002 (0.000) & 0.014 (0.000) & 0.001 (0.000) \\
\rowcolor[HTML]{F2F2F2} 
 & 5 & 0.213494 & 0.001 (0.000) & -0.005 (0.000) & -0.001 (0.000) & 0.002 (0.001) & 0.013 (0.000) & 0.001 (0.000) \\
\rowcolor[HTML]{FFFFFF} 
Ventilate early & 1 & 0.71684 & -0.000 (0.000) & -0.001 (0.000) & -0.000 (0.000) & -0.001 (0.000) & -0.006 (0.000) & -0.000 (0.000) \\
\rowcolor[HTML]{FFFFFF} 
 & 2 & 0.598069 & 0.000 (0.000) & -0.000 (0.000) & 0.001 (0.000) & -0.000 (0.000) & -0.012 (0.000) & -0.000 (0.000) \\
\rowcolor[HTML]{FFFFFF} 
 & 3 & 0.558623 & -0.000 (0.000) & 0.000 (0.000) & 0.002 (0.000) & -0.001 (0.000) & -0.015 (0.000) & -0.000 (0.000) \\
\rowcolor[HTML]{FFFFFF} 
 & 4 & 0.545213 & -0.000 (0.000) & 0.000 (0.000) & 0.003 (0.000) & -0.001 (0.000) & -0.025 (0.000) & -0.000 (0.000) \\
\rowcolor[HTML]{FFFFFF} 
 & 5 & 0.537898 & -0.000 (0.000) & -0.003 (0.000) & 0.003 (0.000) & -0.001 (0.000) & -0.038 (0.000) & -0.000 (0.000) \\
\rowcolor[HTML]{F2F2F2} 
Wait to ventilate & 1 & 0.65644 & 0.000 (0.000) & 0.001 (0.000) & -0.000 (0.000) & 0.000 (0.000) & -0.002 (0.000) & 0.000 (0.000) \\
\rowcolor[HTML]{F2F2F2} 
 & 2 & 0.431714 & -0.000 (0.000) & -0.002 (0.000) & -0.000 (0.000) & -0.000 (0.000) & 0.007 (0.000) & -0.000 (0.000) \\
\rowcolor[HTML]{F2F2F2} 
 & 3 & 0.320735 & -0.000 (0.000) & -0.004 (0.000) & -0.001 (0.000) & 0.000 (0.000) & 0.014 (0.000) & -0.000 (0.000) \\
\rowcolor[HTML]{F2F2F2} 
 & 4 & 0.272515 & 0.001 (0.000) & -0.003 (0.000) & -0.001 (0.000) & 0.001 (0.000) & 0.016 (0.000) & 0.001 (0.000) \\
\rowcolor[HTML]{F2F2F2} 
 & 5 & 0.25257 & 0.002 (0.000) & 0.000 (0.000) & -0.001 (0.000) & 0.003 (0.000) & 0.018 (0.000) & 0.003 (0.000)

% ---------------------------------
% end of section to replace
% ---------------------------------

\end{tabular}
\end{adjustbox}
\caption{Bias (with corresponding Monte Carlo standard errors) from 1000 simulations for simulation scenario 5: no effect of past monitoring on current treatment.}
\label{tab:sim_results_scenario5}
\end{table}

\newpage

\begin{figure}[H]
\centering
\hspace*{-1.3cm}
\includegraphics[width=1\columnwidth]{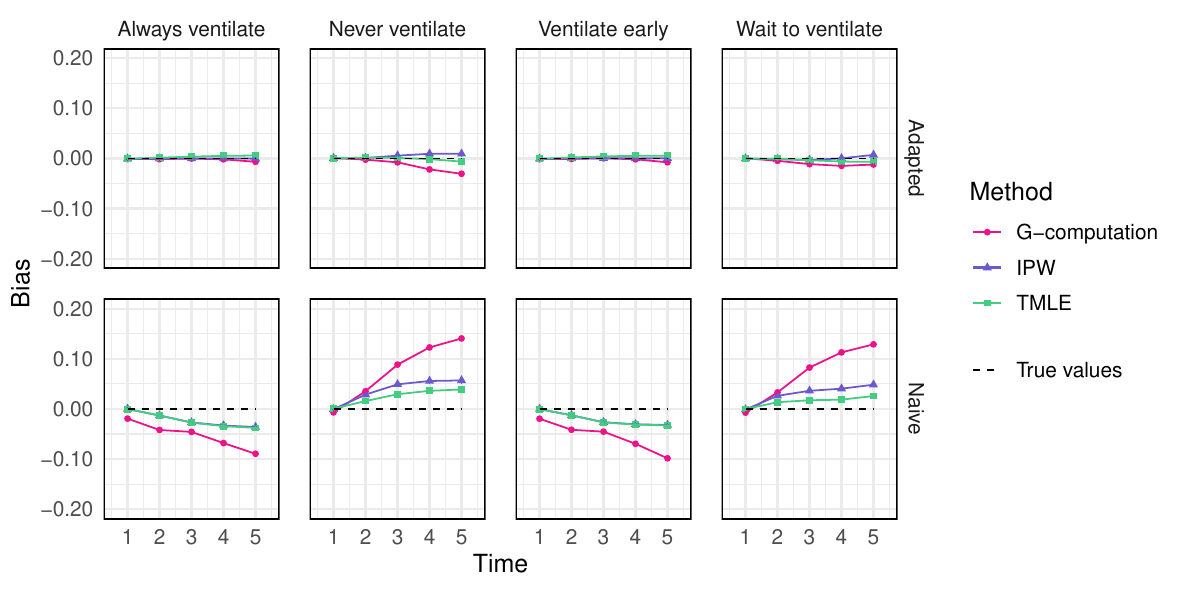}
\caption{Simulation results for scenario 2: stronger effect of past treatment, covariate and monitoring on current monitoring.}
\label{fig:scenario2}
\end{figure}

\begin{figure}[H]
\centering
\hspace*{-1.3cm}
\includegraphics[width=1\columnwidth]{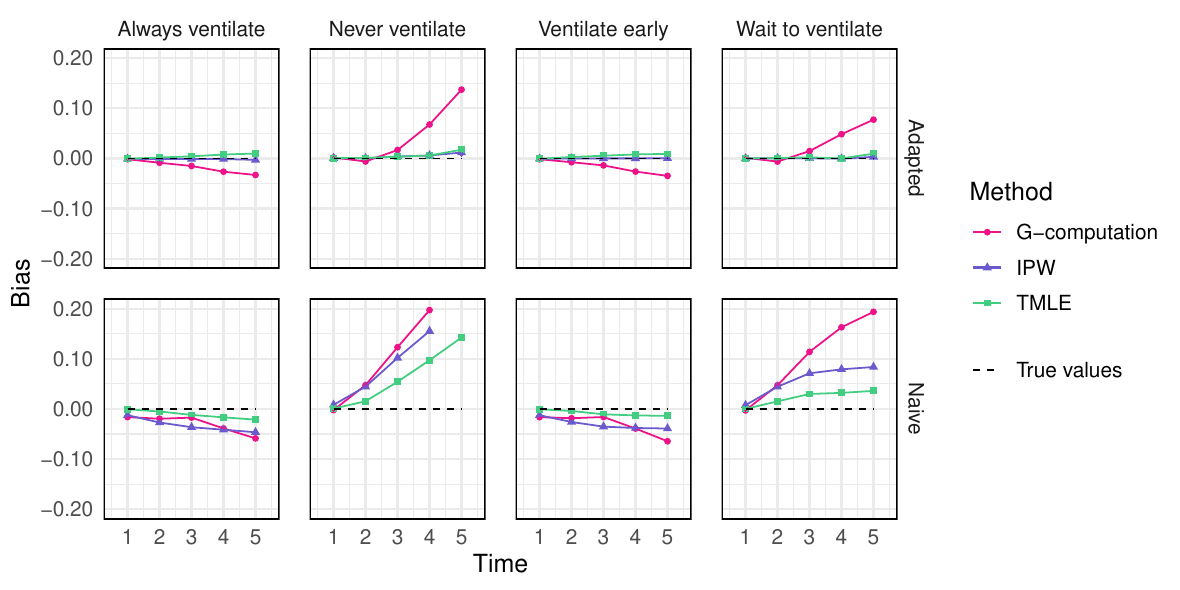}
\caption{Simulation results for scenario 3: no effect of past treatment, covariate and monitoring on current monitoring (i.e. monitoring random).}
\label{fig:scenario3}
\end{figure}

\begin{figure}[H]
\centering
\hspace*{-1.3cm}
\includegraphics[width=1\columnwidth]{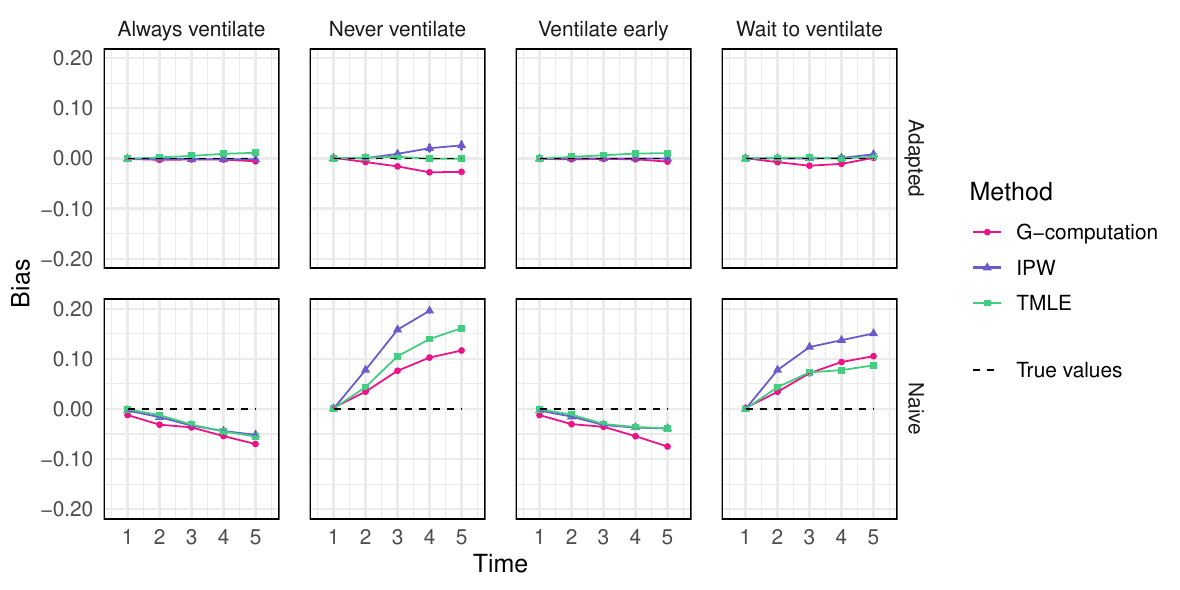}
\caption{Simulation results for scenario 4: stronger effect of past monitoring on current treatment.}
\label{fig:scenario4}
\end{figure}

\begin{figure}[H]
\centering
\hspace*{-1.3cm}
\includegraphics[width=1\columnwidth]{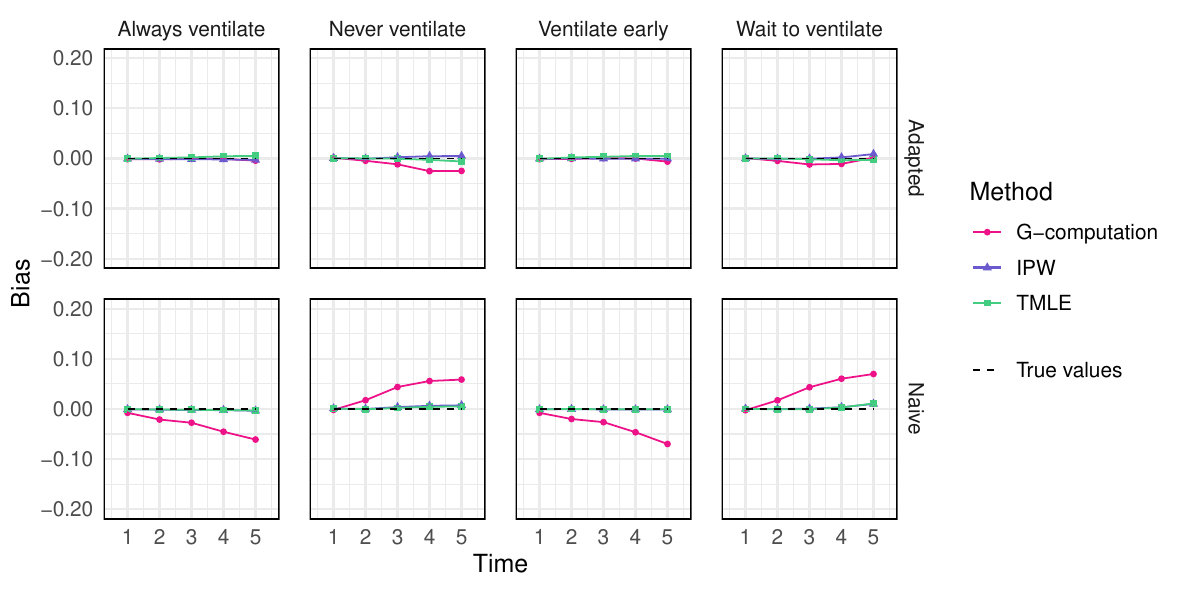}
\caption{Simulation results for scenario 5: no effect of past monitoring on current treatment.}
\label{fig:scenario5}
\end{figure}

\newpage

\subsection{Application to UCLH ventilation}

% --------------------------------------
% table of time-dependent confounders
% --------------------------------------

\renewcommand{\arraystretch}{1.5}
\begin{table}[H]
\begin{adjustbox}{max width=\textwidth}
\small
\begin{tabular}{
>{\raggedright\arraybackslash}m{4.5cm}
>{\raggedright\arraybackslash}m{10cm}
}

\rowcolor[HTML]{D9D9D9} 
\textbf{Confounder} & \textbf{Summary measure} \\

\rowcolor[HTML]{FFFFFF}
Glasgow Coma Score (GCS)  & Maximum GCS score recorded in 24h period  \\

\rowcolor[HTML]{F2F2F2}
Respiratory rate (RR)  & Mean RR measurement in 24h period \\

\rowcolor[HTML]{FFFFFF}
$F_I O_2$  & Maximum $F_I O_2$ measurement in 24h period \\

\rowcolor[HTML]{F2F2F2}
$SpO_2$  & Minimum $SpO_2$ measurement in 24h period \\

\rowcolor[HTML]{FFFFFF}
pH  & Minimum pH measurement in 24h period \\

\rowcolor[HTML]{F2F2F2}
Steroid use & Binary indicator of whether steroid was administered at any point during   24h period \\

\rowcolor[HTML]{FFFFFF}
Non-invasive ventilation (NIV)  & Binary indicator of whether patient received NIV at any point during 24h   period     \\

\rowcolor[HTML]{F2F2F2}
ICU status  & Binary indicator of whether patient was in the ICU at start of 24h period             
\end{tabular}
\end{adjustbox}
\caption{List of time-varying confounders and details of how measurements in the raw data are summarised within each 24-hour period to obtain a single value of $L_k$ in the discretised data}
\end{table}

\vspace{3cm}

% --------------------------------------
% figure 1: ventilation status variable
% --------------------------------------

\begin{figure}[H]
    \centering
    \includegraphics[width=1\linewidth]{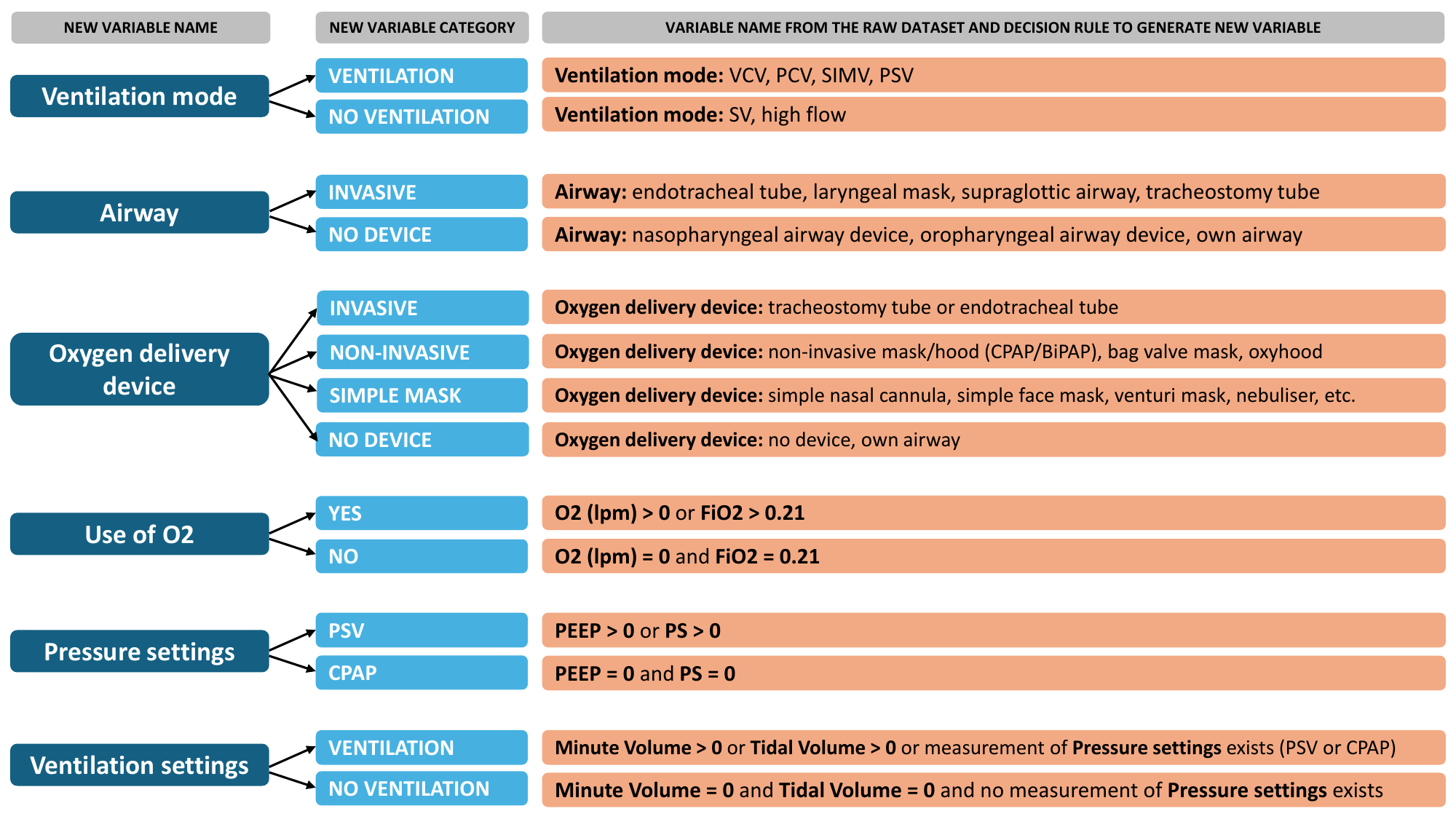}
    \caption{Step 1 to define ventilation status variable. From the raw data we identify the complete set of measurements for variables relating to ventilation status. Using these variables we create a cleaned set of six variables describing ventilation: \textit{ventilation mode}, \textit{airway}, \textit{oxygen delivery device}, \textit{use of $O_2$}, \textit{pressure settings} and \textit{ventilation settings}. Dark blue boxes are the newly created variable names. Light blue boxes are the categories of the newly created variables. Orange boxes identify the existing variables within the raw dataset and how these are categorised. For \textit{ventilation mode}, \textit{airway} and \textit{oxygen delivery device}, we take existing categorical variables of the same names and recategorise into a restricted number of categories. For \textit{use of $O_2$}, \textit{pressure settings} and \textit{ventilation settings} an observation is generated in the cleaned data when a measurement of the relevant variable or variables in the raw data is identified at that observation time. For example, an observation of the variable \textit{use of $O_2$} is generated at observation time $t$ when a measurement of $O_2$ or a measurement of $F_I O_2$ is identified in the raw data at time $t$, with its value being determined according to the decision rules above. VCV = volume controlled ventilation, PCV = pressure controlled ventilation, SIMV = , PSV = pressure support ventilation, SV = , CPAP = , PEEP = , PS = , lpm = litres per minute.}
    \label{fig:vent_status1}
\end{figure}

\vspace{3cm}

% --------------------------------------
% figure 2: ventilation status variable
% --------------------------------------

\begin{figure}[H]
    \centering
    \includegraphics[width=1\linewidth]{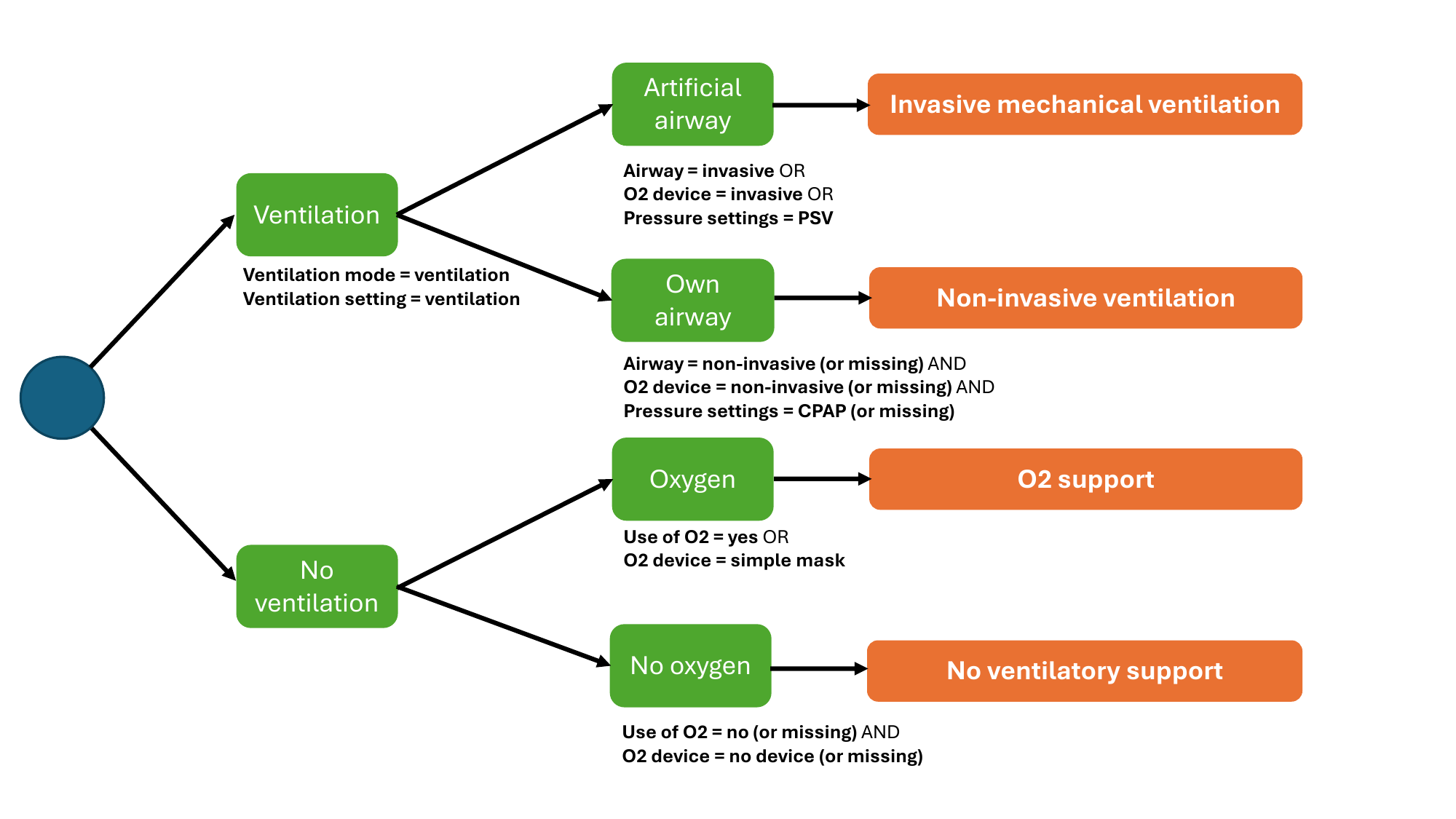}
    \caption{Step 2 to define ventilation status variable. We create a dataset with rows equal to the unique set of person IDs and measurement datetimes available from the six newly created ventilation variables from step 1, with a column for each ventilation variable. The dataset is ordered by person ID and then by measurement datetime, and we use last observation carried forward to populate missing values. We then use the above decision rules to create a single variable identifying ventilation status with categories: \textit{invasive mechanical ventilation} (IMV), \textit{non-invasive ventilation} (NIV), \textit{$O_2$ support}, and \textit{no ventilatory support}. From this variable we also create the binary IMV and NIV status variables. When we discretise the data into 24-hour periods, an individual is assigned IMV status $=1$ (i.e. $A_k=1$) if an observation of IMV exists at any point during the 24-hour period. If no observations of ventilation status exist during the 24-hour period, this is assumed to be indicative of receiving no ventilatory support.}
    \label{fig:vent_status2}
\end{figure}

\vspace{3cm}

% -----------------------------------------
% table of eligibility criteria and numbers
% -----------------------------------------

\renewcommand{\arraystretch}{1.2}
\begin{table}[H]
\caption{Table describing the number of patient-spells and number of excluded patient-spells after application of each eligibility criteria.}
\begin{adjustbox}{max width=\textwidth}
\Large
\begin{threeparttable}
\begin{tabular}{>{\raggedright\arraybackslash}m{18cm}
                >{\raggedright\arraybackslash}m{2.5cm}
                >{\raggedright\arraybackslash}m{2.5cm}}

\rowcolor[HTML]{D9D9D9} 
\textbf{Eligibility description} & \textbf{No. patient-spells}\tnote{a} & \textbf{No. excluded} \\

\rowcolor[HTML]{FFFFFF}
- & \textbf{8676}\tnote{b} & - \\

\rowcolor[HTML]{F2F2F2}
Exclude ICU visits $<$ 24 hours & \textbf{8488} & 188 \\

\rowcolor[HTML]{FFFFFF}
Exclude patients under the age of 18 years on ICU admission & \textbf{8486} & 2 \\

\rowcolor[HTML]{F2F2F2}
Exclude patients with a DNAR issued in the first 24 hours of ICU admission & \textbf{8115} & 371 \\

\rowcolor[HTML]{FFFFFF}
Exclude subsequent ICU visits occurring within the study period & \textbf{7050} & 1065 \\

\rowcolor[HTML]{F2F2F2}
Exclude patients with less than 30 days of follow-up due to data cut-off & \textbf{6972} & 78 \\

\rowcolor[HTML]{FFFFFF}
Exclude patients who are not receiving at least supplemental oxygen at baseline (i.e. within the first 24 hours of ICU admission) & \textbf{4398} & 2574\tnote{c} \\

\rowcolor[HTML]{F2F2F2}
Exclude patients with respiratory rate outside of the range {[}20,40{]} breaths per minute at baseline (i.e. within the first 24 hours of ICU admission) & \textbf{827} & 3571 \\

\end{tabular}

\begin{tablenotes}
\item[a] A spell refers to an ICU stay. The same patient may have multiple ICU stays until we restrict the population to only the first ICU stay.
\item[b] The starting point is the number of patient-spells in the CC-HIC database with ICU admission date falling within the study period.
\item[c] 97\% of the excluded patients had no available measurements of either $O_2$ or $F_I O_2$ (which is likely to be indicative of receiving no supplemental oxygen). The remaining 3\% of excluded patients have a measurement of $O_2=0$ litres per minute and $F_I O_2=0.21$.
\end{tablenotes}

\end{threeparttable}

\end{adjustbox}

\label{tab:eligibility_criteria_no_patients}
\end{table}

\vspace{3cm}

% ------------------------------------------------------------
% table of baseline characteristics for the eligible patients
% ------------------------------------------------------------

\renewcommand{\arraystretch}{1.5}
\begin{table}[H]
\begin{adjustbox}{max width=\textwidth}
\begin{tabular}{>{\raggedright\arraybackslash}m{5cm}ccc}

\rowcolor[HTML]{D9D9D9}
\multicolumn{1}{c}{} & \textbf{\begin{tabular}[c]{@{}c@{}}Overall\\      (N = 827)\end{tabular}} & \textbf{\begin{tabular}[c]{@{}c@{}}Receiving IMV at baseline\\      (N = 255)\end{tabular}} & \textbf{\begin{tabular}[c]{@{}c@{}}Not receiving IMV at baseline\\      (N = 572)\end{tabular}} \\

\rowcolor[HTML]{F2F2F2}
\textbf{Age (years)} &  &  &  \\
Median (IQR) & 60.6 (47.6, 70.5) & 58.1 (47.3, 67.0) & 62.4 (47.7, 71.4) \\

\rowcolor[HTML]{F2F2F2}
\textbf{Sex} &  &  &  \\
Female & 316 (38.2\%) & 74 (29.0\%) & 242 (42.3\%) \\
Male & 511 (61.8\%) & 181 (71.0\%) & 330 (57.7\%) \\

\rowcolor[HTML]{F2F2F2}
\textbf{Ethnicity} &  &  &  \\
Non-white & 239 (28.9\%) & 80 (31.4\%) & 159 (27.8\%) \\
White & 365 (44.1\%) & 93 (36.5\%) & 272 (47.6\%) \\
(Missing) & 223 (27.0\%) & 82 (32.2\%) & 141 (24.7\%) \\

\rowcolor[HTML]{F2F2F2}
\textbf{Index of multiple deprivation   (quintile)} &  &  &  \\
1 & 213 (25.8\%) & 72 (28.2\%) & 141 (24.7\%) \\
2 & 134 (16.2\%) & 35 (13.7\%) & 99 (17.3\%) \\
3 & 194 (23.5\%) & 62 (24.3\%) & 132 (23.1\%) \\
4 & 122 (14.8\%) & 38 (14.9\%) & 84 (14.7\%) \\
5 & 142 (17.2\%) & 40 (15.7\%) & 102 (17.8\%) \\
(Missing) & 22 (2.7\%) & 8 (3.1\%) & 14 (2.4\%) \\

\rowcolor[HTML]{F2F2F2}
\textbf{Body mass index ($kg / m^2$)} &  &  &  \\
Median (IQR) & 26.7 (22.8, 32.2) & 26.4 (22.8, 33.3) & 26.8 (22.9, 31.9) \\
(Missing) & 262 (31.7\%) & 78 (30.6\%) & 184 (32.2\%) \\

\rowcolor[HTML]{F2F2F2}
\textbf{Smoking status} &  &  &  \\
Current smoker & 94 (11.4\%) & 31 (12.2\%) & 63 (11.0\%) \\
Ex-smoker & 206 (24.9\%) & 43 (16.9\%) & 163 (28.5\%) \\
Never smoker & 230 (27.8\%) & 65 (25.5\%) & 165 (28.8\%) \\
(Missing) & 297 (35.9\%) & 116 (45.5\%) & 181 (31.6\%) \\

\rowcolor[HTML]{F2F2F2}
\textbf{Clinical frailty score} &  &  &  \\
Median (IQR) & 3 (2, 4) & 2 (2, 4) & 3 (2, 4) \\
(Missing) & 786 (95.0\%) & 246 (96.5\%) & 540 (94.4\%) \\

\rowcolor[HTML]{F2F2F2}
\textbf{Asthma} & 111 (13.4\%) & 29 (11.4\%) & 82 (14.3\%) \\

\rowcolor[HTML]{F2F2F2}
\textbf{Chronic respiratory disease} & 250 (30.2\%) & 85 (33.3\%) & 165 (28.8\%) \\

\rowcolor[HTML]{F2F2F2}
\textbf{Chronic kidney disease} & 82 (9.9\%) & 20 (7.8\%) & 62 (10.8\%) \\

\rowcolor[HTML]{F2F2F2}
\textbf{Diabetes} & 209 (25.3\%) & 59 (23.1\%) & 150 (26.2\%) \\

\rowcolor[HTML]{F2F2F2}
\textbf{Heart failure} & 168 (20.3\%) & 62 (24.3\%) & 106 (18.5\%) \\

\rowcolor[HTML]{F2F2F2}
\textbf{Hypertension} & 402 (48.6\%) & 120 (47.1\%) & 282 (49.3\%) \\

\rowcolor[HTML]{F2F2F2}
\textbf{Pneumonia} & 73 (8.8\%) & 24 (9.4\%) & 49 (8.6\%) \\

\rowcolor[HTML]{F2F2F2}
\textbf{Baseline steroid use} & 233 (28.2\%) & 70 (27.5\%) & 163 (28.5\%) \\

\rowcolor[HTML]{F2F2F2}
\textbf{Baseline Glasgow coma score (scale   3:15)} &  &  &  \\
Median (IQR) & 15 (15, 15) & 14.5 (6, 15) & 15 (15, 15) \\
(Missing) & 29 (3.5\%) & 5 (2.0\%) & 24 (4.2\%) \\

\rowcolor[HTML]{F2F2F2}
\textbf{Baseline pH} &  &  &  \\
Median (IQR) & 7.37 (7.28, 7.42) & 7.31 (7.20, 7.40) & 7.38 (7.32, 7.42) \\
(Missing) & 33 (4.0\%) & 3 (1.2\%) & 30 (5.2\%) \\

\rowcolor[HTML]{F2F2F2}
\textbf{Baseline respiratory rate (breaths per   minute)} &  &  &  \\
Median (IQR) & 22.5 (21.0, 25.2) & 22.6 (20.8, 26.2) & 22.5 (21.1, 24.7) \\

\rowcolor[HTML]{F2F2F2}
\textbf{Baseline $SpO_2$ ($\%$)} &  &  &  \\
Median (IQR) & 92 (88, 94) & 90 (86, 93) & 92 (89, 94) \\

\rowcolor[HTML]{F2F2F2}
\textbf{Baseline $F_I O_2$ ($\%$)} &  &  &  \\
Median (IQR) & 50 (33.5, 70) & 70 (50, 99) & 36 (28, 60) \\
(Missing) & 101 (12.2\%) & 0 (0.0\%) & 101 (17.7\%) \\

\rowcolor[HTML]{F2F2F2}
\textbf{Baseline $O_2$ (litres per minute)} &  &  &  \\
Median (IQR) & 0 & 0 & 0 \\
(Missing) & 281 (34.0\%) & 123 (48.2\%) & 158 (27.6\%) \\
\end{tabular}
\end{adjustbox}
\caption{Baseline characteristics for the eligible patients. Continuous variables are summarised as median (IQR) and categorical variables are summarised as number (\%).}
\end{table}

\newpage
\printbibliography

\end{document}

% **********************
% end of document
% **********************